\documentclass[a4paper,11pt]{article}
\pdfoutput=1
\usepackage{jheppub}

\usepackage{slashed}
\usepackage{bm,wasysym}
\usepackage[normalem]{ulem}

\usepackage{amsmath}
\usepackage{amssymb,dsfont,bbold}

\usepackage[utf8x]{inputenc}
\usepackage{color}
\usepackage{graphicx,epsf}

\usepackage{booktabs}
\usepackage{hhline}

\def\beq{\begin{eqnarray}}
\def\eeq{\end{eqnarray}}

\def\cblue{}

\title{$D \to \rho \,\ell^+\ell^-$ Decays in the QCD Factorization Approach}

\author{Thorsten Feldmann,}
\author{Bastian M\"uller,}
\author{Dirk Seidel}

\affiliation{Theoretische Physik 1, 
Universit\"at Siegen, Walter-Flex-Stra\ss{}e 3, D-57068 Siegen, Germany}

\emailAdd{mueller@physik.uni-siegen.de}
\emailAdd{thorsten.feldmann@uni-siegen.de}
\emailAdd{dirk.seidel@uni-siegen.de}

\abstract{We consider rare semileptonic 
decays of a heavy $D$-meson into a light vector meson 
in the framework of QCD factorization. 
In contrast to the corresponding $B$-meson decays,
the naive factorization hypothesis does not even
serve as a first approximation. Rather, the decay amplitudes appear 
to be dominated by non-factorizable dynamics, e.g.\ 
through annihilation topologies, which are particularly
sensitive to long-distance hadronic contributions.
We therefore pay 
particular attention to intermediate vector-meson resonances
appearing in quark-loop and annihilation topologies.
Compared to the 
analogous $B$-meson decays, we identify 
a number of effects that result in very large theoretical uncertainties 
for differential decay rates. Some of these effects are found to cancel
in the ratio of partially integrated decay rates for transversely and
longitudinally polarized $\rho$ mesons.
On the phenomenological side this implies 
a very limited potential to constrain physics beyond 
the Standard Model by means of these decays.}


\note{Revised Version, 18 July 2017. \hfill 
Preprint:  SI-HEP-2017-09.}

\begin{document}

\maketitle

\section{Introduction}

In view of the persistent non-observation of any \emph{direct} signals for 
new particles or interactions at the high-energy frontier, which is currently
explored at the Large Hadron Collider (LHC),
\emph{indirect} probes of physics beyond the Standard Model (BSM) from 
low-energy observables gain ever more importance. In particular, depending 
on the specific model, precision measurements of 
rare flavour decays together with reliable theoretical predictions constrain 
the BSM parameter space for masses and coupling constants 
(for reviews and further references,
  see e.g.\ the according sections in \cite{Antonelli:2009ws,Bediaga:2012py,Buchalla:2008jp}).
This is particularly true for rare decays induced by 
flavour-changing neutral currents in the down-quark sector, i.e.\ 
$b \to s,d$ and $s \to d$ transitions, as well as $B$-$\bar B$ and $K$-$\bar K$-mixing.
On the other hand, rare $c \to u$ transitions are known to be plagued by 
serious theoretical uncertainties related to long-distance hadronic effects 
which are prominent because, due to
the small Yukawa coupling $y_b \ll y_t$, the GIM cancellation
is more efficient for $b$-quarks in the loops than for top quarks. 
Nevertheless, there have been 
a number of phenomenological studies on the new-physics (NP) sensitivities 
of rare charm decays, including radiative and 
rare semileptonic $D$-meson decays induced by $c \to u\gamma$
and $c \to u \ell^+\ell^-$ transitions, see e.g.\ 
\cite{Fajfer:2001sa,Burdman:2001tf,Fajfer:2005ke,Fajfer:2007dy,Paul:2011ar,Cappiello:2012vg,Isidori:2012yx,Lyon:2012fk,Fajfer:2012nr,Fajfer:2015mia,deBoer:2015boa,deBoer:2017que,Biswas:2017eyn}.

On the theoretical side,
as a first approximation, one may employ the naive factorization
  hypothesis which expresses the decay amplitudes in terms of 
  perturbatively calculable Wilson coefficients for $c \to u\gamma$ 
  and $c \to u \ell^+\ell^-$ transitions, multiplied by hadronic
  form factors for $D \to \pi$ or $D \to \rho$ transitions. 
  A simple model to estimate the non-factorizable long-distance effects is to  
  assume vector-meson dominance, i.e.\ to describe the radiative 
  decays via $ D \to \pi(\rho) \, V(\to \gamma/\ell^+\ell^-)$ with 
  suitable vector mesons $V$ that couple to the corresponding
  hadronic current in the weak effective Hamiltonian and decay
  into a charged lepton pair via electromagnetic interactions.\footnote{We 
  ignore in the following the contributions of intermediate 
  pseudoscalar resonances.} In such an approach, however, the 
  separation of short- and long-distance dynamics is no longer 
  manifest.

To proceed, the systematic inclusion of strong-interaction effects in the 
relevant hadron\-ic amplitudes requires additional approximations. In particular,
one may consider an expansion in inverse powers of the charm-quark mass, 
which -- however -- is expected to work less effectively than in the corresponding $B$-meson decays
because $m_c < m_b$. In such an approach, short-distance corrections in Quantum Chromodynamics (QCD)
related to distances $\Delta x \leq 1/m_c$ will be calculated perturbatively.
Radiative corrections between the electroweak scale 
and the charm-quark mass will be included in Wilson coefficients of the effective 
Hamiltonian for $|\Delta C|=|\Delta U| = 1$ transitions \cite{Greub:1996wn}. Recent next-to-leading order 
calculations for the relevant Wilson coefficients can be found in \cite{deBoer:2016dcg}.

It is known from the analogous radiative and semileptonic $b \to s,d$ decays that 
corrections to the hadronic matrix elements from higher orders in
a simultaneous expansion in the strong coupling and the inverse
heavy-quark mass lead to sizeable effects, in particular in the region
of large hadronic recoil (i.e.\ small invariant lepton mass $q^2$)
\cite{Beneke:2001at,Bosch:2001gv,Ali:2001ez,Kagan:2001zk,Feldmann:2002iw,Beneke:2004dp,Ali:2007sj,Bartsch:2009qp,Jager:2012uw}. 
Furthermore 
it seems a rather non-trivial task to 
construct realistic models for the effect of 
intermediate vector-meson resonances 
contributing to the $q^2$ spectrum in $B \to K^{(*)} \ell^+\ell^-$
{\cblue decays above and below the $c\bar c$ threshold, see e.g.\ the discussions in  \cite{Buchalla:1998mt,Grinstein:2004vb,Beylich:2011aq,Khodjamirian:2010vf,Lyon:2014hpa}.}\footnote{Notably, 
facing current experimental data in rare semileptonic $B$ decays \cite{Aaij:2013pta},
the region above the $c\bar c$ threshold behaves somewhat differently than
expected.} 
Of course, we expect these issues to be even more pronounced in 
radiative and rare semileptonic $D$-meson decays. On the one
hand, this severely limits the sensitivity to generic NP scenarios.
On the other hand,
the investigation of exclusive $c \to u\gamma^*$ transitions may 
help to better understand the hadronic uncertainties in 
exclusive $b \to s(d) \gamma^*$ decays. 

The aim of this paper therefore is to critically assess the theoretical
control on $D \to \rho\,\ell^+\ell^-$ (and also $D \to \pi\ell^+\ell^-$)
decays within the framework of QCD factorization (QCDF) \cite{Beneke:1999br,Beneke:2001ev},
closely following the analyses for the analogous $B$-meson decays 
in \cite{Beneke:2001at,Beneke:2004dp}. As a new ingredient we propose 
a simplified approach which allows to estimate the {\cblue potential} effect of light 
vector resonances on the level of the individual decay topologies that 
appear in QCDF (for simplicity, we restrict ourselves to the leading-order 
expressions in the strong coupling). 
To this end we model the tower of vector-meson resonances 
as in \cite{Blok:1997hs,Shifman:2000jv,Shifman:2003de} and 
connect it to the QCDF expressions via dispersion relations such 
that the asymptotic behaviour (i.e.\ far away from the resonances)
of the perturbative result is reproduced.
{\cblue Clearly, in this way we 
    ignore additional (non-factorizable) hadronic rescattering effects 
    that would potentially lead to a more complicated decay spectrum.
    As our main goal is \emph{not} to get a fully realistic description 
    of the differential decay width, our simple procedure should be 
    sufficient to get a rough estimate on the associated hadronic 
    uncertainties.
}

Our paper is organized as follows. In the next section, we 
give a brief overview over the theoretical framework, specifying
the operator basis in the weak effective Hamiltonian and providing
the definitions and factorization formulas for the generalized
form factors and coefficient functions appearing in the QCDF approach. In the following 
Section~\ref{sec:three} we provide detailed formulas for the 
contributions from the different decay topologies within QCDF.
Here, we remind the reader that in the ``naive factorization approximation'', only  
contributions from the electromagnetic dipole operator ${\cal O}_7$ 
and the semi-leptonic operators ${\cal O}_{9,10}$ appear.
Non-trivial contributions from the hadronic operators can be 
obtained by either closing two quark lines to a loop, or 
annihilating/pair-creating the valence quarks in the initial-
and final-state mesons. Radiative corrections at first order
of the strong coupling $\alpha_s$ are included by adapting the 
results in \cite{Beneke:2001at,Beneke:2004dp} to the 
corresponding $c \to u$ transitions; in particular, this includes non-factorizable spectator scattering 
effects. 
Notice that the $\alpha_s$
corrections to the annihilation topologies are presently unknown.
In Section~\ref{sec:four} we provide some numerical results 
for the individual contributions to the coefficient functions 
describing the decay amplitudes for transversely and longitudinally
polarized $\rho$ mesons for neutral and charged decay modes,
respectively. On that basis we give numerical estimates for
the central values and the dominating theoretical uncertainties
for the partially integrated transverse and longitudinal decay
rates, as well as for their ratio.
We summarize and conclude in Section~\ref{sec:five}.
In Appendix~\ref{app:Dtopi} we provide the explicit formulas
that allow to deduce the decay amplitudes for $D \to \pi\ell^+\ell^-$
decays from the corresponding expressions with longitudinally
polarized $\rho$ mesons. Appendix~\ref{app:model} gives a 
detailed derivation of the hadronic model that we have used to estimate 
the effect of light vector resonances. Finally, Appendix~\ref{app:input}
specifies a number of input parameters that have been used in 
the numerical analysis.

\section{Theoretical Framework}

In the following section, we briefly summarize the theoretical framework
and fix the notation used for the computation of the decay amplitudes.

\subsection{Effective Hamiltonian}

The low-energy effective Hamiltonian for $c \to u$ transitions
is written as follows,
\begin{align}
 \mathcal{H}_{\text{eff}} &= -\frac{4 G_F}{\sqrt{2}} \Big( \lambda_b \,
 \mathcal{H}_{\text{eff}}^{(b)} + \lambda_d \,  \mathcal{H}_{\text{eff}}^{(d)}\Big) \,,
\end{align}
with 
\begin{align}
  \mathcal{H}_{\text{eff}}^{(b)} &= C_1 \, \mathcal{O}_1^s + C_2 \, \mathcal{O}_2^s + \sum_{i=3}^9 C_i \, \mathcal{O}_i \,, \nonumber \\[0.2em]
   \mathcal{H}_{\text{eff}}^{(d)} &= C_1 (\mathcal{O}_1^s - \mathcal{O}_1^d) + C_2 (\mathcal{O}_2^s - \mathcal{O}_2^d)\,,
\end{align}
and 
\begin{align}
   \lambda_q &= V_{cq}^* V_{uq} \,.
\end{align}
The operators are defined in the CMM basis \cite{Chetyrkin:1997gb}. 
The current-current operators read 
\begin{align}
&{\cal O}^q_1 = (\bar{u}_L \gamma_{\mu} T^a q_L)(\bar{q}_L \gamma^{\mu} T^a c_L)\, ,\qquad 
{\cal O}^q_2 = (\bar{u}_L \gamma_{\mu} q_L) (\bar{q}_L  \gamma^{\mu} c_L)\, ,
\end{align}
where $q = d, s$.
The strong penguin operators are written as 
\begin{eqnarray}
{\cal O}_3 & = &(\bar{u}_L \gamma_\mu c_L) \sum_{q=u,d,s,c} (\bar{q} \gamma^\mu q) \,,
\qquad \quad \, {\cal O}_5  =  (\bar{u}_L \gamma_\mu \gamma_\nu \gamma_\rho c_L)
\sum_{q=u,d,s,c} (\bar{q} \gamma^\mu \gamma^\nu \gamma^\rho q) \,  ,\cr 
{\cal O}_4 & = &(\bar{u}_L \gamma_\mu T^a c_L) \sum_{q=u,d,s,c} (\bar{q}\gamma^\mu T^a q) \,,
\quad 
{\cal O}_6  = (\bar{u}_L \gamma_\mu \gamma_\nu \gamma_\rho T^a c_L)
\sum_{q=u,d,s,c} (\bar{q} \gamma^\mu \gamma^\nu \gamma^\rho T^a q) \,.
\cr &&
\end{eqnarray}
The electro- and chromomagnetic penguin operators are given by 
\begin{align}
& 
{\cal O}_7  = -\frac{g_{\rm em}m_c}{16\pi^2} (\bar{u}_L \sigma^{\mu \nu} c_R) F_{\mu \nu}\, ,
\qquad 
{\cal O}_8  = -\frac{g_s m_c}{16\pi^2} (\bar{u}_L \sigma^{\mu \nu} T^a c_R) G_{\mu
  \nu}^a \, ,
\end{align}
and, finally, the semi-leptonic operators are chosen as 
\begin{align}
& 
{\cal O}_9  = \frac{\alpha_{\rm em}}{4\pi} (\bar{u}_L \gamma_\mu c_L) (\bar{\ell} \gamma^\mu \ell)\,,
\qquad 
{\cal O}_{10}  =  \frac{\alpha_{\rm em}}{4\pi} (\bar{u}_L \gamma_\mu c_L)(\bar{\ell} \gamma^\mu 
\gamma_5 \ell)\, .
\end{align}
The higher-order QCD corrections to the various short-distance
Wilson coefficients $C_i$ 
have recently been calculated in \cite{deBoer:2016dcg}.
For convenience, we have summarized in Table~\ref{tab:Wilson} 
the SM values for the Wilson coefficients 
at LL and NLL (NNLL for $C_9$)
at a reference scale $\mu=\mu_c=1.5$~GeV. Notice that, contrary to $B$-meson decays,
the Wilson coefficient ${C}_{10}$ vanishes for $c\to u$ transitions
because of the perfect GIM cancellation at the electroweak matching scale
(using $m_b^2/M_W^2 \to 0$).

\begin{table}[t!!pb]
\begin{center}
\renewcommand{\arraystretch}{1.0}
\begin{tabular}{|c|c|c|c|c|c|c|}
 \hline
 \hbox{} & ${C}_1$ & ${C}_2$ & ${C}_3$ & ${C}_4$ & ${C}_5$ & ${C}_{6}$ \\
 \hline
 LL & -0.948 & 1.080 & -0.003 & -0.049 & 0.000 & 0.001 \\
 NLL &  -0.647 & 1.033 & -0.004 & -0.076 & 0.000 & 0.000 \\
 \hline \hline
 \hbox{} &  ${C}_{7,\rm eff}$ & ${C}_{8, \rm eff}$ & ${C}_9$ & ${C}_{10}$ & \hbox{} & \hbox{}\\
 \hline
 LL & 0.066 & -0.047 & -0.098 & 0 & \hbox{} & \hbox{} \\
 NLL & 0.042 & -0.052 & -0.288 & 0 & \hbox{} & \hbox{}\\
 NNLL & -- & -- & -0.445 & 0 & \hbox{} & \hbox{}\\
 \hline
 \end{tabular}
\end{center}
\caption{\label{tab:Wilson} SM values for the Wilson coefficients in $\mathcal{H}_{\rm eff}$ 
at LL and NLL (NNLL for $C_9$)
at a reference scale $\mu_c=1.5$~GeV, following the 
results of \cite{deBoer:2016dcg}.}
\end{table}

\subsection{Generalized form factors}

As has been shown in \cite{Beneke:2001at}, the hadronic matrix elements 
of the weak effective Hamiltonian simplify in the large-recoil limit,\footnote{Notice 
that for $D$-meson decays the 
large-recoil limit only refers to a rather restricted portion of phase space 
compared to $B$ decays, and since $m_D$ is not very large the convergence of the 
$\Lambda_{\rm QCD}/m_D$ expansion is expected to be rather slow.}
where 
$$
E \equiv \frac{p \cdot p'}{m_D} = \frac{m_D^2+m_\rho^2-q^2}{2 m_D} \ \to \ \frac{m_D}{2} \gg \Lambda_{\rm QCD} \,.
$$
In the following,
we will adopt the notation used in \cite{Seidel:2004jh,Beneke:2004dp}
(see also references therein).
For radiative decays of a $D$-meson into light vector mesons, we express 
the hadronic transition matrix elements 
in terms of generalized form factors ${\cal T}_a^{(i)}(q^2)$ 
which are defined via
\begin{align}
  & \langle \gamma^*(q,\mu) \, \rho^+(p',\varepsilon^*)| \mathcal{H}_{\rm eff}^{(i)} | D^+(p)\rangle 
\cr 
&= 
  \frac{i g_{\rm em} m_c}{4\pi^2} 
  \Bigg\{ 2 \, {\cal T}_\perp^{(i)}(q^2) \, \epsilon^{\mu\nu\rho\sigma} 
  \varepsilon_\nu^* p_\rho p'_\sigma  
  \cr 
  & \qquad -  2 i \, {\cal T}_\perp^{(i)}(q^2) \left[ 
  \left( p \cdot p' - \frac{m_\rho^2 m_D^2}{p\cdot p'} \right)  \varepsilon^{*\mu} 
  - (\varepsilon^*\cdot p) \left( p'{}^{\mu} - \frac{m_\rho^2}{p\cdot p'} \, p^\mu \right) 
  \right] 
  \cr 
  & \qquad -  i \, {\cal T}_\parallel^{(i)}(q^2) \, (\varepsilon^* \cdot p) 
  \left[q^\mu - \frac{q^2}{m_D^2-m_\rho^2} \, (p^\mu + p'{}^{\mu}) \right] \Bigg\}\,.
  \label{master}
  \end{align}
Here for each set of operators ($i=b,d$) 
only two independent generalized form factors appear, referring to transversely or
longitudinally polarized vector mesons ($a=\perp,\parallel$), respectively. 
They can be further factorized in the form 
\begin{align}
 \mathcal{T}_a^{(i)} &\simeq \xi_a \, C_a^{(i)} + \frac{\pi^2}{N_c} \,
 \frac{f_D f_{\rho,a}}{m_D} \, \Xi_a  \,
 \sum_\pm \int \frac{d\omega}{\omega} \, \phi_{D,\pm}(\omega) \,
 \int_0^1 du \, \phi_a(u) \, T_{a,\pm}^{(i)}(u,\omega)
 \label{Tadecomp}
\end{align}
with $\Xi_\perp \equiv 1$, $\Xi_\parallel \equiv \tfrac{m_\rho}{E}$.
In the first term,
the functions $\xi_a$ denote the universal (``soft'') form factors for 
$D \to \rho$ transitions in the large-recoil limit \cite{Beneke:2000wa},
and the coefficient functions $C_a^{(i)}$ contain the QCD corrections to 
the partonic $c \to u\ell^+\ell^-$ amplitude, dubbed ``form factor corrections''
in the following. 
The second term contains the spectator effects (including annihilation topologies)
and is proportional to 
the meson decay constants ($f_D$ and $f_{\rho,\perp(\parallel)}$).
It contains a convolution integral of the relevant 
light-cone distribution amplitudes (LCDAs), $\phi_{D,\pm}(\omega)$ and $\phi_{\perp,\parallel}(u)$,
and a hard-scattering kernel denoted as $T_{a,\pm}^{(i)}(u,\omega)$.
It is to be stressed already at this point that the spectator interactions 
involve typical gluon virtualities of order $\sqrt{m_D \Lambda_{\rm QCD}}\sim 1$~GeV, and
therefore the corresponding value of the strong coupling constant $\alpha_s$ 
and the associated perturbative uncertainty from scale variations 
will be large.

Furthermore, we define the following coefficient functions, which are independent of the conventions 
chosen to renormalize the weak effective Hamiltonian,
\begin{align}
 \mathcal{C}^{(i)}_{9,\perp}(q^2) &\equiv \delta^{ib} \, C_9 + \frac{2 m_c m_D}{q^2} \frac{\mathcal{T}_\perp^{(i)}(q^2)}{\xi_\perp(q^2)} \,, \quad 
 \mathcal{C}^{(i)}_{9,\parallel}(q^2) \equiv \delta^{ib} \, C_9 - \frac{2 m_c}{m_D} \frac{\mathcal{T}_\parallel^{(i)}(q^2)}{\xi_\parallel(q^2)} \,.
\label{eq:calC}
\end{align}
In terms of these, we obtain the double-differential decay rate as \cite{Beneke:2004dp}
\begin{align}
 \frac{d^2\Gamma}{dq^2 d\rm{cos} \theta} &= \frac{G_F^2}{128 \pi^3} \, m_D^3 \, S \, \lambda_D(q^2,m_\rho^2)^3 \, \big(\frac{\alpha_{\rm{em}}}{4 \pi} \big)^2 \times \Bigg[\cr
                                         & (1 + \rm{cos}^2 \theta) \, \frac{2 q^2}{m_D^2} \, \xi_\perp(q^2)^2 \left|\lambda_d \, \mathcal{C}_{9,\perp}^{(d)} + \lambda_b \, \mathcal{C}_{9,\perp}^{(b)}\right|^2\cr 
                                         & (1 - \rm{cos}^2 \theta) \, \Big( \frac{E \, \xi_\parallel(q^2)}{m_\rho} \Big)^2 \left|\lambda_d \, \mathcal{C}_{9,\parallel}^{(d)} + \lambda_b\, \mathcal{C}_{9,\parallel}^{(b)}\right|^2 \Bigg] \,,
\label{eq:diffrate}
\end{align}
with $S=1$ for $\rho^-$ and $S = 1/2$ for $\rho^0$. Here 
\begin{align}
\lambda_D(q^2,m_\rho^2) &= \left[\left(1- \frac{q^2}{m_D^2} \right)^2 - \frac{2m_\rho^2}{m_D^2}
\left( 1+ \frac{q^2}{m_D^2} \right) + \frac{m_\rho^4}{m_D^4} \right]^{1/2} 
\end{align}
is the standard kinematic prefactor.
Note, that we have used $C_{10}=0$, which also implies that 
the forward-backward asymmetry with respect to the angle $\theta$ vanishes.\footnote{In 
that respect the foward-backward asymmetry provides a null test of the SM. However,
in specific NP models, the forward-backward asymmetry may still remain too small
to be measured with reasonable experimental sensitivity, see e.g.\ the discussion in \cite{Paul:2011ar,Fajfer:2005ke}.}

For decays into light pseudoscalar mesons, analogous functions ${\cal T}_p^{(i)}$ 
can be defined \cite{Beneke:2001at,Buchalla:2010jv}. The explicit formulas will 
be provided in Appendix~\ref{app:Dtopi} for completeness.

\section{Detailed Analysis of Decay Topologies}

\label{sec:three}

\subsection{Naive factorization}

\begin{figure}[t!!pbh]
\begin{center}
  \fbox{\includegraphics[width=0.46\textwidth]{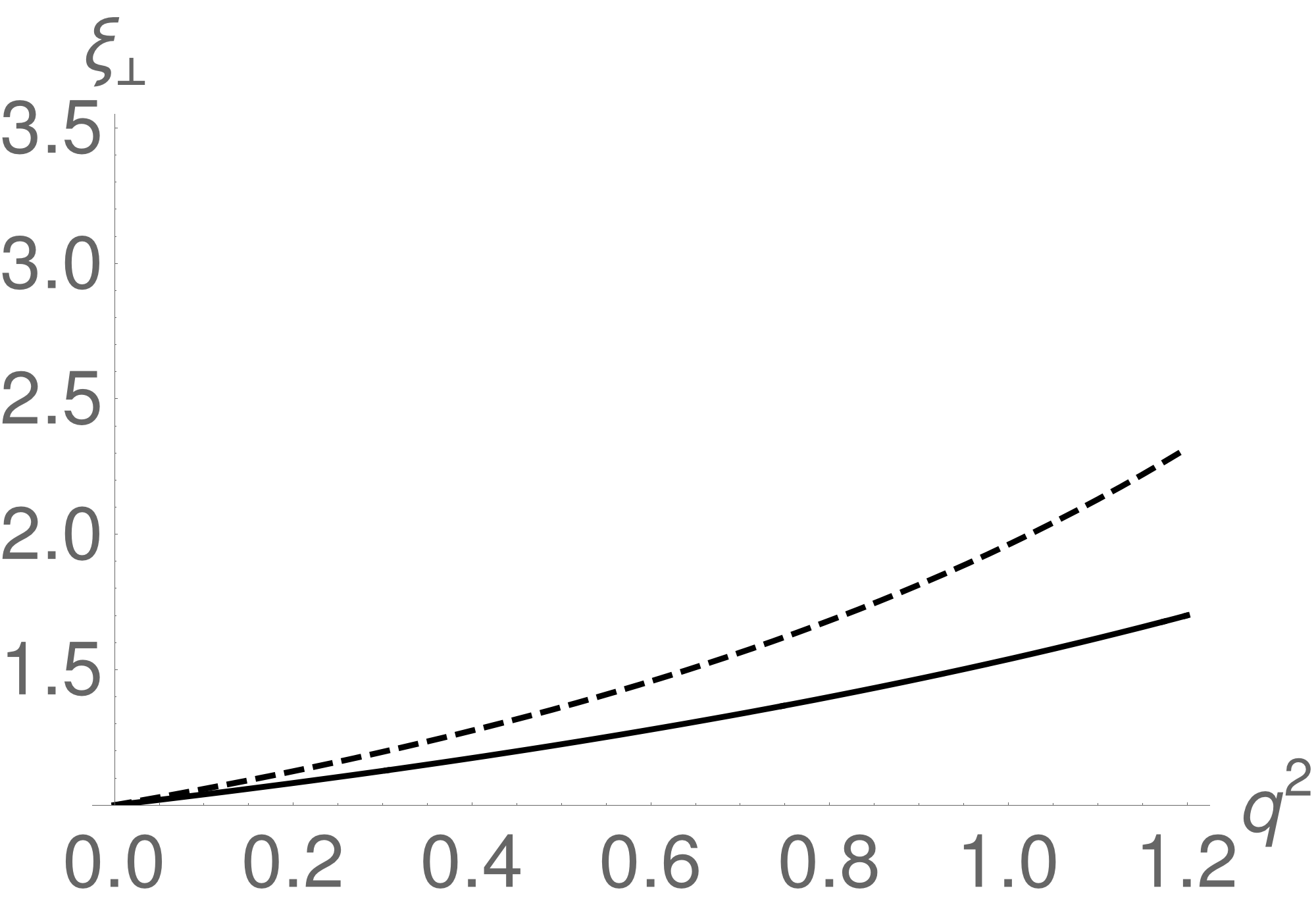}} \quad 
  \fbox{\includegraphics[width=0.46\textwidth]{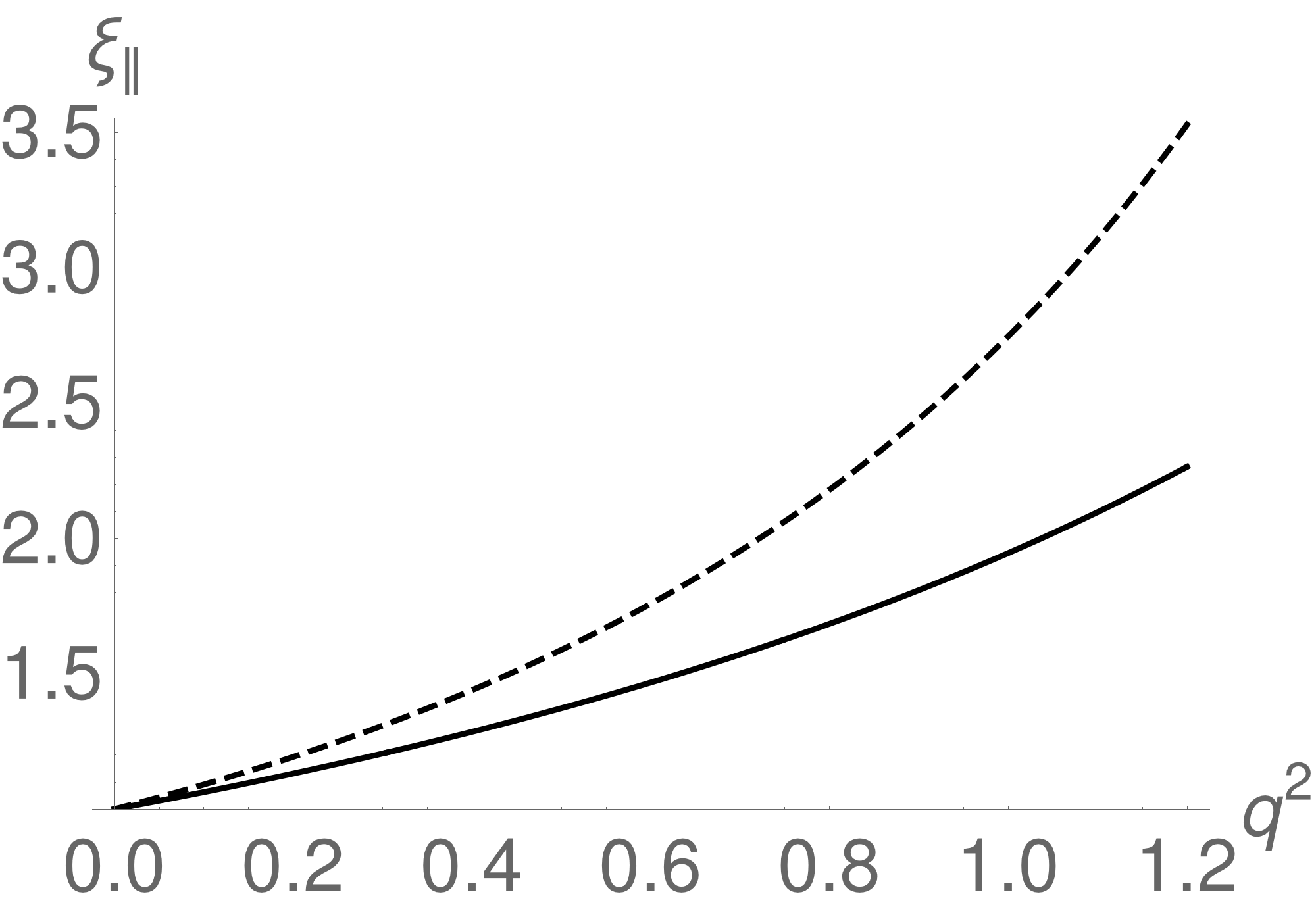}} 
\end{center}
\caption{\label{fig:FFcomp} The dependence of the $D\to \rho$ 
transition form factors as a function of momentum transfer $q^2$
(in units of GeV$^2$, normalized to $q^2=0$), following from (\ref{eq:ffscaling}) [dashed line], 
compared  to the parametrization determined in \cite{Fajfer:2005ug} [solid line].}
\end{figure}

In naive factorization, only the Wilson coefficients associated to the 
operators ${\cal O}_{7,9,10}$ contribute,
multiplied by the corresponding vector, axial-vector and tensor form factors
for $D \to \rho$ transitions. In the large recoil limit, one could further 
reduce these form factors to the two ``soft'' form factors,
\begin{align}
  \xi_\perp(q^2) &\equiv \frac{m_D}{m_D + m_\rho} \, V(q^2)
\,, \cr   
  \xi_\parallel(q^2) & \equiv \frac{m_D \, (m_D + m_\rho)}{m_D^2-q^2} \, A_1(q^2) 
  - \frac{m_D - m_\rho}{m_D} \, A_2(q^2) \,,
\end{align} 
appearing in (\ref{master}). 
The normalization of the form factors at zero momentum transfer
has been measured experimentally by the CLEO collaboration
\cite{CLEO:2011ab}, as summarized in Table~\ref{tab:hadronic} below.
For the $q^2$ dependence, we adopt the 
parametrization that has been proposed in 
\cite{Fajfer:2005ug}, where 
\begin{align}
A_{1,2}(q^2) &= \frac{A_{1,2}(0)}{1- b' x} \,, \qquad 
V(q^2) = \frac{V(0)}{(1 - x) \, (1 - a x)} \,,
\label{eq:ffmodel}
\end{align}
with $x=q^2/M_{D^*}^2$ and $a=0.55$, $b'= 0.69$.
The $q^2$-dependence of the soft form factors is shown 
in Fig.~\ref{fig:FFcomp} and compared to the 
naive scaling behaviour 
\begin{align}
  &\xi_\perp(q^2) 
  \simeq \frac{\xi_\perp(0)}{(1-q^2/m_D^2)^2} \,, 
  \qquad  
  \xi_\parallel(q^2) \simeq   \frac{\xi_\parallel(0)}{(1-q^2/m_D^2)^3} \,, 
  \label{eq:ffscaling}
\end{align}
which follows 
in the framework of SCET/QCDF at tree level \cite{Beneke:2000wa}.
As one can observe, the latter systematically leads to a somewhat too steep
$q^2$-dependence. We will therefore stick to the parametrization (\ref{eq:ffmodel})
in our numerical analysis below.

\subsubsection{Corrections to the {\cblue $D \to \rho$} form factors in the large recoil limit}

Adapting the terminology of \cite{Beneke:2001at,Beneke:2004dp},
\emph{factorizable form-factor corrections}
are accounted for by 
\cite{Beneke:2004dp}
\begin{align}
 C_\perp^{({\rm f},b)} &= C_7^{\text{eff}} \Big(\ln \frac{m_c^2}{\mu^2} - L + \Delta M \Big) \,, 
 \qquad 
 C_\parallel^{({\rm f},b)} = -C_7^{\text{eff}} \Big(\ln \frac{m_c^2}{\mu^2} + 2 L + \Delta M\Big) \,;
\end{align}
whereas $C_{\perp,\parallel}^{({\rm f},d)}=0$.
Here 
\begin{align}
L &= - \frac{m_c^2-q^2}{q^2} \, \ln \left(1 - \frac{q^2}{m_c^2} \right)
\,,
\end{align}
and $\Delta M$ depends on the convention to define the charm-quark mass.
In this work, we choose the {$\overline{\rm MS}$} scheme for simplicity, 
which amounts to setting $\Delta M = 0$.
With our convention for defining the ``soft'' $D \to \rho$ form
factors, the \emph{factorizable spectator corrections} to the form factors 
are reflected in the contributions
\begin{align}
 T_{\perp,+}^{(\text{f},b)} =
 T_{\parallel,+}^{(\text{f},b)} &= C_7^{\text{eff}} \, \frac{4 m_D}{\bar{u} E} \,,
\end{align}
while $ T_{a,-}^{(\text{f},b)} =0$ 
and $T_{a,\pm}^{(\text{f},d)}=0$ in the large-recoil limit.

\subsection{Quark-loop topologies (w/o spectator effects)}
\label{sec:quarkloopLO}

\begin{figure}[t!!pb]
\begin{center}
 \fbox{\includegraphics[width=0.46\textwidth]{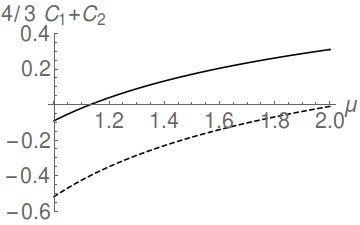}}
\end{center}
\caption{\label{fig:43c1c2}
The combination of Wilson coefficients $(4/3 \, C_1 + C_2)$
as a function of the renormalization scale $\mu$ (in units of GeV)
at LL (dashed line) and NLL (solid line) accuracy.
}
\end{figure}

Closing two quark lines from the 4-quark operators and radiating
a (virtual) photon from the quark loop results in form-factor
corrections to naive factorization that contribute to the 
\emph{effective} Wilson coefficients in (\ref{eq:calC}) as follows,
\begin{align}
 C_\perp^{(0,i)} &= \delta^{ib} \, C_7^{\text{eff}} + \frac{q^2}{2 m_c \, m_D} \,Y^{(i)}(q^2) \,, \qquad 
 C_\parallel^{(0,i)} = -\delta^{ib} \, C_7^{\text{eff}} - \frac{m_D}{2 m_c} \, Y^{(i)}(q^2) \,.
\end{align}
Here, the 1-loop functions $Y^{(i)}(s)$
can be decomposed as follows \cite{deBoer:2016dcg},
\begin{align}
 Y^{(b)}(s) & = \left[h(s, m_c) + h(s,m_u) \right]
 \Big(7 \, C_3 + \frac{4}{3} \, C_4 + 76 \, C_5 + \frac{64}{3} \, C_6\Big) \cr 
            & \quad {}- h(s, m_s) \Big(\frac{2}{3} \, C_1 + \frac{1}{2} \, C_2 + 3 \, C_3 + 30 \, C_5\Big) \cr 
            & \quad {}- h(s, m_d) \Big( 3 \, C_3 + 30 \, C_5\Big)
            + \frac89 \left( 3 \, C_3 + 16 \, C_5 + \frac{16}{3} \, C_6 \right) \,, 
            \label{Yb}
\end{align}
and
\begin{align}
  Y^{(d)}(s) &= -\Big( \frac{2}{3} \, C_1 +  \frac12 \, C_2\Big) \lbrack h(s, m_s) - h(s, m_d) \rbrack \,,
\label{Yd}
\end{align}
where the function $h(s,m)$ can be found, for instance, in \cite{Buchalla:1995vs},
and -- with our normalization convention\footnote{Our definition of $h(s,m)$
differs from the one in \cite{deBoer:2016dcg}
by a relative factor $(-1/2)$. We also remark that the definition of the constant piece in
the function $Y^{(b)}(s)$ depends on the operator basis, and only the sum $C_9+Y^{(b)}(s)$ 
is basis-independent. The quoted results for $C_9$ in Table~\ref{tab:Wilson} and 
$Y^{(b)}(s)$ in (\ref{Yb}) refer to the CMM basis \cite{Chetyrkin:1997gb}.} 
-- is 
given in (\ref{eq:hfunc}) in the appendix. Notice that the contribution 
of the function $Y^{(b)}(s)$ to the decay amplitudes is strongly CKM suppressed.
On the other hand, the function $Y^{(d)}(s)$ vanishes in the limit $m_s \to m_{u,d} \simeq 0$.
Furthermore, at scales of the order of the charm-quark mass one encounters 
numerical cancellations in the combination of Wilson coefficients $(4/3 \, C_1 + C_2)$; 
the effect is illustrated in Fig.~\ref{fig:43c1c2}. In particular, one observes a 
large shift when going from the LL to the NLL result, indicating that this combination
of Wilson coefficients is particularly affected by higher-order radiative corrections,
see also the discussion around Fig.~\ref{fig:FF}.

At this point it is to be stressed that, strictly speaking, the perturbative calculation
of the loop-function $h(s,m)$ is only justified in the deep Euclidean region, where 
$- s = Q^2 \gg \Lambda_{\rm QCD}^2$. In contrast, for time-like momentum transfer, $s=q^2>0$, 
the $q^2$-dependence would rather be given by a \emph{hadronic} spectral function that 
describes the effects of a tower of light vector mesons and multi-hadron final states
with the appropriate quantum numbers. While in the corresponding $B$-meson decays,
the annihilation topologies only provide a correction to the total decay amplitude,
this is no longer the case for $D \to \rho\,\ell^+\ell^-$ decays, 
{\cblue see also earlier estimates in Refs.~\cite{Khodjamirian:1995uc,Lyon:2012fk}.}
In addition, for $B$-meson decays
the region of 
small momentum transfer, $4 m_\ell^2 \leq q^2 \lesssim \mbox{1-2~GeV$^2$}$, could simply be 
excluded from the phenomenological analysis. Due to the restricted phase space this
 is no longer possible in $D$-meson decays. For this reason one
should carefully discuss the associated parametric and systematic hadronic uncertainties.
Inevitably, this requires some hadronic modelling.
In particular, the amount of GIM cancellation in the difference $h(s,m_s)-h(s,m_d)$ 
in the perturbative calculation will be quite different from estimates based on
hadronic models. 

A straightforward approach, which has
already been extensively used in the past \cite{Fajfer:2007dy,Cappiello:2012vg,Fajfer:2012nr,Fajfer:2015mia,deBoer:2017que}, 
assumes that the long-distance 
hadronic effects are completely dominated by the lowest-lying narrow vector states
($\rho,\omega, \phi$ etc.) which are then modelled by Breit-Wigner resonances.
In this work, we propose a more involved (but still oversimplified) picture, following
\cite{Shifman:2000jv,Shifman:2003de},\footnote{This model has also been used to 
estimate the effects of higher charmonium resonances in $B\to K\ell^+\ell^-$ decays 
at large values of $q^2$ \cite{Beylich:2011aq}.} where 
(i) a model for the infinite tower of higher resonances is included, and (ii)
the shape of the Breit-Wigner resonances for the lowest states is modified to be in accordance with
simple (leading-order) analyticity arguments. As explained in more detail in Appendix~\ref{app:model},
our idea amounts to replacing
\begin{align}
\label{eq:hmod}
h(q^2,m_q) & \to h(-\sigma^2,m_q) 
  + \frac49 \, \int_0^\infty ds \, \frac{\sigma^2+q^2}{\sigma^2+s} 
  \, \frac{j_{q}(s)}{s-q^2-i\epsilon} \,,
\qquad (m=0,m_s) \end{align}
where the precise form of the spectral function $j_q(s)$, which models the effect 
of hadronic vector resonances with the corresponding quantum numbers,\footnote{For
simplicity, we do not distinguish between $\rho$ and $\omega$ mesons and set $m_u=m_d=0$.
Moreover, as $q^2 < 4m_c^2$ in open charm decays, we do not have to modify the perturbative
result for $h(s,m_c)$.} and the meaning of the parameter $\sigma^2$
can be found in Appendix~\ref{app:model}.
In Fig~\ref{fig:plothsminushd} we illustrate the effect of the resonance model
on the difference $[h(q^2,m_s) - h(q^2,m_d)]$ in the low-$q^2$ region,
relevant for $D \to \rho$ decays. One observes that neither the sign nor the 
order of magnitude nor the shape of the hadronic model can be reproduced by
the perturbative result (despite the fact  that -- by construction --
for $q^2<0$ one has almost perfect
numerical agreement, see Figs.~\ref{fig:hud},\ref{fig:hs} in the appendix).

It is important to note that the numerical effect of the vector resonances 
in rare semileptonic $D$-decays will be quite different from $B$-decays or $e^+e^-$ 
annihilation. While in the latter case, the cross section for $e^+e^- \to \text{hadrons}$
will be given by the \emph{imaginary} part of the loop function, in $B$-decays the 
dominating correction to the decay width 
will arise from the interference with the short-distance terms (which do not yield
strong phase differences), and therefore one will probe the \emph{real} part of 
the loop function (see e.g.\ \cite{Beylich:2011aq}). 
In $D$-meson decays, however, the short-distance part is heavily suppressed,
and therefore we expect the dominating effect to be given by the \emph{absolute} value of 
the loop function, see Fig.~\ref{fig:plothsminushd}.

\begin{figure}[t!!!pbh]
\begin{center}
\fbox{\includegraphics[width=0.45\textwidth]{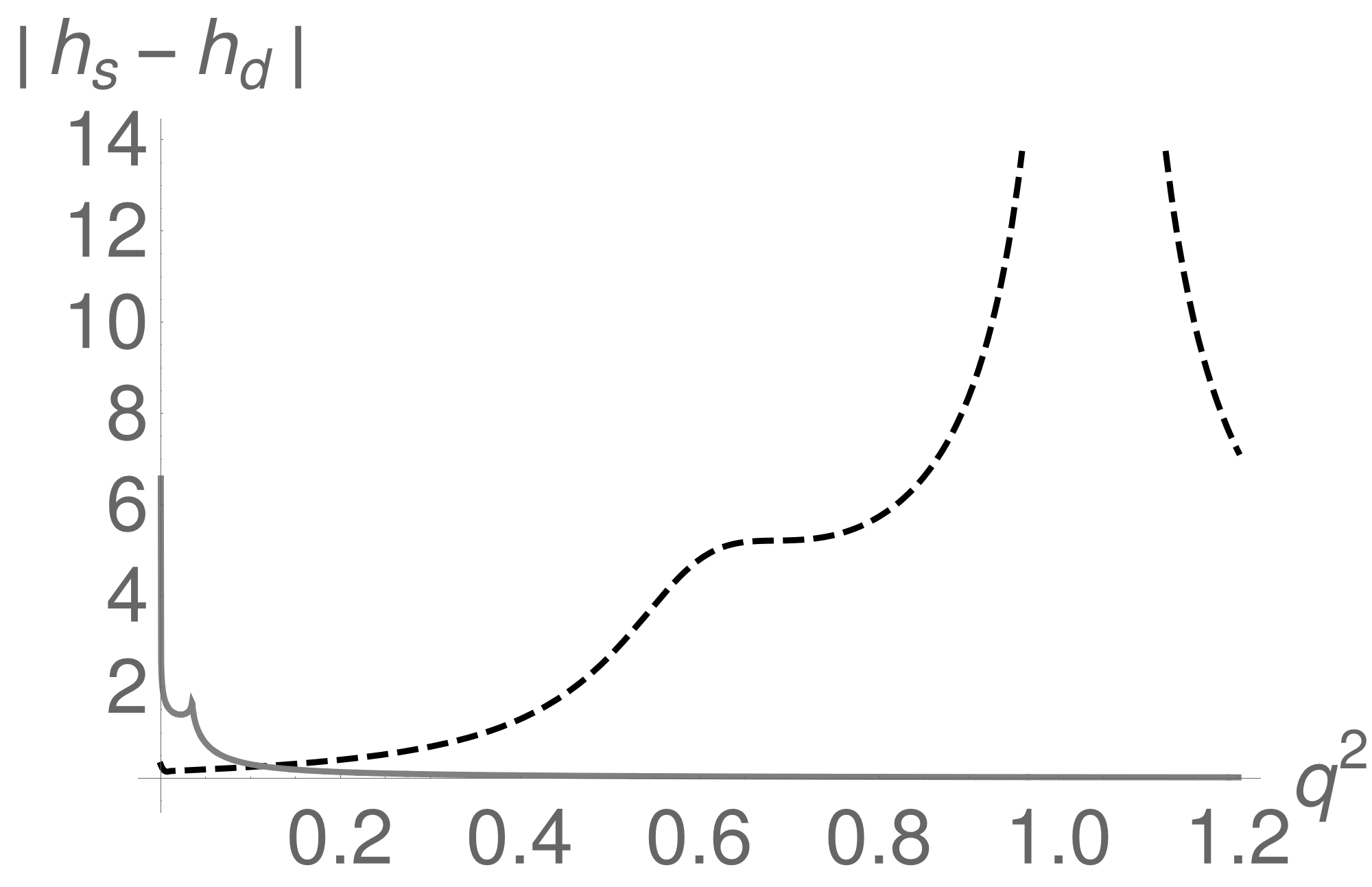}}  \quad 
\fbox{\includegraphics[width=0.47\textwidth]{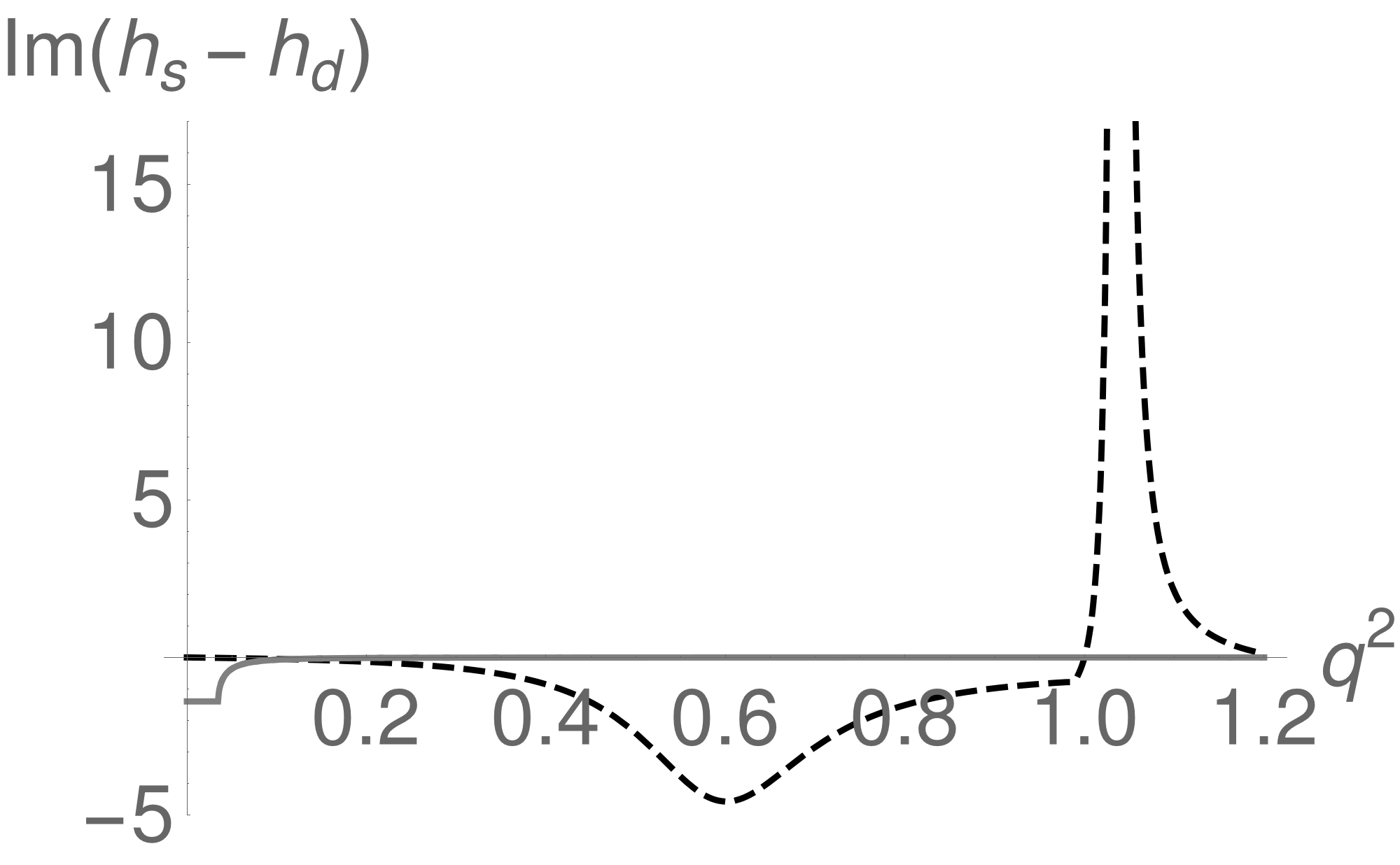}} 
\end{center}
\caption{Comparison of the perturbative result for the absolute value and the imaginary part of 
the function $h(q^2,m_s)-h(q^2,m_d)$ 
(gray solid line) and the model (\ref{eq:hmod}) for its hadronic modification  
(black dashed line), as a function of momentum transfer $q^2$ (in units of GeV$^2$).
The parameter values are chosen as follows. 
For the modelling of the $d$-quark loop ($m_d=0$), we take 
$\sigma^2=2~\text{GeV}^2$, $a=1$, $b=1/6$. 
For the $s$-quark loop ($m_s=100$~MeV), we use 
$\sigma_s^2=2~\text{GeV}^2$, $a_s=1.4$, $b_s=0.1$.
The parameters $n_V=1.94$ and $n_\phi=2.37$ 
are tuned  to reproduce the perturbative result for $h(q^2,m_{s/d})$
in the limit $q^2 \to -\infty$.
For details, see appendix~\ref{app:model}.
\label{fig:plothsminushd}}
\end{figure}

\subsubsection{Non-factorizable $c\to u$ form-factor corrections}

The remaining \emph{non-factorizable} contributions can be expressed in terms 
of $q^2$-dependent short-distance functions $F_i^{(j)}$, which enter
as follows (see \cite{Beneke:2001at,Beneke:2004dp})
\begin{align}
 C_F C_\perp^{({\rm nf},i)} &= -\Big(-C_1 (F_1^{(7)} - \delta_{id} F_{1,d}^{(7)}) - C_2 (F_2^{(7)} - \delta_{id} F_{2,d}^{(7)}) + \delta_{ib} C_8^{\text{eff}} F_8^{(7)}\nonumber \\
                      &- \frac{q^2}{2 m_c m_D} \Big[C_1 (F_1^{(9)} - \delta_{id} F_{1,d}^{(9)}) + C_2 (F_2^{(9)} - \delta_{id} F_{2,d}^{(9)}) - \delta_{ib} C_8^{\text{eff}} F_8^{(9)}\Big]\Big)
                      \,,  \nonumber 
                      \\[0.3em]
C_F C_\parallel^{({\rm nf},i)} &= -\Big(C_1 (F_1^{(7)} - \delta_{id} F_{1,d}^{(7)}) + C_2 (F_2^{(7)} - \delta_{id} F_{2,d}^{(7)}) - \delta_{ib} C_8^{\text{eff}} F_8^{(7)}\nonumber \\
                      &+ \frac{m_D}{2 m_c} \Big[C_1 (F_1^{(9)} - \delta_{id} F_{1,d}^{(9)}) + C_2 (F_2^{(9)} - \delta_{id} F_{2,d}^{(9)}) - \delta_{ib} C_8^{\text{eff}} F_8^{(9)}\Big]\Big) \,.
                      \label{Ffunc}
\end{align}
At this point, a few comments are in order on how to obtain the functions $F_i^{(j)}$ from 
the  expressions which have been calculated for the corresponding $ b \to (s,d)\gamma^*$ amplitudes:
\begin{itemize}
\item The functions $F_8^{(7)}$ and $F_8^{(9)}$ can be obtained 
by changing the corresponding charge factors (which amounts to  multiplying 
by ${Q_u}/{Q_d}$, where $Q_{u,d}$ are the charge factors for 
up- and down-quarks, respectively).

\item In order to get the functions $F_{1,d}^{(7)}, F_{1,d}^{(9)}, F_{2,d}^{(7)}$ and $F_{2,d}^{(9)}$,
one has to go back to the bare (i.e.\ unrenormalized) functions from $b$-decays \cite{Seidel:2004jh}, 
       replace the charge factors and renormalize the functions. 
       Here, the renormalization constant with general charge factors 
       can be derived from the results given in \cite{Gambino:2003zm} and reads 
       \begin{align}
       \setcounter{MaxMatrixCols}{20}
        Z &= \frac{\alpha_s}{4 \pi} \, \frac{1}{\epsilon} \,
        \begin{pmatrix} -2 & \frac{4}{3} & 0 & -\frac{1}{9} & 0 & 0 & 0 & 0 & -\frac{8 Q_u}{9} 
        & 0 & \frac{5}{12} & \frac{2}{9}\\[0.1em]
                                                    6 & 0 & 0 & \frac{2}{3} & 0 & 0 & 0 & 0 & 
                                                    -\frac{2 Q_u}{3} & 0 & 1 & 0
                                    \end{pmatrix} \nonumber \\
          &+ \Big(\frac{\alpha_s}{4 \pi}\Big)^2 \frac{1}{\epsilon} \, 
          \begin{pmatrix}
             0 & 0 & 0 & 0 & 0 & 0 & \frac{4 Q_d}{81} - \frac{Q_u}{3}
             & 0 & \frac{-44 Q_d}{243} - \frac{2 Q_u}{9} & 0 & 0 & 0\\[0.1em]
             0 & 0 & 0 & 0 & 0 & 0 & \frac{-8 Q_d}{27} + 2 Q_u & 0 & 
             \frac{88 Q_d}{81} + \frac{16Q_u}{3} 
              & 0 & 0 & 0
                                                \end{pmatrix}\nonumber \\
          &+ \Big(\frac{\alpha_s}{4 \pi}\Big)^2 \frac{1}{\epsilon^2} \, 
          \begin{pmatrix}
             0 & 0 & 0 & 0 & 0 & 0 & 0
             & 0 &  - \frac{4(-2 Q_d + 3(-69 + 4 n_f) Q_u)}{81} & 0 & 0 & 0\\[0.1em]
             0 & 0 & 0 & 0 & 0 & 0 & 0 & 0 & 
             - \frac{2(8 Q_d + 3(-21 + 2 n_f) Q_u)}{27} & 0 & 0 & 0
                                                \end{pmatrix}.
       \end{align}
 
\item The results for the 
functions $F_{1}^{(7)}, F_{1}^{(9)}, F_{2}^{(7)}$ and $F_{2}^{(9)}$ 
with general charge factors have been reconstructed 
from Mathematica notebooks that have been kindly provided by
Christoph Greub (related to the work in \cite{Greub:2008cy}).
 
\item For the limit $q^2 \to 0$, 
the relevant functions $F_{1}^{(7)}$ and $F_{2}^{(7)}$ 
can be directly extracted from \cite{Greub:1996tg}.

\end{itemize}

\begin{figure}[tpb]
\begin{center}
 \fbox{\includegraphics[width=0.46\textwidth]{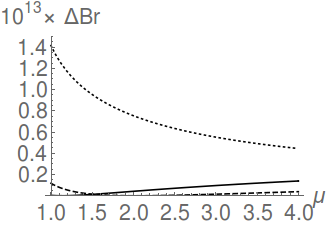}}
\end{center}
\caption{\label{fig:FF}
Contribution of quark-loop topologies to the differential branching ratio,
integrated over a $q^2$-bin $[0.5,0.7]~{\rm GeV}^2$,
as a function of the renormalization scale $\mu$ (in units of GeV),
restricted to the contributions 
proportional to $\lambda_d$. 
The LO result with LL (NLL) Wilson coefficients is shown 
as the dashed (solid) line. The NLO contributions with NLL Wilson coefficients
is shown as the dotted line. 
}
\end{figure}
As explained in Section~\ref{sec:quarkloopLO}, the leading contributions from
quark-loop topologies in $D \to \rho$ transitions suffer from
a renormalization-scale uncertainty due to partial numerical cancellations
in the combination of Wilson coefficients $ (\frac43 \, C_1 +C_2)$.
The higher-order corrections encoded in (\ref{Ffunc}) are expected to
reduce this uncertainty.
To numerically investigate the effect, we plot in Fig.~\ref{fig:FF} the 
contribution to the partially integrated branching ratio, $0.5~{\rm GeV}^2 \leq q^2 \leq 0.7~{\rm GeV}^2$,
of the quark-loop topologies as a function of the renormalization scale $\mu$
(where we have restricted ourselves to the CKM-favoured contributions proportional 
to $\lambda_d$). 
We observe that the NLO corrections actually dominate over the LO contribution.
On the one hand, this removes the issue with the accidental numerical
cancellations at LO. On the other hand, it implies a very slow convergence of
the perturbative series, i.e.\ large renormalization-scale uncertainties even
at NLO. (We remind the reader that annihilation and spectator-scattering topologies
will induce additional and formally independent scale uncertainties.)

\subsection{Annihilation}

 As already mentioned, the so-called annihilation topologies will turn out to give
large contributions to the decay rate, and therefore
the associated hadronic uncertainties will
be essential for
phenomenological studies. The fact that 
the quark propagators in the annihilation diagrams involve 
\emph{time-like} virtualities implies a particular sensitivity to
the modelling of hadronic resonance effects. Furthermore, 
annihilation diagrams with the photon radiated from the 
quarks in the final-state meson are formally power-suppressed 
in the $1/m_c$ expansion, but may still be phenomenologically
important. In particular, radiative corrections to these
topologies have not been systematically computed in QCDF so far.
As a consequence, the ambiguities related to the renormalization-scale 
setting in the relevant Wilson coefficients will remain a major 
source of theoretical uncertainties.

In the heavy-quark limit, the leading contributions from quark annihilation topologies
originate from photon radiation off the light-quark in the $D$-meson. The results thus 
depends on the charge factor $e_q$ of the spectator quark.
Translating the results from $b \to s,d$ to $c \to u$ transitions,
we obtain 
\begin{align}
 m_c \, T_{\parallel,-}^{(0,b)} &= -e_q \, 
 \frac{ 4  m_D^2 \, \omega}{m_D \, \omega- q^2 - i\epsilon} \,
 (\delta_{qd} - \delta_{qu}) \Big[C_3 +
 \frac{4}{3} \Big(C_4 + 12 \, C_5 + 16 \, C_6\Big)\Big] \,, \nonumber  \\[0.2em]
 m_c \, T_{\parallel,-}^{(0,d)} &= e_q \,
 \frac{ 4 m_D^2 \, \omega}{m_D \,\omega- q^2 - i\epsilon}  \Big[\delta_{qd} \, 3  C_2 
 + \delta_{qu} \Big(\frac{4}{3} \, C_1 + C_2\Big)\Big] \,.
\label{anni:LO}
\end{align}
Again, the contributions from $T_{\parallel,-}^{(0,b)}$ will 
be strongly CKM suppressed. Therefore,
for charged $D$-meson decays ($q=d$) the annihilation contribution
is triggered by the Wilson coefficient $C_2$,
while for   neutral $D$-meson decays ($q=u$) it is proportional
to the same combination $\frac{4 C_1}{3} + C_2$ as appearing in the
quark-loop topologies. As a consequence,
we expect that charged $D$-meson decays will actually 
be dominated by annihilation topologies, since the partial numerical
cancellation in Wilson coefficients does not occur here. In contrast, 
in neutral $D$-meson decays annihilation and quark-loop topologies will enter
with similar magnitudes.

\begin{figure}[tpbh]
 \begin{center}
\fbox{\includegraphics[width=0.46\textwidth]{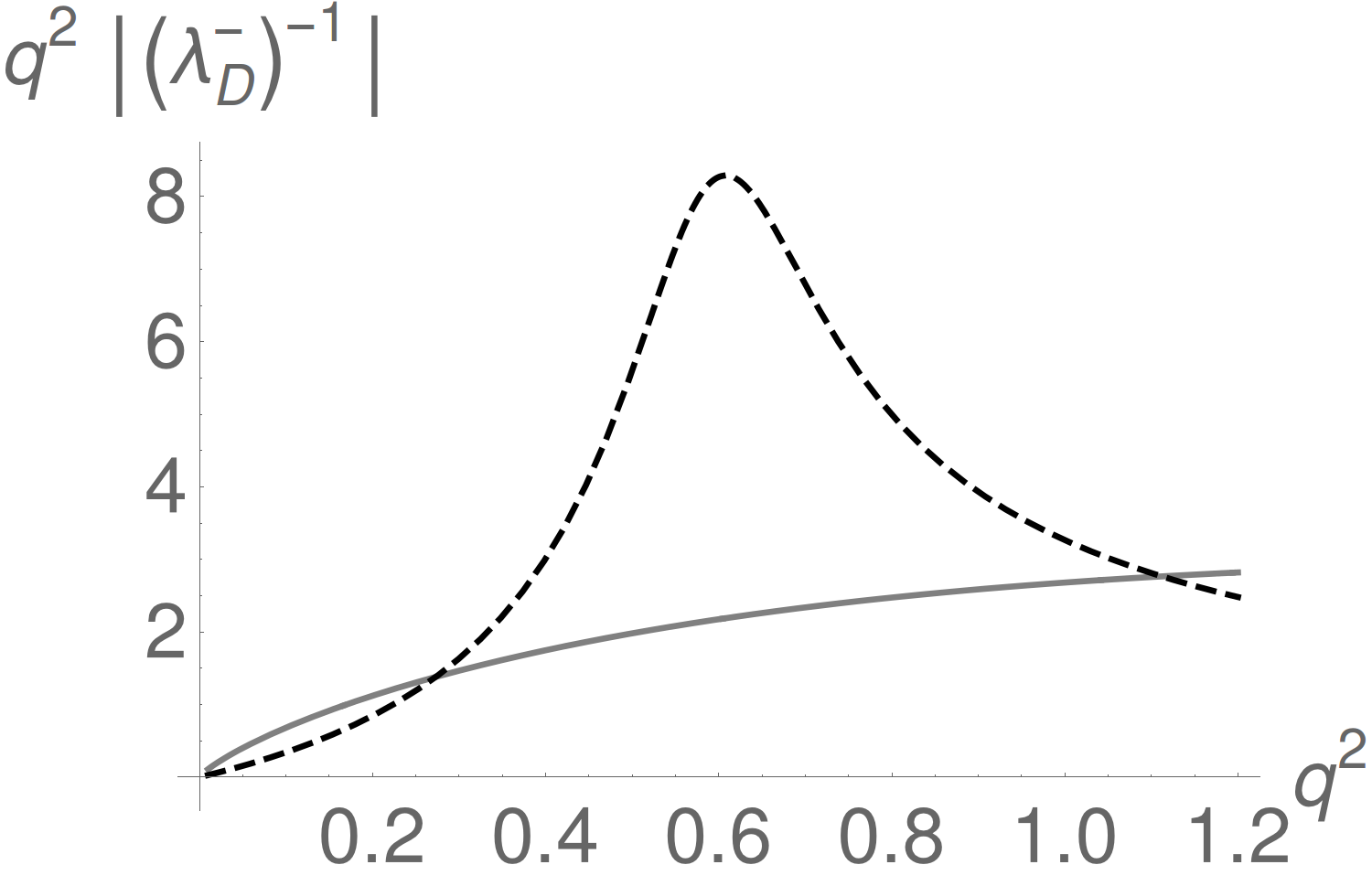}} 
\end{center}
\caption{Comparison of the perturbative result for the absolute value of 
the function $\left(\lambda_D^-(q^2)\right)^{-1}$ 
(massless spectator, gray solid line) and the model (\ref{eq:lamminusmod}) for its hadronic modification  
(black dashed line). Notice that the result has been multiplied by $q^2$ (in units of GeV$^2$).
The parameter values are chosen as follows:
$\sigma^2=2~\text{GeV}^2$, $a=1$, $b=1/6$.
The $D$-meson LCDA is modelled by an exponential (\ref{eq:phim}) 
with $\omega_0=0.45$~GeV. 
The parameter $n_V=2.40$ is tuned to reproduce the result for deep Euclidean values of $q^2$.
For details, see appendix~\ref{app:model}. 
\label{fig:lammin}}
\end{figure}

The convolution of the expressions in (\ref{anni:LO}) with the $D$-meson LCDA 
leads to the $q^2$-dependent ``moment''
\begin{align}
  \left( \lambda_D^-(q^2) \right)^{-1} 
  &= \int_0^\infty \frac{d\omega}{\omega- q^2/m_D - i \epsilon} \, \phi_D^-(\omega) \,,
  \label{lammindef}
\end{align} 
whose analytic properties are further discussed in appendix~\ref{app:model_anni}.
Notice that the limit $q^2 \to 0$ does not exist in (\ref{lammindef}) which limits
the applicability of QCDF for that part of  the amplitude to hard-collinear values of the 
momentum transfer that formally scale as $q^2 \sim {\cal O}(\Lambda m_D) \sim 1$~GeV$^2$.
Moreover, as we have already discussed for the quark-loop contributions to 
the form-factor--like terms in the previous subsection, the physical $q^2$ spectrum
in that region will be significantly influenced by light vector-meson resonances and looks
quite different from the partonic result following from (\ref{lammindef}).
Applying the same kind of model for the hadronic spectrum,
as explained in appendix~\ref{app:model_anni}, we end up with an estimate for 
the hadronic effects in  $\lambda_D^-(q^2)$. Here, the $q^2$-spectrum
(given by the imaginary part of $(\lambda_D^-)^{-1}$) is assumed to factorize into the 
$D$-meson LCDA and a hadronic model for the spectrum associated to the light
vector current,
\begin{align}  
  \left( \lambda_D^-(q^2) \right)^{-1}  &  \to  
  \int_0^\infty  \frac{ds}{s-q^2-i\epsilon} \, 
  \phi_D^-(s/m_D) \,  j_{q}(s) 
  \,.
  \label{eq:lamminusmod}
\end{align}
The parameters in the function $j_{q}(s)$ are adjusted to reproduce 
the perturbative result in the limit $q^2 \ll - m_D \, \Lambda$. 
To illustrate the numerical effect, we plot in Fig.~\ref{fig:lammin}
the absolute value of $\left[\lambda_D^-(q^2)\right]^{-1}$
(the real and imaginary part are plotted in Fig.~\ref{fig:plot9bis} 
in the appendix).
Here we have taken a simple  exponential model \cite{Grozin:1996pq}, 
\begin{align}
  \phi_D^-(\omega) &= \frac{1}{\omega_0} \, e^{-\omega/\omega_0} \,,
  \qquad 
  \phi_D^+(\omega) = \frac{\omega}{\omega_0^2} \, e^{-\omega/\omega_0} \,,
  \label{eq:phim}
\end{align}
for the LCDAs $\phi_D^\pm(\omega)$.
The resulting picture looks quite
similar as for the quark-loop functions $h(s,m)$.
In particular the asymptotic behaviour for large values of $|q^2|$
is unchanged (by construction), 
while the typical  modifications  from  the resonances occur in the region 
below $5$~GeV$^2$.

\subsubsection{Power corrections of order $1/m_D$}

\begin{figure}[t!pbh]
\begin{center}
\fbox{\includegraphics[width=0.46\textwidth]{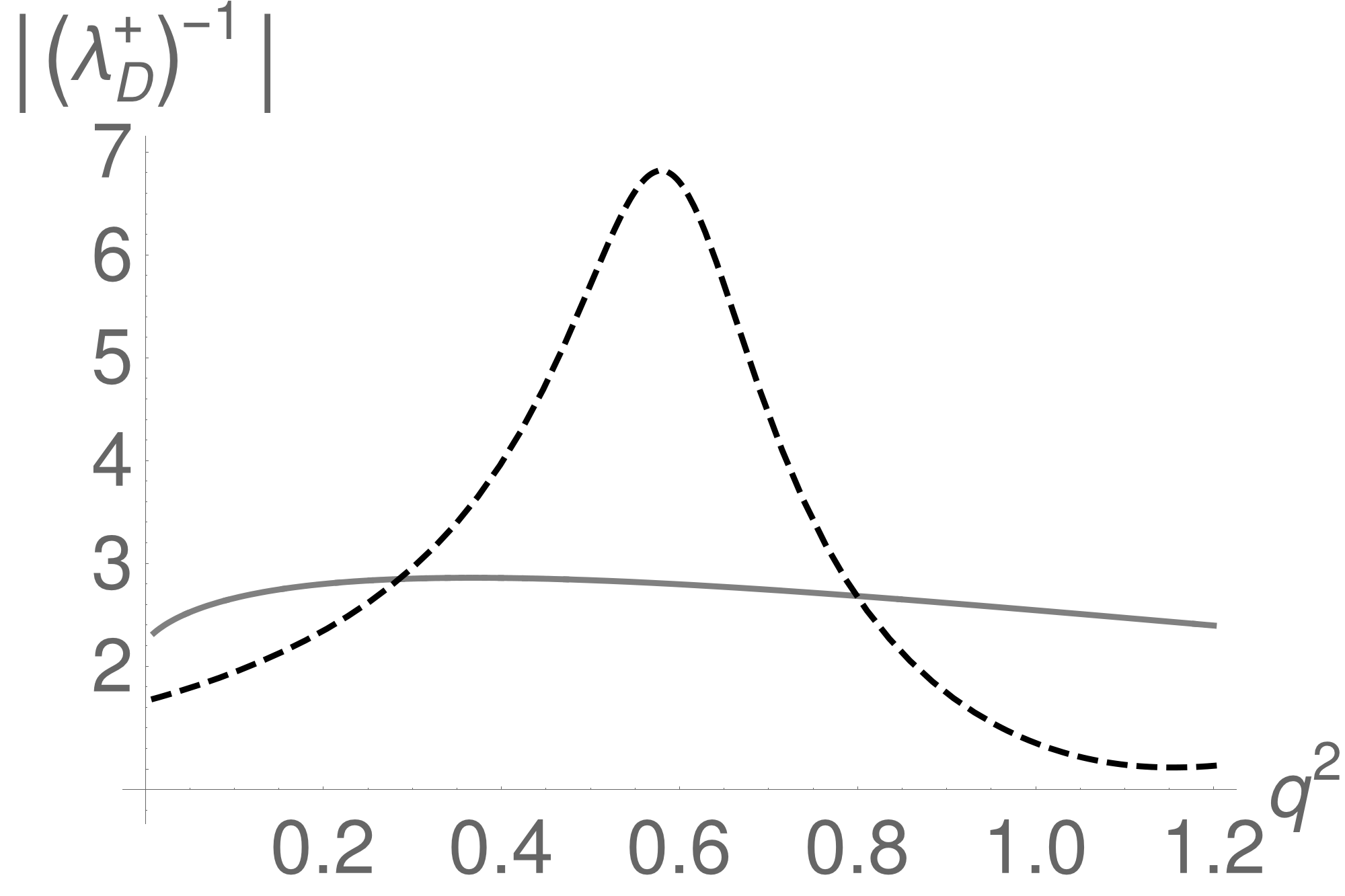}} 
\end{center}
\caption{Comparison of the perturbative result for the absolute value of 
the function $\left(\lambda_D^+(q^2)\right)^{-1}$ 
(massless spectator, gray solid line) and the model (\ref{eq:lamplusmod}) for its hadronic modification  
(black dashed line).
The parameter values are chosen as follows:
$\sigma^2=2~\text{GeV}^2$, $a=1$, $b=1/6$.
The $D$-meson LCDA is modelled by an exponential with $\omega_0=0.45$~GeV. 
The parameter $n_V=1.75$ is tuned to reproduce the result for deep Euclidean values of $q^2$.
For details, see appendix~\ref{app:model}. 
\label{fig:lamplus}}
\end{figure}

The contribution of annihilation topologies to the amplitudes
for transversely polarized vector mesons only start at relative
order $1/m_D$. It is known from the analysis of the 
analogous $B$-meson decays that these terms should not be neglected,
in particular for observables that are sensitive to the transverse 
decay amplitudes or isospin asymmetries \cite{Bosch:2001gv,Ali:2001ez,Kagan:2001zk,Feldmann:2002iw,Beneke:2004dp}. 
Translating again the known results from 
the $B$-meson sector, we end up with the following 
expressions,
\begin{align}
 m_c \, 
 T_{\perp,+}^{(0,b)} &= -e_q \, \frac{4\omega}{\bar{u} + u s / m_D^2} (\delta_{qd} - \delta_{qu}) \Big[ C_3 + \frac{4}{3} \Big( C_4 + 3 C_5 + 4 C_6 \Big) \Big]\cr 
                     & \quad 
                     + e_q \, \frac{2f_\parallel}{f_\perp} \frac{M_V}{1 - s / m_D^2} \frac{m_D \omega}{m_D \omega - s - i \epsilon} (\delta_{qd} - \delta_{qu}) \Big[ C_3 + \frac{4}{3} \Big( C_4 + 12 C_5 + 16 C_6 \Big) \Big] \,,
                     \nonumber \\[0.2em]
 m_c \, T_{\perp,+}^{(0,d)} &= -e_q \, \frac{2f_\parallel}{f_\perp} \frac{M_V}{1 - s / m_D^2} \frac{m_D \omega}{m_D \omega - s - i \epsilon} \Big[\delta_{qd}  3 C_2 + \delta_{qu} \Big( \frac{4}{3} C_1 + C_2 \Big) \Big] \,.
\label{eq:annipower}
\end{align}
Here, the first term arises from photon radiation off the constituents of the $\rho$-meson\footnote{Following the notation in \cite{Beneke:2001at}, we have written 
a factor of the spectator momentum $\omega$ in the numerator which is cancelled by the definition 
of the amplitudes $T_x$ in (\ref{Tadecomp}). The notation is slightly inconsistent, as the normalization 
integral of the $D$-meson LCDAs is not defined beyond LO. In these terms, it is thus to be understood 
that the appearance of an $\omega$-independent kernel implies that one only needs the matrix element 
of the \emph{local} $c \to q$ current which is simply given by the 
$D$-meson decay constant $f_D$.} 
and therefore contains a non-trivial convolution with respect to the corresponding 
momentum fractions $u$ and $\bar u=1-u$. The other terms stem from sub-leading 
contributions from photon radiation off the $D$-meson constituents, which now give rise to
the $q^2$-dependent ``moment''
\begin{align}
  \left( \lambda_D^+(q^2) \right)^{-1} 
  &= \int_0^\infty \frac{d\omega}{\omega- q^2/m_D - i \epsilon} \, \phi_D^+(\omega) \,.
  \label{lamplusdef}
\end{align} 
As before, our hadronic model for the vector resonances can be implemented
by the replacement
\begin{align}  
  \left( \lambda_D^+(q^2) \right)^{-1}  &  \to  
  \int_0^\infty  \frac{ds}{s-q^2-i\epsilon} \, 
  \phi_D^+(s/m_D) \,  j_{q}(s) 
  \,.
  \label{eq:lamplusmod}
\end{align}
The numerical effect is illustrated in 
Fig.~\ref{fig:lamplus} (see also Fig.~\ref{fig:plot5bis}
in the appendix).
Notice that at $q^2=0$ the function $\lambda_D^+(0)$ describes the ``soft'' contribution to 
the $D \to \gamma$ transition form factor. The hadronic effects in our simple-minded
model lead to a \emph{reduction} of the form factor by $20-30$\% which is in 
qualitative agreement with the findings in \cite{Braun:2012kp}
for the $B \to \gamma$ form factor.

\subsection{Non-factorizable spectator scattering}

Non-factorizable spectator scattering effects arise 
from matrix elements of the hadronic operators 
${\cal O}_{1-6,8}$ where the (would-be) spectators
in the $D \to \rho$ transitions take part in the 
short-distance scattering process.
The various contributions that appear to order $\alpha_s$, again,
can be adapted from \cite{Beneke:2001at,Beneke:2004dp}
with the appropriate modifications for $c \to u$ transitions.\footnote{Notice that, as 
for the corresponding analyses in rare semileptonic $B$-decays, we do not
take into account radiative corrections to the annihilation
topologies which are presently unknown.}
We find 
\begin{align}
 T_{\perp,+}^{(\text{nf},b)} &= 
 -\frac{4 Q_u C_8^{\text{eff}}}{u + \bar{u} s / m_D^2}
 \cr  & \quad 
 {} + \frac{m_D}{2 m_c} \Bigg[
   Q_u \big( t_\perp(u, m_c) + t_\perp(u,m_u) \big) 
  \left(C_3 - \frac{1}{6} \, C_4 + 16 \, C_5 + \frac{10}{3} \, C_6 \right)
 \cr & \qquad \phantom{\frac{m_D}{2 m_c} \Bigg[}  
 {} - Q_u \, t_\perp(u, m_c) \, \frac{4 m_c}{m_D} \Big(
 C_3 - \frac{1}{6} \, C_4 + 4 \, C_5 - \frac{2}{3} \, C_6\Big)
 \cr & \qquad \phantom{\frac{m_D}{2 m_c} \Bigg[}  
 {} + Q_d \, t_\perp(u, m_s) \left(C_2 - \frac{1}{6} \, C_1 + 6 \, C_6\right) + Q_d \, t_\perp(u, m_d)\, 6 \, C_6\Bigg] \,,
 \end{align}
and 
\begin{align}
 T_{\parallel,+}^{(\text{nf},b)} &= \frac{m_D}{m_c} \Bigg[
 Q_u \big( t_\parallel(u, m_c) + t_\parallel(u,m_u) \big) 
 \left( C_3 - \frac{1}{6} \, C_4 + 16 \, C_5 + \frac{10}{3} \, C_6 \right)\cr 
                          & \quad \phantom{\frac{m_D}{m_c} \Bigg[ }
                          + Q_d \, t_\parallel(u, m_s) \left( C_2 - \frac{1}{6} \, C_1 + 6 \, C_6 \right)
                           +  Q_d \, t_\parallel(u,m_d)  \, 6 \, C_6  \Big)\Bigg] \,,
\end{align}
together with 
\begin{align}
 T_{\perp,+}^{(\text{nf},d)} &= Q_d \, \frac{m_D}{2 m_c} \left( C_2 - \frac{1}{6} \, C_1 \right) 
 \big(t_\perp(u, m_s) - t_\perp(u,m_d) \big) \,, \\   
 T_{\parallel,+}^{(\text{nf},d)} &= Q_d \, \frac{m_D}{m_c} \left( C_2 - \frac{1}{6} \, C_1 \right) \big(t_\parallel(u, m_s) - t_\parallel(u,m_d)\big) \,,
 \label{leadtrans}
 \end{align}
 and
 \begin{align}
 T_{\parallel,-}^{(\text{nf},i)} &= e_q \, \frac{m_D \, \omega}{m_D \,\omega - s - i \epsilon} 
 \left[ \delta^{ib} \frac{8 C_8^{\text{eff}}}{\bar{u} + u s / m_D^2} 
 + \frac{6 m_D}{m_c} \, F_V^{(i)}(\bar{u} m_D^2 + u s) \right] \,.
\end{align}
Here the quark-loop functions describing the relevant sub-diagrams are given by
\begin{align}
 F_V^{(b)}(s) &= \big(h(s, m_c) +h(s,m_u) \big) 
                 \left( C_3 + \frac{5}{6} \, C_4 + 16 \, C_5 + \frac{22}{3} \, C_6 \right) \cr
              &\quad {} + h(s, m_s) \left( C_2-\frac{1}{6} \, C_1   + C_4 + 10 \,C_6 \right)\cr
              &\quad {} + h(s,m_d) \big( C_4 + 10 \, C_6 \big) 
              +\frac89 \left( 2\, C_4 - \frac{16}{5} \, C_5 + \frac{128}{15} \, C_6 \right) \,,
              \\
 F_V^{(d)}(s) &= \left( C_2 - \frac{1}{6} \, C_1 \right) \big( h(s, m_s) - h(s, m_d) \big) \,,
\end{align}
while the functions $t_{\perp,\parallel}(u,m_q)$ and the definition of $C_8^{\rm eff}$ 
can be found in \cite{Beneke:2001at}.

Notice that the contribution to the CKM-favoured amplitudes for transversely polarized 
vector mesons, $T_{\perp,+}^{(nf),d}$
in (\ref{leadtrans}), are again GIM-suppressed. We therefore also include power
corrections of relative order $1/m_D$ to the transverse amplitudes, 
which again can be adapted
from the corresponding expressions for $B$-meson decays. We find
\begin{align}
 T_{\perp,+}^{(0,i)} &= 4 e_q \, \delta^{ib} \,
 C_8^{\text{eff}} \, \frac{\omega}{m_D} 
 \Big( \frac{1}{\bar{u} + u s / m_D^2} + \frac{1}{(\bar{u} + u s / m_D^2)^2} \Big)\cr 
                     & \quad {} 
                     + 6 e_q \, \frac{\omega}{m_c(\bar{u} + u s / m_D^2)} \, F_V^{(i)}(\bar{u} m_D^2 + u s)\cr 
                     & \quad {} - 3 e_q \, \frac{M_V}{m_c(1 - s / m_D^2)} \, \frac{f_\parallel}{f_\perp \phi_\perp(u)} \, \frac{m_D  \,\omega}{m_D \, \omega - s - i \epsilon} \ F_V^{(i)}(\bar{u} m_D^2 + u s) \, \int_0^u dv \, \frac{\phi_\parallel(v)}{v} \,.
                     \cr & 
\end{align}

\section{Numerical Results}

\label{sec:four}

In this section we present some numerical estimates following 
from our theoretical analysis in the previous section. 
The theoretical predictions for the differential decay rates depend on
a number of parameters. The most important hadronic input parameters 
are listed in Table~\ref{tab:hadronic}. The remaining input parameters
that have been used for the numerical analysis are listed in 
Table~\ref{tab:rest} in the appendix for completeness.
Here a comment is in order about our choice for the parameter $\omega_0$
which determines the average value for the light-cone momentum of the
spectator quark in the $D$-meson. While the analogous parameter 
for $B$-meson decays has been studied in some detail in the past 
(see e.g.\ \cite{Beneke:2011nf,Braun:2012kp} and references therein), there is practically no 
theoretical or phenomenological information on the $D$-meson LCDAs.
We have therefore used an ad-hoc range for $\omega_0$ which reflects 
the naive expectation from heavy-quark symmetry 
with a sufficiently conservative uncertainty.

\begin{table}[t!!pbh]
\centering
\renewcommand{\arraystretch}{1.2}
\begin{tabular}{|c|c|c|}
\hline
 $A_1(0)$ & $0.56 \pm 0.01^{+0.02}_{-0.03}$ & \cite{CLEO:2011ab} \\
 $A_2(0)$ & $0.47 \pm 0.06^{+0.04}_{-0.04}$ & \cite{CLEO:2011ab} \\
 $V(0)$ & $0.84 \pm 0.09^{+0.05}_{-0.06}$ & \cite{CLEO:2011ab} \\
\hline 
 $f_{\rho, \parallel}$ & $209 \pm 1$ MeV & \cite{PDG} \\
 $f_{\rho, \perp}(1 \, \rm{GeV})$ & $165 \pm 9$ MeV & \cite{Ball:2006nr} \\
 $a_2(\rho)_{\perp, \parallel}$ & $0.15 \pm 0.07$ &\cite{Ball:2006nr}  \\
\hline 
 $f_D$ & $ (209 \pm 3)$ MeV & \cite{Vladikas:2015bra} \\
 $\omega_0$ & $(450 \pm 150$ MeV & (ad-hoc) \\
\hline
\end{tabular}
\caption{
\label{tab:hadronic}
Summary of the most relevant hadronic input parameters.}
\end{table}

\subsection{Detailed breakdown of contributions to ${\cal C}_{9,\perp}$ and ${\cal C}_{9,\parallel}$}

We first summarize the individual contributions 
to the coefficient functions ${\cal C}_{9,\perp}(q^2)$ and 
${\cal C}_{9,\parallel}(q^2)$ as defined in Eq.~(\ref{eq:calC})
at NLO,\footnote{We remind the reader that $\alpha_s$
corrections to annihilation topologies are not included.}
at a benchmark value $q^2=0.5$~GeV$^2$ for the momentum transfer,
see Tables~\ref{tab:calCneutral} and \ref{tab:calCcharged}.
Compared to the situation in the analogous $B$-meson decays
(see the discussion in \cite{Beneke:2004dp}), one observes a 
number of important differences:
\begin{itemize}
  \item As is well known, the purely short-distance contribution
    from the Wilson coefficient $C_9$ 
    is heavily suppressed by two effects:  (i) the GIM cancellation between down-type
    quarks running in the loops, leading 
    to a factor-10 smaller value for $C_9$ in $D$-decays compared to $B$-decays;
    (ii) the CKM suppression for $c\to u$ transitions, reflected by $\lambda_b \ll \lambda_d$.
    The decay amplitudes are therefore dominated by long-distance contributions 
    proportional to the CKM structure $\lambda_d$.
    
  \item Among these, it turns out that for the neutral decay mode --
    at least at the considered value of $q^2$ --
    non-factorizable form-factor corrections (FFnf) 
    and annihilation topologies (Ann) enter with the same order 
    of magnitude, and also non-factorizable spectator effects (Specnf) give
    a non-negligible contribution. In the case of transverse vector mesons this requires 
    to include (numerically unsuppressed) power corrections to annihilation 
    spectator topologies ($1/M$ Ann) as well.
    
   \item The charged decay modes are completely dominated by contributions 
     from annihilation topologies, where again the case of transverse vector mesons
     requires to include terms that are formally suppressed by $1/M_D$.

     
   \item We stress again that in each case, the estimates for the 
     relevant long-distance contributions within the QCDF approach 
     are very sensitive to hadronic resonance effects which adds to the 
     systematic theoretical uncertainties.

\end{itemize}

\begin{table}[t!pbh]
 \renewcommand{\arraystretch}{1.3}
 \centering
\begin{tabular}{|c||c|c|c|}
 \hline
 Decay & Contr. & $\propto \lambda_b$ & $\propto \lambda_d$ \\
 \hline \hline 
 $D^0\to \rho_\perp^0 \ell^+ \ell^-$ & $C_9$ & $-0.445$ & $0$ \\
 \cline{2-4}
 \hbox{} & FFf & $0.186 - 0.252 i$ & $-0.006+0.000 i$  \\
 \cline{2-4}
 \hbox{} & FFnf & $1.373 + 1.254 i$ & $0.071-0.044 i$ \\
 \cline{2-4} 
 & Specf & $0.071$ & $0$ \\
 \cline{2-4}
 \hbox{} & Specnf & $-0.114 + 0.009 i$ & $-0.013+0.009 i$ \\
 \cline{2-4}
 \hbox{} & $1 / M$ Ann & $-0.191 + 0.209 i$ & $-0.271-0.372 i$ \\
 \cline{2-4}
 \hbox{} & $1 / M$ Spec & $0.243 + 0.082 i$ & $0.002-0.006 i$ \\
 \hline \hline 
 ${\cal C}_{9,\perp}$  & Sum & $1.123 + 1.302 i$ & $-0.217-0.413 i$ \\
 \hline \hline 
 $D^0 \to \rho^0_\parallel \ell^+ \ell^-$ & $C_9$ & $-0.445$ & $0$\\
 \cline{2-4}
 \hbox{} & FFf & $-0.106 - 0.252 i$ & $-0.006+0.000 i$ \\
 \cline{2-4}
 \hbox{} & FFnf & $0.267 + 0.570 i$ & $0.029+0.020 i$ \\
 \cline{2-4}
 & Ann & $-0.050 + 0.206 i$ & $0.088-0.366 i$ \\
 \cline{2-4}
 \hbox{} & Specf & $-0.027$ & $0$ \\
 \cline{2-4}
 \hbox{} & Specnf & $0.146 + 0.019 i$ & $-0.008+0.004 i$ \\
 \hline \hline 
 ${\cal C}_{9,\parallel}$  & Sum & $-0.215 + 0.543 i$ & $0.103-0.342 i$ \\
 \hline
\end{tabular}
\caption{\label{tab:calCneutral} Breakdown of individual 
contributions to coefficient functions ${\cal C}_{9,\perp}$ and 
${\cal C}_{9,\parallel}$ at NLO
for the neutral decay mode, 
$D^0\to \rho^0 \ell^+\ell^-$ at $q^2 = 0.5$~GeV$^2$. 
The renormalization scale is set to $\mu = 1.5$~GeV. All other
input parameters are set to their default values. Here $C_9$ 
denotes the purely short-distance contribution. We further 
list factorizable (FFf) and non-factorizable (FFnf) form-factor
corrections, annihilation (Ann) at leading power, factorizable (Specf)
and non-factorizable (Specnf) spectator interactions at leading power,
as well as the included power corrections ($1/M$) 
to annihilation and spectator topologies.}
\end{table}

\begin{table}[t!pbh]
 \renewcommand{\arraystretch}{1.3}
 \centering
\begin{tabular}{|c||c|c|c|}
 \hline
 Decay & Contr. & $\propto \lambda_b$ & $\propto \lambda_d$\\
 \hline \hline 
$D^+  \to \rho_\perp^+ \ell^+ \ell^-$ & $C_9$ & $-0.445$ & $0$  \\
 \cline{2-4}
 \hbox{} & FFf & $0.187 - 0.252 i$ & $-0.006+0.000 i$  \\
 \cline{2-4}
 \hbox{} & FFnf & $1.376 + 1.256 i$ & $0.071-0.044 i$  \\
 \cline{2-4}  & Specf & $0.071$ & $0$  \\
 \cline{2-4}  & Specnf & $-0.114 + 0.009 i$ & $-0.013+0.009 i$  \\
 \cline{2-4}
 \hbox{} & $1 / M$ Ann& $-0.095 + 0.104 i$ & $2.472+3.381 i$  \\
 \cline{2-4} & $1 / M$ Spec& $-0.121 - 0.041 i$ & $-0.001+0.003 i$  \\
 \hline \hline 
 ${\cal C}_{9,\perp}$ & Sum & $0.859 + 1.077 i$ & $2.523+3.349 i$  \\
 \hline \hline 
 $D^+\to \rho^+_\parallel \ell^+ \ell^-$ & $C_9$ & $-0.445$ & $0$  \\
 \cline{2-4}
 \hbox{} & FFf & $-0.106 - 0.252 i$ & $-0.006+0.000 i$  \\
 \cline{2-4}
 \hbox{} & FFnf & $0.266 + 0.570 i$ & $0.029+0.020 i$  \\
 \cline{2-4} & Ann & $-0.025 + 0.103 i$ & $-0.799+3.335 i$  \\
 \cline{2-4}
 \hbox{} & Specf & $-0.027$ & $0$  \\
 \cline{2-4}
 \hbox{} & Specnf & $-0.186 + 0.001 i$ & $-0.009+0.009 i$  \\
 \hline \hline  
 ${\cal C}_{9,\parallel}$ & Sum & $-0.523 + 0.422 i$ & $-0.785+3.364 i$ \\
 \hline
\end{tabular}
\caption{\label{tab:calCcharged} Same as Table~\ref{tab:calCneutral}
for the charged mode, $D^+ \to \rho^+\ell^+\ell^-$.}
\end{table}

\clearpage 

\subsection{Differential decay rates}

\begin{figure}[t!pbh]
\begin{center}
 {\cblue \fbox{\includegraphics[width=0.46\textwidth]{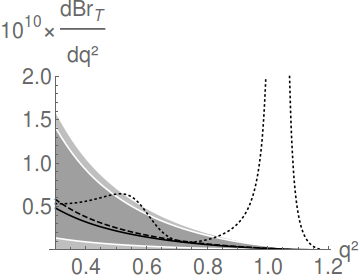}}}
\quad  \fbox{\includegraphics[width=0.46\textwidth]{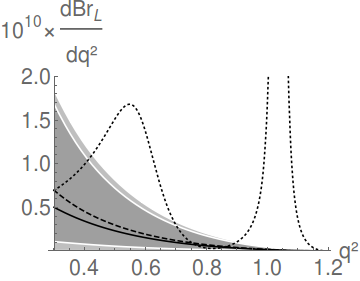}}
\end{center}
\caption{\label{fig:TLneutral}
The differential branching ratios  for 
$D^0\to \rho^0\ell^+\ell^-$ decays (in units of GeV$^{-2}$) with 
transversely or longitudinally polarized $\rho$ mesons 
as a function of momentum transfer $q^2$ (in GeV$^2$).
The LO QCDF result (including quark annihilation only at LO) is shown
as the dashed line;
the NLO result (without NLO corrections to annihilation)
is shown as the solid line. 
The uncertainty bands (dark grey for LO; light grey for NLO)
refer to the variation of the 
renormalization scale, \emph{only}, with 
$\mu \in [1.3,2.0]~{\rm GeV}$. The dotted line illustrates the 
LO result for the hadronic resonance model.
}
\end{figure}

\begin{figure}[t!pbh]
\begin{center}
 {\cblue \fbox{\includegraphics[width=0.46\textwidth]{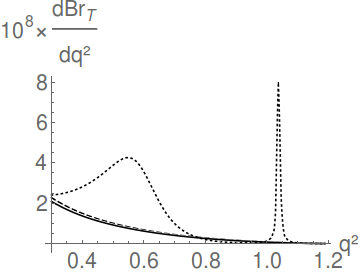}}}
\quad  \fbox{\includegraphics[width=0.46\textwidth]{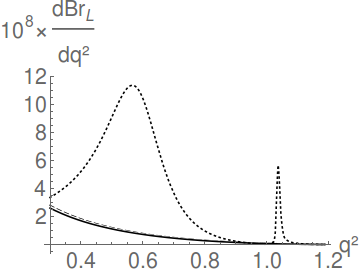}}
\end{center}
\caption{\label{fig:TLcharged}
The same as Fig.~\ref{fig:TLneutral} for 
$D^+\to \rho^+\ell^+\ell^-$ decays,
with $\mu \in [1.0,2.0]~{\rm GeV}$.
}
\end{figure}

We next turn to the differential decay rates, 
where we distinguish between the contributions of
transverse and longitudinal $\rho$ mesons, $d\Gamma_{T,L}$,
which are obtained by
projection onto the terms with $(1 \pm \cos^2\theta)$ in Eq.~(\ref{eq:diffrate}).
In Figs.~\ref{fig:TLneutral} and \ref{fig:TLcharged}
we show our result for the differential branching fractions in the case 
of neutral and charged mesons, respectively. 
Here, we compare the LO and NLO results, displaying only the 
uncertainties from scale variation for simplicity.
The following comments can be made:
\begin{itemize}
\item The difference between the central values for the 
  LO and NLO predictions (for our default choice of $\mu$) is not very pronounced.
 
\item Still, at least for the neutral decay mode, we observe a large renormalization-scale 
 dependence. This can be traced back to 
 the issues with the
 combination of Wilson coefficients $(4/3 C_1+C_2)$
 (see the discussion around 
 Fig.~\ref{fig:43c1c2}) appearing in the relevant annihilation contributions 
 in (\ref{anni:LO}) and (\ref{eq:annipower}). 
 Such cancellations do not occur in the charged decay mode. 
 
\item  One should be aware that 
(presently unknown) NLO corrections to annihilation are not included.
 This particularly concerns the charged decay modes which are
completely dominated by annihilation topologies.

\item The hadronic-resonance model mainly leads to an enhancement of the 
 differential rates. This is related to the fact that the rates are sensitive
 to the \emph{absolute} values of the modelled complex functions, which has 
 already been pointed out above. As a consequence, in contrast to the well-known
 $R$-ratio, the pseudo-realistic $q^2$-spectrum does not show the naively
 expected oscillations around the perturbative result.
 
\item There is, however, a small region around $q^2 \simeq 0.75$~GeV$^2$,
    where the perturbative result does not seem to be very much affected 
    by resonance effects, at least within our simplified hadronic model.
    However, in that region the differential branching fractions are small,
    and the relative uncertainties are large.

\end{itemize}

Apart from the renormalization-scale dependence, the main parametric 
uncertainties stem from the LCDA of the $D$-meson, while the modelling of
the hadronic resonances yields an estimate for the systematic hadronic
uncertainties. These are summarized in Tables~\ref{tab:BRTneutral}--\ref{tab:BRLcharged},
where we display our predictions for partially integrated branching ratios in 
specific $q^2$ bins. As one can see, the uncertainties in the chosen $q^2$ bins 
can easily exceed 100\%. 
In addition, we also expect substantial contributions from higher
orders in the $1/m_D$ expansion which, however, are difficult to quantify.

\clearpage 

\begin{table}[t!pbh]
\centering
\renewcommand{\arraystretch}{1.2}
\begin{tabular}{|c |l||c|c|c|c|}
 \hline
 $\Delta q^2$& \hbox{} & Br$_{\rm{T}} \times 10^{-12}$ & $\mu$ & $\omega_0$& hadr.\\
 \hline \hline
(0.3--0.5)~{GeV}$^2$
& LO & $7.861$ & $^{+11.218}_{-7.833}$ & $^{+9.010}_{-3.428}$ & \cblue $+3.574$ \\
& NLO & $6.637$ & $^{+15.209}_{-4.726}$ & $^{+8.574}_{-3.149}$ & \cblue$+1.752$ \\
\hline
(0.5--0.7)~{GeV}$^2$ & LO & $3.487$ & $^{+4.977}_{-3.475}$ & $^{+3.354}_{-1.437}$ & \cblue$+4.713$ \\
& NLO & $3.039$ & $^{+6.838}_{-2.153}$ & $^{+3.299}_{-1.367}$ & \cblue$+3.820$ \\
\hline
(0.7--0.9)~{GeV}$^2$
& LO & $1.419$ & $^{+2.026}_{-1.414}$ & $^{+1.123}_{-0.547}$ &\cblue $+1.450$ \\
& NLO & $1.266$ & $^{+2.815}_{-0.893}$ & $^{+1.140}_{-0.536}$ & \cblue$+0.865$ \\
\hline
\end{tabular}
\caption{Branching fractions for the neutral decay 
mode into transversely polarized vector mesons, partially integrated over
different bins $\Delta q^2$.
\label{tab:BRTneutral}}
\end{table}

\begin{table}[t!pbh]
\centering
\renewcommand{\arraystretch}{1.2}
\begin{tabular}{|c |l||c|c|c|c|}
 \hline
 $\Delta q^2$& \hbox{} & Br$_{\rm{L}} \times 10^{-12}$ & $\mu$ & $\omega_0$& hadr.\\
 \hline \hline
(0.3--0.5)~{GeV}$^2$
& LO & $9.334$ & $^{+13.326}_{-9.302}$ & $^{+5.157}_{-2.756}$ & $+11.300$ \\
& NLO & $6.982$ & $^{+18.243}_{-5.418}$ & $^{+4.524}_{-2.321}$ & $+8.759$ \\
\hline
(0.5--0.7)~{GeV}$^2$ & LO & $3.927$ & $^{+5.608}_{-3.914}$ & $^{+1.643}_{-1.026}$ & $+19.766$ \\
& NLO & $3.103$ & $^{+7.793}_{-2.360}$ & $^{+1.509}_{-0.903}$ & $+16.051$ \\
\hline
(0.7--0.9)~{GeV}$^2$
& LO & $1.467$ & $^{+2.095}_{-1.462}$ & $^{+0.446}_{-0.334}$ & $+0.198$ \\
& NLO & $1.194$ & $^{+2.939}_{-0.899}$ & $^{+0.430}_{-0.304}$ & $+0.115$ \\
\hline
\end{tabular}
\caption{Same as Table~\ref{tab:BRTneutral} for longitudinally polarized vector mesons.
\label{tab:BRLneutral}}
\end{table}

\clearpage 

\begin{table}[t!pbh]
\centering
\renewcommand{\arraystretch}{1.2}
\begin{tabular}{|c |l||c|c|c|c|}
 \hline
 $\Delta q^2$& \hbox{} & Br$_{\rm{T}} \times 10^{-9}$ & $\mu$ & $\omega_0$& hadr.\\
 \hline \hline
(0.3--0.5)~{GeV}$^2$
& LO & $3.085$ & $^{+0.288}_{-0.123}$ & $^{+3.592}_{-1.360}$ & \cblue$+2.763$ \\
& NLO & $2.827$ & $^{+0.000}_{-0.000}$ & $^{+3.261}_{-1.237}$ &\cblue $+2.937$ \\
\hline
(0.5--0.7)~{GeV}$^2$ & LO & $1.384$ & $^{+0.128}_{-0.054}$ & $^{+1.354}_{-0.577}$ &\cblue $+4.836$ \\
& NLO & $1.265$ & $^{+0.000}_{-0.000}$ & $^{+1.226}_{-0.523}$ & \cblue$+5.417$ \\
\hline
(0.7--0.9)~{GeV}$^2$
& LO & $0.573$ & $^{+0.052}_{-0.022}$ & $^{+0.462}_{-0.224}$ & \cblue$+0.030$ \\
& NLO & $0.523$ & $^{+0.000}_{-0.000}$ & $^{+0.417}_{-0.202}$ &\cblue $+0.350$ \\
\hline
\end{tabular}
\caption{Same as Table~\ref{tab:BRTneutral} for the charged decay mode.
\label{tab:BRTcharged}}
\end{table}

\begin{table}[t!pbh]
\centering
\renewcommand{\arraystretch}{1.2}
\begin{tabular}{|c |l||c|c|c|c|}
 \hline
 $\Delta q^2$& \hbox{} & Br$_{\rm{L}} \times 10^{-9}$ & $\mu$ & $\omega_0$& hadr.\\
 \hline \hline
(0.3--0.5)~{GeV}$^2$
& LO & $3.753$ & $^{+0.332}_{-0.140}$ & $^{+2.101}_{-1.120}$ & $+7.365$ \\
& NLO & $3.497$ & $^{+0.000}_{-0.000}$ & $^{+1.929}_{-1.032}$ & $+7.523$ \\
\hline
(0.5--0.7)~{GeV}$^2$ & LO & $1.598$ & $^{+0.140}_{-0.059}$ & $^{+0.677}_{-0.421}$ & $+16.022$ \\
& NLO & $1.481$ & $^{+0.000}_{-0.000}$ & $^{+0.619}_{-0.387}$ & $+17.028$ \\
\hline
(0.7--0.9)~{GeV}$^2$
& LO & $0.607$ & $^{+0.053}_{-0.022}$ & $^{+0.187}_{-0.139}$ & $+1.959$ \\
& NLO & $0.561$ & $^{+0.000}_{-0.000}$ & $^{+0.170}_{-0.127}$ & $+2.703$ \\
\hline
\end{tabular}
\caption{Same as Table~\ref{tab:BRLneutral} for the charged decay mode.
\label{tab:BRLcharged}}
\end{table}

\clearpage

\subsection{Ratio of transverse and longitudinal rates}

\begin{table}[t!pbh]
\centering
\renewcommand{\arraystretch}{1.2}
\begin{tabular}{|c |l||c|c|c|c|}
 \hline
 $\Delta q^2$& \hbox{} & ${\cal R}_{TL}$ & $\mu$ & $\omega_0$& hadr.\\
 \hline \hline 
(0.3--0.5)~{GeV}$^2$
& LO & $0.842$ & $^{+0.009}_{-0.000}$ & $^{+0.322}_{-0.168}$ & \cblue$-0.288$ \\
& NLO & $0.950$ & $^{+0.271}_{-0.363}$ & $^{+0.371}_{-0.202}$ & \cblue$-0.417$ \\
\hline 
(0.5--0.7)~{GeV}$^2$ & LO & $0.888$ & $^{+0.011}_{-0.000}$ & $^{+0.340}_{-0.181}$ & \cblue$-0.542$ \\
& NLO & $0.979$ & $^{+0.330}_{-0.608}$ & $^{+0.395}_{-0.219}$ &\cblue $-0.621$ \\
\hline 
(0.7--0.9)~{GeV}$^2$
& LO & $0.968$ & $^{+0.012}_{-0.000}$ & $^{+0.361}_{-0.198}$ & \cblue$+0.755$ \\
& NLO & $1.060$ & $^{+0.789}_{-0.726}$ & $^{+0.421}_{-0.240}$ &\cblue $+0.568$ \\
\hline 
\end{tabular}
\caption{Ratio of transverse and longitudinal rate as defined in Eq.~(\ref{eq:RTL}) 
for the neutral decay mode.
\label{tab:RTLneutral}
}
\end{table}

\begin{table}[t!pbh]
\centering
\renewcommand{\arraystretch}{1.2}
\begin{tabular}{|c|l||c|c|c|c|}
 \hline
$\Delta q^2$ & \hbox{} & ${\cal R}_{TL}$ & $\mu$ & $\omega_0$& had\\
 \hline \hline 
(0.3--0.5)~{GeV}$^2$
&  LO & $0.822$ & $^{+0.004}_{-0.002}$ & $^{+0.319}_{-0.167}$ &\cblue $-0.296$ \\
& NLO & $0.808$ & $^{+0.003}_{-0.009}$ & $^{+0.314}_{-0.163}$ & \cblue$-0.285$ \\
\hline 
(0.5--0.7)~{GeV}$^2$
& LO & $0.866$ & $^{+0.004}_{-0.002}$ & $^{+0.338}_{-0.180}$ & \cblue$-0.513$ \\
& NLO & $0.854$ & $^{+0.003}_{-0.009}$ & $^{+0.332}_{-0.176}$ & \cblue$-0.493$ \\
\hline 
(0.7--0.9)~{GeV}$^2$
&  LO & $0.944$ & $^{+0.004}_{-0.002}$ & $^{+0.359}_{-0.197}$ & \cblue$-0.709$ \\
&  NLO & $0.932$ & $^{+0.003}_{-0.009}$ & $^{+0.353}_{-0.193}$ & \cblue$-0.665$ \\
\hline 
\end{tabular}
\caption{Ratio of transverse and longitudinal rate as defined in Eq.~(\ref{eq:RTL}) 
for the charged decay mode.
\label{tab:RTLcharged}
}
\end{table}

Given the large parametric and systematic uncertainties, we 
do not expect to obtain reasonably reliable predictions for 
the $D \to \rho\,\ell^+\ell^-$ differential decay rates within the QCDF
framework. 
We will therefore briefly investigate to what extent at least
some of the theoretical uncertainties might be reduced in 
\emph{ratios} of decay widths. To this end we consider 
the ratio of partially integrated rates for 
transversely and longitudinally polarized 
vector mesons, 
\begin{align}
{\cal R}_{TL} &:= \left( \, \int\limits_{\Delta q^2} dq^2 \, 
\frac{d\Gamma_T}{dq^2} \right) / \left( \, \int\limits_{\Delta q^2} dq^2 \, 
\frac{d\Gamma_L}{dq^2} \right)
\label{eq:RTL}
\,.
\end{align}
{\cblue In a generic NP scenario with sizeable short-distance
contributions this ratio would be sensitive to the following
combinations of Wilson coefficients in the large-recoil limit 
(see e.g.\ the discussion in \cite{Kruger:2005ep}),
$$
  \frac{
   {|C_{10}|}^2+{|C_{10}'|}^2+{|C_9|}^2+{|C_9'|}^2+\frac{4
   {\hat m_c}^2}{\hat s^2} \left({|C_7|}^2+{|C_7'|}^2\right)+\frac{4 {\hat m_c}}{\hat s}
   \,\text{Re} \left[{C_7} {C_9^*}+{C_7'} {C_9'{}^*}\right]}{
   |{C_{10}}-{C_{10}'}|^2+|{C_9}-{C_9'}|^2
   +\frac{4{\hat m_c}^2}{\hat s^2} \, |{C_7}-{C_7'}|^2
   +\frac{4{\hat m_c}}{\hat s^2} \, \text{Re}\left[({C_7}-{C_7'}) ({C_9}-{C_9'})^*\right]} \,,
 $$
where the primed coefficients refer to the operators with
flipped quark chiralities, $\hat m_c = m_c/M_D$, $\hat s=q^2/M_D^2$,
and possible contributions from scalar or tensor operators have been neglected.
}
We present estimates for the quantity ${\cal R}_{TL}$ 
in specific regions of momentum transfer $\Delta q^2$
in Tables~\ref{tab:RTLneutral} and 
\ref{tab:RTLcharged}. 
We observe that, indeed, the parametric and systematic uncertainties are {\cblue somewhat}
reduced, at least for the lower two $q^2$ bins:
\begin{itemize}

\item For the neutral decay modes the scale dependence at NLO still 
 can reach several 10 percent.
 
\item The uncertainties induced by the parameter $\omega_0$ in the $D$-meson LCDA 
  typically reaches up to 30\%. Notice that this estimate has been obtained 
  using a particular (simple) model function (\ref{eq:phim})
  for the two 2-particle $D$-meson LCDAs.

\item As our model correlates the hadronic-resonance effects in the different 
decay topologies, we also found a {\cblue slight}
reduction of the associated uncertainty in ${\cal R}_{TL}$.
We emphasize again that the hadronic model is only meant for illustration,
and therefore one should not draw  specific conclusions about the 
resonance effects from this.
\end{itemize}

We have also investigated to what extent hadronic
uncertainties may cancel in the leptonic forward-backward asymmetry. 
Normalizing to the transverse rate, depending on the relative phase
of a possible NP contribution to the Wilson coefficient $C_{10}$,
this observable is determined by the ratio 
${\rm Re}\left[{\cal C}_{9,\perp}(q^2) \, C_{10}^* \right]/|{\cal C}_{9,\perp}^2(q^2)|$.
As the real or imaginary part of the resonant contributions to ${\cal C}_{9,\perp}$ 
look rather different from the square of the absolute value, we did not observe 
a significant reduction of the associated theoretical uncertainties in that case.

\section{Conclusions}

\label{sec:five}

In this paper we have investigated the rare semileptonic
decays $D \to \rho \, \ell^+\ell^-$ in the framework of 
QCD factorization (QCDF). 
Here, our primary goal was \emph{not} 
to obtain very precise predictions for the SM decay rates, 
but rather to achieve a sufficiently realistic (i.e.\ conservative) 
estimate of hadronic uncertainties 
related to long-distance QCD effects.

The QCDF framework allows one to separate contributions 
from various decay topologies via a simultaneous expansion
in the strong coupling constant and inverse powers of the
heavy charm-quark mass. This includes factorizable and 
non-factorizable effects from quark loops, annihilation 
topologies, as well as spectator-scattering effects.
Within the perturbative analysis, we have found that 
in general the contributions from non-factorizable 
effects dominate. In the case of neutral 
mesons form-factor corrections, annihilation and 
spectator effects enter with similar magnitude, whereas
the decay modes with charged mesons are dominated
by annihilation topologies alone.

We have also seen that -- not surprisingly -- the 
convergence of the QCDF predictions for the considered
decay rates is relatively poor, which is clearly to be
attributed to the relatively small charm-quark mass,
but also -- in the case of neutral mesons -- due to 
accidental cancellations between the Wilson coefficients
multiplying the dominant expressions.
Even more importantly, the restricted phase
space for $D \to \rho$ transitions together with 
the suppression of the purely short-distance effects 
(i.e.\ the ones encoded in the semileptonic and 
electromagnetic operators) in the Standard Model 
imply a much stronger
sensitivity to hadronic-resonance effects as compared 
to the analogous $B$-meson decays.
In order to get a quantitative estimate of these
effects, we have extended a model, that has been
originally designed to describe the effect of 
vector resonances in $e^+e^-$ annihilation
or hadronic $\tau$ decays. 
This allows us to study aspects of 
quark-hadron duality within the QCDF approach 
via dispersion relations and to estimate 
the related systematic uncertainties in 
 the $q^2$ spectrum. (We repeat that we do not claim
a realistic description of the $q^2$-spectrum itself.) 
  
Together with the remaining parametric uncertainties 
coming from non-factorizable contributions and (yet unknown)
higher-order effects in the QCDF approach (notably NLO corrections
to the annihilation topologies), we have found that reliable
theoretical predictions for the differential decay widths 
in $D \to \rho\,\ell^+\ell^-$ within the QCDF approach 
are almost impossible. On the other hand, part of the 
systematic and parametric uncertainties tend to partially cancel
in ratios of decay widths, at least in certain regions
of phase space. As an example, we have studied the ratio of
transverse and longitudinal decay rates. 
Still, our conclusion about possible new-physics sensitivity of
rare semileptonic $D$-meson decays tends to be  
{\cblue somewhat pessimistic, unless one concentrates on 
bins in the lepton invariant mass far below (or,  in case of 
radiative $D \to \pi \ell^+\ell^-$ decays, also far above)
the light resonances, or one considers observables that (practically) vanish in the Standard Model.
In this way one can still look for new-physics scenarios with large 
Wilson coefficients which are not excluded by current exprerimental data, 
see e.g.\ the discussion in the recent literature \cite{Fajfer:2015mia,deBoer:2015boa}.}
%
(To a certain extent, this also applies to the purely radiative decays 
$D\to \rho\gamma$ which have recently been seen by the {\sc Belle} experiment 
\cite{Abdesselam:2016yvr}.)

On the other hand, one may exploit the fact that non-factorizable
hadronic effects are more pronounced in $D$-meson decays,
and use $D \to \rho \, \ell^+\ell^-$ as a playground for QCD studies 
that are also relevant to estimate sub-leading effects in 
the corresponding $B$-meson decays. 

\section*{Acknowledgements}
 
We thank Oscar Cata for helpful discussions
on the modelling of the hadronic resonances. 
We further thank Svjetlana Fajfer and Jernej Kamenik 
for helpful comments on the manuscript.
This work is
supported in parts by the Bundesministerium for Bildung und Forschung (BMBF FSP-105), and
by the Deutsche Forschungsgemeinschaft (DFG FOR 1873).

\appendix 

\section{Explicit Formulas for $D \to \pi \ell^+\ell^-$ Decays}

\label{app:Dtopi}

Defining generalized form factors for $D \to \pi\gamma^*$ transitions as
\begin{align}
   \langle \gamma^*(q,\mu) \, \pi^+(p')|\mathcal{H}_{\rm eff}^{(i)} | D^+(p)\rangle &=
  \frac{g_{\rm em} m_c}{4\pi^2} \, \frac{{\cal T}_{P=\pi}^{(i)}(q^2)}{m_D} \, \left[ q^2 \, (p^\mu + p'^\mu) - (m_D^2 - M_\pi^2) \, q^\mu \right] \,,
\end{align}
following \cite{Beneke:2001at}, we obtain 
the factorization formula 
\begin{align}
 \mathcal{T}_P^{(i)} &\simeq - \xi_P \, C_\parallel^{(i)} + \frac{\pi^2}{N_c} \,
 \frac{f_D f_{P}}{m_D} \, 
 \sum_\pm \int \frac{d\omega}{\omega} \, \phi_{D,\pm}(\omega) \,
 \int_0^1 du \, \phi_P(u) \, T_{\parallel,\pm}^{(i)}(u,\omega)\,,
 \label{Tpdecomp}
\end{align}
where the same functions $T_{\parallel,\pm}^{(i)}$ appear as in the 
case of decays into longitudinally polarized vector mesons, and $\xi_P(q^2)$
is the soft form factor for $D \to \pi$ transitions as defined in \cite{Beneke:2001at}.
If we define again the coefficient function\footnote{Notice the
different relative sign compared to Eq.~(\ref{eq:calC}).}
\begin{align}
 \mathcal{C}^{(i)}_{9,P}(q^2) &= \delta^{ib} \, C_9 + \frac{2 m_c}{m_D} \frac{\mathcal{T}_P^{(i)}(q^2)}{\xi_P(q^2)} \,,
\end{align}
the twofold differential decay rate can be written as 
\begin{align}
 \frac{d^2\Gamma(D \to \pi \ell^+ \ell^-)}{dq^2 d\rm{cos} \theta} &= 
 \frac{G_F^2}{128 \pi^3}\, m_D^3 \, S \, \lambda_D^3(q^2,M_\pi^2) \, \left(\frac{\alpha_{\rm{em}}}{4 \pi} \right)^2 \times \\
                                         & (1 - \rm{cos}^2 \theta) \, \xi_\pi^2(q^2) \left|\lambda_d \, \mathcal{C}_{9,\pi}^{(d)} + \lambda_b \, \mathcal{C}_{9,\pi}^{(b)}\right|^2 \,.
\end{align}


\section{Semi-naive Model for Duality violation}

\label{app:model}

In the perturbative analysis of non-factorizing effects 
from hadronic operators, the virtual photon is treated 
as a point-like particle. However, at time-like virtualities,
$q^2>0$, the coupling of the photon to a (perturbative)
sum of intermediate \emph{partonic} states has to be
replaced by the coupling to an (infinite) sum 
over physical \emph{hadronic} states.
Parton-hadron duality (PHD) is expected to hold if one 
averages over a ``sufficiently'' large phase-space 
region of $q^2$. In order to estimate the numerical
effect of violation of PHD, we will construct 
a simple model which reflects the main theoretical 
features and phenomenological constraints, following 
the ideas discussed in \cite{Blok:1997hs,Shifman:2000jv,Shifman:2003de}
(see also \cite{Cata:2005zj,Cata:2008ye,Cata:2008ru,Boito:2011qt,Beylich:2011aq}).

\subsection*{Preliminaries: (a) Modelling of vector resonances}

The standard procedure to model a vector-meson resonance 
is to consider a Breit-Wigner (BW) ansatz,
\begin{align}
 f_{\rm BW}(q^2) &= \frac{n_V \, M_V^2}{M_V^2-q^2 - i \, M_V \Gamma_V} \,, \qquad  
\frac{1}{\pi} \, {\rm Im}\left[f_{\rm BV} \right]
= \frac{n_V}{\pi} \, \frac{M_V^3 \Gamma_V}{(q^2-M_V^2)^2 + M_V^2 \, \Gamma_V^2}\,.
\end{align}
In the narrow-width approximation, one would further neglect 
$\Gamma_V/M_V \ll 1$. The above form gives a successful description
of the resonance shape in the vicinity of $q^2 \simeq M_V^2$;
however, for our purposes we would also need to control the 
effect away from the resonance peak. A more sophisticated ansatz
for a modified BW-like shape has been suggested by Shifman \cite{Shifman:2000jv,Shifman:2003de}, 
where 
\begin{align}
f_{\rm mod}(q^2) &= \hat n_V 
\left( 1 + z_V  \, \frac{\sigma_V^2}{M_V^2} \right)^{-1} , \quad 
z_V = \left(\frac{- q^2 -i \epsilon}{\sigma_V^2} \right)^{1-b_V/\pi} , \quad 
b_V = \frac{\Gamma_V}{M_V}
\,.
\end{align}
Among others, this form has the correct analytic behaviour, i.e.\ a branch cut 
at $q^2=0$ (we neglect the pion mass in the following).
On the other hand,
it reproduces the BW-ansatz when $q^2 \simeq M_V^2 \approx \sigma_V^2$ and $b_V \ll 1$.
The imaginary part of the modified BW ansatz can then be approximated as ,
\begin{align}
\frac{1}{\pi} \, {\rm Im}\left[f_{\rm mod} \right]
&= \frac{\hat n_V\, \theta(q^2)}{\pi}  \,
\frac{  |z_V| \, \hat \sigma_V^2 \, \sin b_V }{
1 - 2 \, |z_V| \, \hat \sigma_V^2 \,\cos b_V + |z_V|^2 \, \hat\sigma_V^4} 
\cr
& \simeq \frac{n_V\, \theta(q^2)}{\pi} \, \frac{q^2 M_V \Gamma_V}{(q^2-M_V^2)^2 + q^2 \, \Gamma_V^2}
\,,
\end{align}
where in the last line we have taken $\hat\sigma_V^2=\sigma_V^2/M_V^2 = 1$ for simplicity
and used $b_V \ll 1$.

For vector resonances
in the $s\bar s$ channel, the definition of $z_V$ has to be modified 
to include the threshold for $K\bar K$ production. Neglecting mixing effects,
this amounts to setting 
\begin{align}
f_{\rm mod}^{(s\bar s)}(q^2) &= \hat n_V 
\left( 1 + \tilde z_V  \, \frac{\tilde\sigma_V^2}{\tilde M_V^2} \right)^{-1} \,,
\end{align}
with 
\begin{align} 
\tilde z_V = \left(\frac{4 m_K^2- q^2 -i \epsilon}{\tilde\sigma_V^2} \right)^{1-\tilde b_V/\pi} , \quad 
\tilde b_V = \frac{\tilde \Gamma_V}{\tilde M_V}\left(1 - \frac{4 m_K^2}{\tilde\sigma_V^2} \right)
\,.
\end{align}
The imaginary part of the modified BW ansatz now reads (for $\tilde \sigma_V^2\simeq M_V^2$) 
\begin{align}
\frac{1}{\pi} \, {\rm Im}\left[f_{\rm mod}^{(s\bar s)} \right]
& \simeq \frac{n_V}{\pi}
\, \frac{(q^2 - 4 m_K^2) \, \theta(q^2-4 m_K^2)}{\tilde M_V^2- 4 m_K^2} \, 
\frac{q^2 \tilde M_V^3 \tilde \Gamma_V}{(q^2-\tilde M_V^2)^2 + 
\frac{q^2 - 4 m_K^2}{\tilde M_V^2-4 m_K^2} \, \tilde M_V^2 \, \tilde \Gamma_V^2}
\,.
\end{align}
Notice that in this form, the original BW ansatz is recovered in the limit 
$m_K^2 \to - \infty$.

\subsection*{Preliminaries: (b) Resumming an infinite tower of vector resonances}

Shifman \cite{Shifman:2000jv,Shifman:2003de} also gives a simple model to describe the effect of 
an infinite tower of equidistant vector resonances with masses $M_n^2 = (n + a_0) \, \sigma^2 $
and widths $\Gamma_n = b M_n$, for $n = \{0,1,2,\ldots\}$ and $a_0 = {\rm const}$. 
To this end, one considers the
function
\begin{align}
  \pi(q^2) 
  & = \frac{1}{1-b/\pi} \, \sum_{n=0}^\infty \frac{1}{n + a_0 +z}
  = - \frac{1}{1-b/\pi} \, \Psi(z+a_0) \,, \qquad z = \left(\frac{-q^2-i\epsilon}{\sigma^2}\right)^{1-b/\pi}
  \,.
  \label{exact}
\end{align}
Each individual resonance contributes with a modified BW term as discussed above.
However, the infinite sum over \emph{all} resonances reproduces the asymptotic result 
\begin{align}
  \lim_{-q^2 \to \infty} \pi(q^2) = -\ln \frac{-q^2}{\sigma^2} + {\cal O}(\sigma^2/q^2) \,.
\end{align}
In \cite{Shifman:2000jv,Shifman:2003de}, a simplified model has been constructed,
taking $a_0=1$. In this case, one has 
\begin{align}
  \psi(z+1) = \psi(z) + \frac{1}{z} &= \psi(-z) - \pi \, \cot (\pi z) \,,
\end{align}
and for time-like momentum transfer, $q^2>0$, 
the imaginary part of $\pi(q^2)$ receives an oscillatory contribution 
from the second term, with 
\begin{align}
  \frac{1}{\pi} \, {\rm Im}\left[\pi(q^2)\right]_{\rm osc.} 
  &\simeq  {\rm Im}\left[ \cot(\pi z) \right] 
  \cr 
  &= -  \frac{\sinh(2 \pi  |z| \sin b)\, \theta(q^2)}{\cos(2 \pi  |z| \cos b)-\cosh(2 \pi  |z|  \sin b)} 
  \cr 
  & \approx \left( 1 + 2 \, \exp\left(- \frac{2 \pi \, q^2 \, b}{\sigma^2} \right) \cos\left( \frac{2\pi q^2}{\sigma^2} \right) \right) \theta(q^2) \,, 
  \label{mod}
\end{align}
where the last approximation is valid for 
$q^2 > s_0 \equiv \frac{\sigma^2}{2 \pi  \, b}$ 
(and $\frac{b}{\pi} \left|\ln \frac{q^2}{\sigma^2} \right|\ll 1 $).
The general idea is then to start from a perturbative result 
in the OPE/factorization approach, where the leading $q^2$ dependence is logarithmic, such 
that the spectrum is given by 
$$j_{\rm OPE}(q^2) = \theta(q^2) + \ldots$$
and to replace it by the hadronic model (\ref{exact})
-- which reproduces the oscillatory behaviour in (\ref{mod}) --
and to add a 
\emph{finite} number of vector resonances in the region $0 < q^2 < s_0$.
In \cite{Shifman:2000jv,Shifman:2003de}
a good fit to $\tau$ decay data in the vector channel
has been obtained for 
$\sigma^2=2$~GeV$^2$ and $b=1/6$.
 (More sophisticated analyses with qualitatively similar 
parameter values can be found in \cite{Cata:2005zj,Cata:2008ye,Cata:2008ru,Boito:2011qt}).

Our model ansatz for the hadronic spectrum with up/down or strange quarks in the loop 
therefore looks as follows.\footnote{For
simplicity, we combine the $\rho$ and $\omega$ resonance into one 
effective expression, where $M_V \simeq m_\rho \simeq m_\omega$ and 
the width is dominated by the $\rho$-meson, $\Gamma_V \simeq \Gamma_\rho$.}
\begin{align}
  j_{(u/d)}(q^2) &= 
\frac{ n_V}{\pi}  \,
\frac{ q^2 \, M_V \Gamma_V}{(q^2-M_V^2)^2 + q^2 \, \Gamma_V^2}
\cr & 
\quad {} - \frac{1}{\pi} \, {\rm Im}\left[ 
\frac{1}{1-b/\pi} \, \Psi(z+a) \right] \,, \qquad 
z = \left( \frac{-q^2-i\epsilon}{\sigma^2} \right)^{1-b/\pi} \,,
\label{jud}
\end{align}
and 
\begin{align}
  j_{(s)}(q^2) &= \frac{n_\phi}{\pi}
\, \frac{(q^2 - 4 m_K^2) \, \theta(q^2-4 m_K^2)}{m_\phi^2- 4 m_K^2} \, 
\frac{m_\phi^3 \Gamma_\phi}{(q^2-m_\phi^2)^2 + 
\frac{q^2 - 4 m_K^2}{m_\phi^2-4 m_K^2} \, m_\phi^2 \, \Gamma_\phi^2}
\cr & 
\quad {} - \frac{1}{\pi} \, {\rm Im}\left[ 
\frac{1}{1-b_s/\pi} \, \Psi(z_s+a_s) \right] \,, \qquad 
z_s =  \left(\frac{4 m_K^2- q^2 -i \epsilon}{\sigma_s^2} \right)^{1-b_s/\pi}\,.
\label{js}
\end{align}
The real part of the function under consideration can then be recovered from 
an appropriate dispersion relation.
For the numerical illustration, we fix the parameters in the above ansatz as follows 
\begin{align}
  & M_V \equiv m_\rho= 775~{\rm MeV}\,,\quad \Gamma_V \equiv \Gamma_\rho= 149~{\rm MeV} \,, 
  \cr 
  & m_\phi= 1019~{\rm MeV}\,, \quad \Gamma_\phi = 4.27~{\rm MeV}\,, \qquad m_K = 497~{\rm MeV}\,, 
\end{align}
together with
\begin{align}
  &  b \equiv 1/6 \,, \quad \sigma^2 \equiv 2~{\rm GeV}^2 \,, \quad a \equiv 1\,, 
  \cr 
  &  b_s \simeq \frac{\Gamma_{\phi(1680)}}{m_{\phi(1680)}} \simeq 0.1 \,, 
  \quad \sigma_s^2 \simeq \sigma_0^2 \,, \quad a_s = \frac{m_{\phi(1680)}^2}{\sigma_s^2} \simeq 1.4 \,.
\end{align}
(Notice that for $\sigma^2$ and $a$ we have taken the same values as for the simplified
model that has been fitted to data; the assumed flavour-invariance to fix the parameter 
$\sigma_s^2$ is consistent with the mass splitting $m_{\phi(2175)}^2-m_{\phi(1680)}^2 \simeq 1.9~$GeV$^2$.)
With this input, the normalization factors $n_V$ and $n_\phi$ will be 
fitted by requiring that the relevant integrals over $j_{(i)}(s)$
are identical for the OPE and the hadronic result.
We emphasize again that our aim is \emph{not} to provide a sophisticated and fully realistic
model for the hadronic resonances, but rather to illustrate the systematic uncertainties
associated to our ignorance about long-distance QCD effects.

\subsection{Application to quark-loop topology}

\begin{figure}[tpbh]
\begin{center}
\fbox{\includegraphics[width=0.45\textwidth]{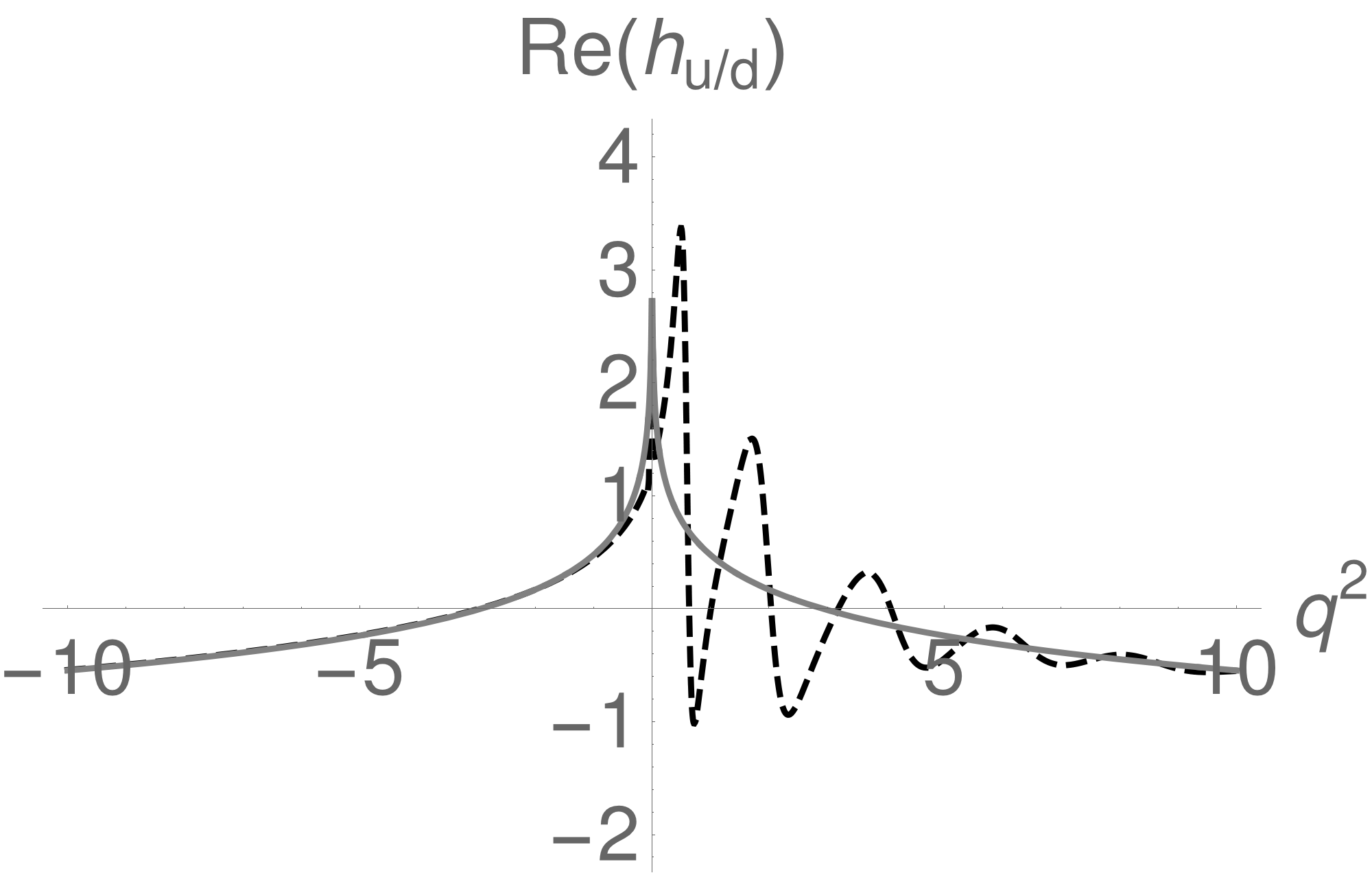}} \quad 
\fbox{\includegraphics[width=0.47\textwidth]{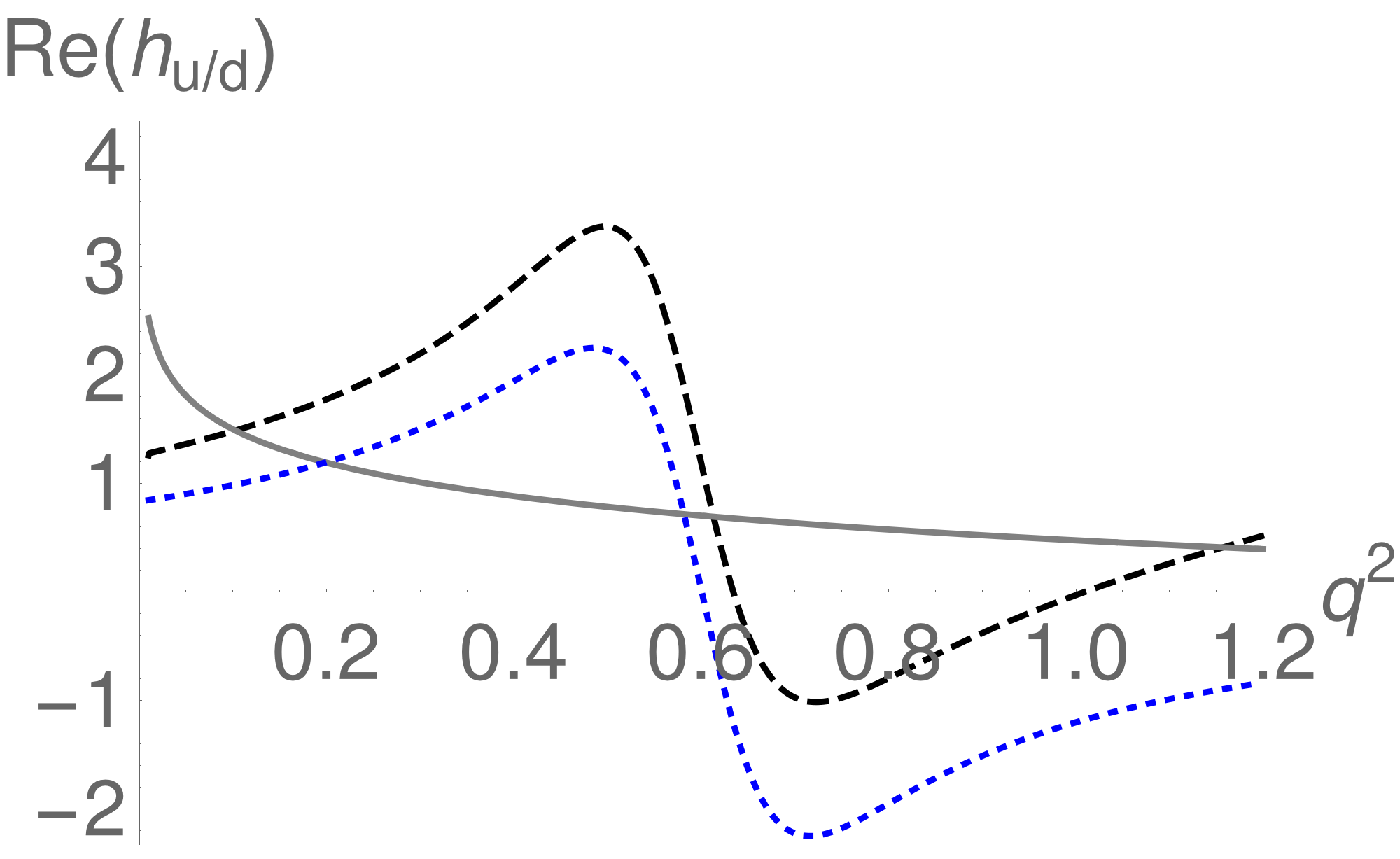}} \\[1em]
\fbox{\includegraphics[width=0.45\textwidth]{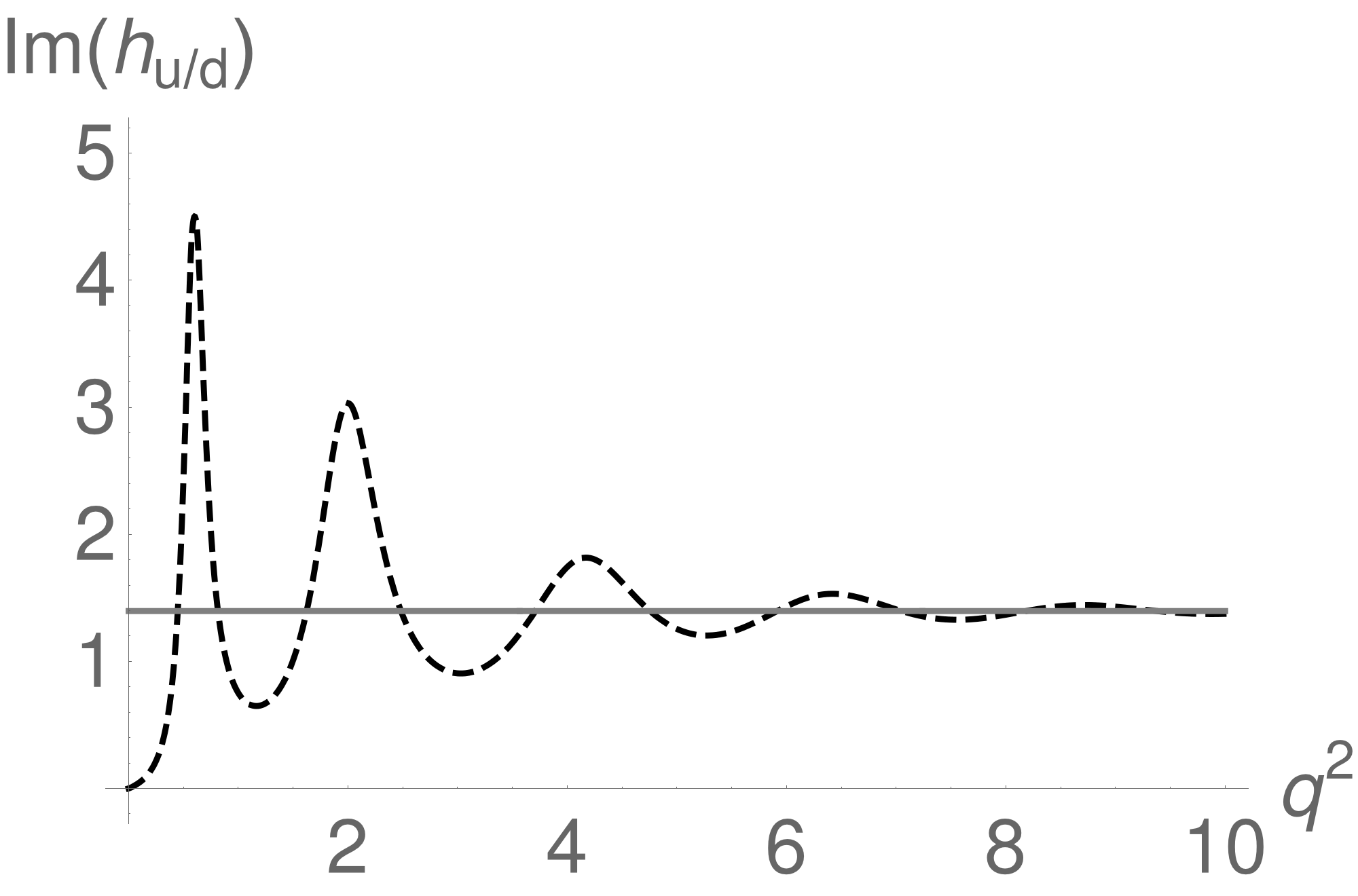}} \quad 
\fbox{\ \includegraphics[width=0.45\textwidth]{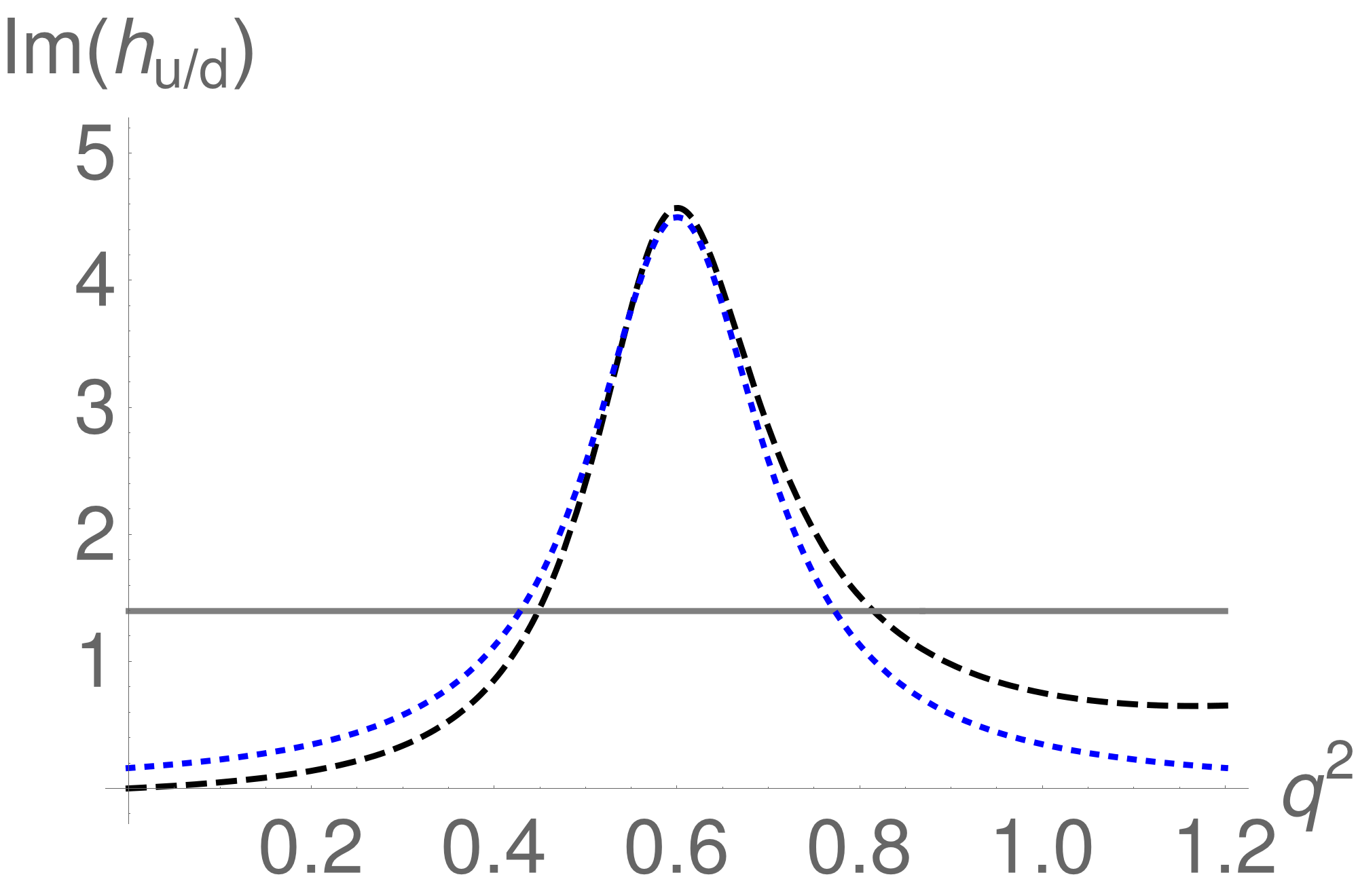}\ } 
\end{center}
\caption{Comparison of the perturbative result for the real and imaginary part of 
the function $h(q^2,0)$ 
(gray solid line) and the model (\ref{eq:hmod}) using (\ref{jud})
for its hadronic modification  (black dashed line) as a function of $q^2$ (in units of GeV$^2$). 
On the right-hand side, we also compare to the simple 
Breit-Wigner ansatz (blue dotted line). The parameter values are chosen as follows:
$\sigma^2=2~\text{GeV}^2$, $a=1$, $b=1/6$, 
$\mu^2=1.5$~GeV$^2$. The value of $n_V=1.94$ is tuned to reproduce the perturbative
result in the limit $q^2 \to -\infty$.
\label{fig:hud}}
\end{figure}

\begin{figure}[t!!!pbh]
\begin{center}
\fbox{\includegraphics[width=0.45\textwidth]{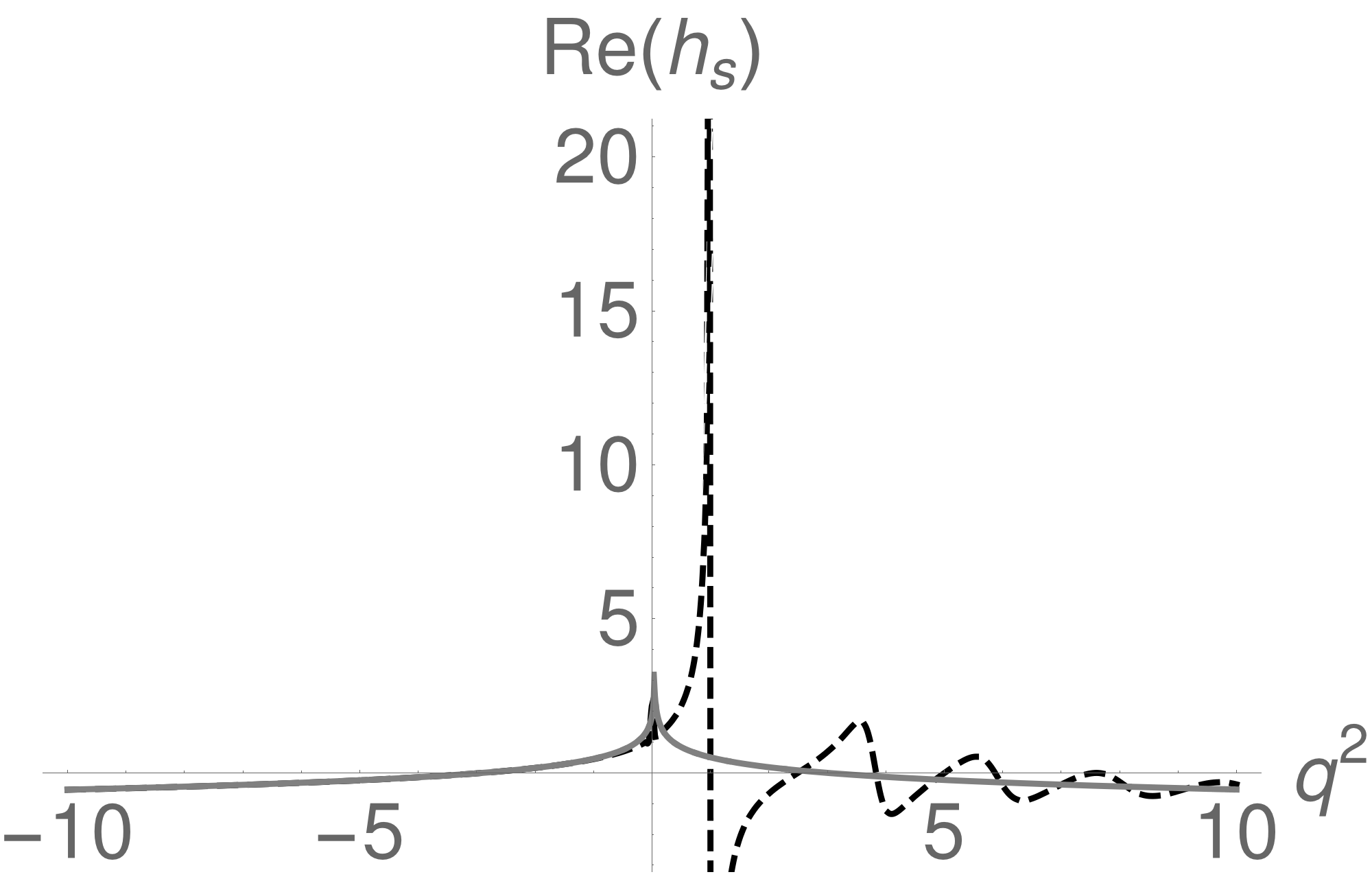}} \quad 
\fbox{\includegraphics[width=0.46\textwidth]{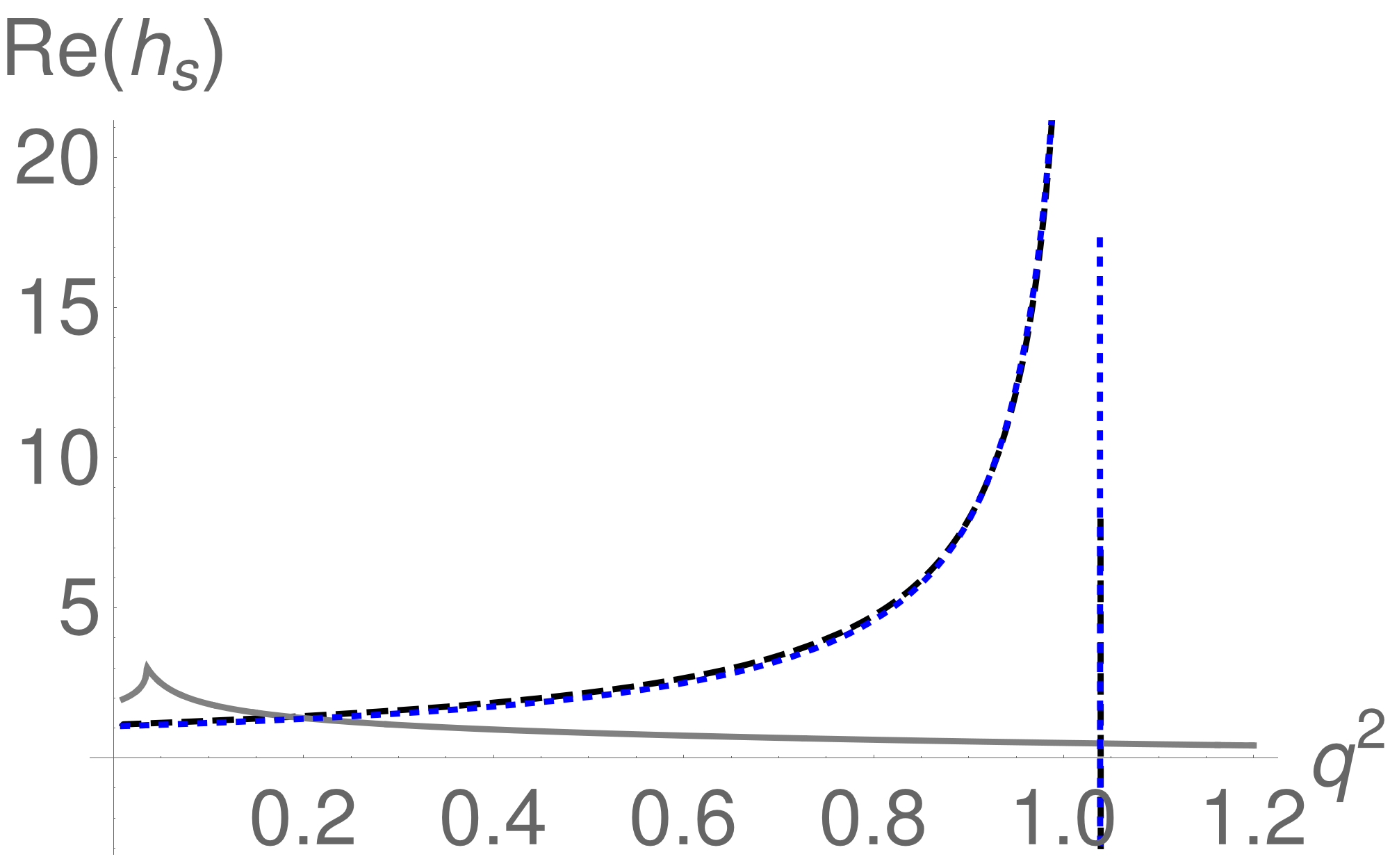}} \\[1em]
\fbox{\includegraphics[width=0.45\textwidth]{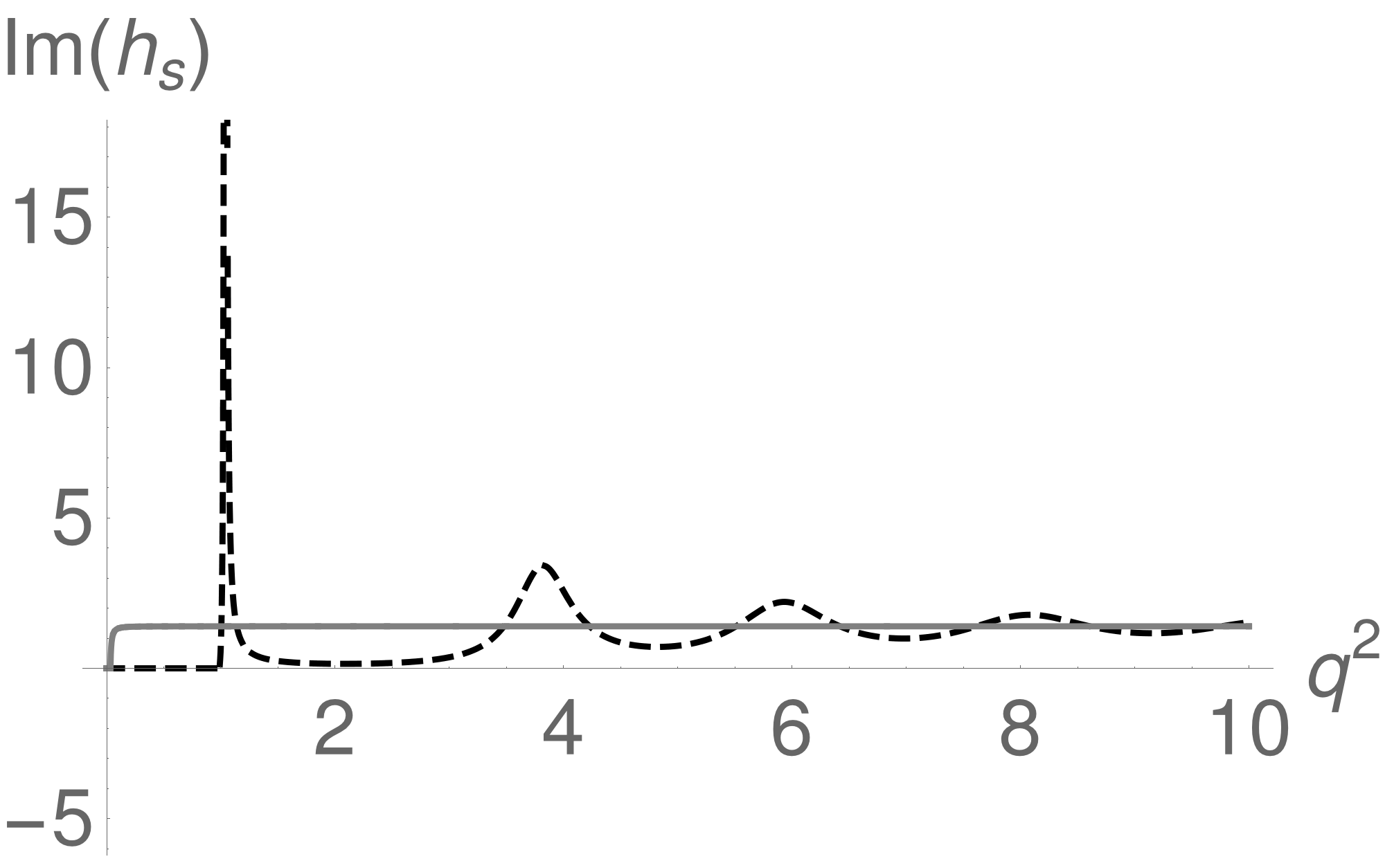}} \quad 
\fbox{\ \includegraphics[width=0.45\textwidth]{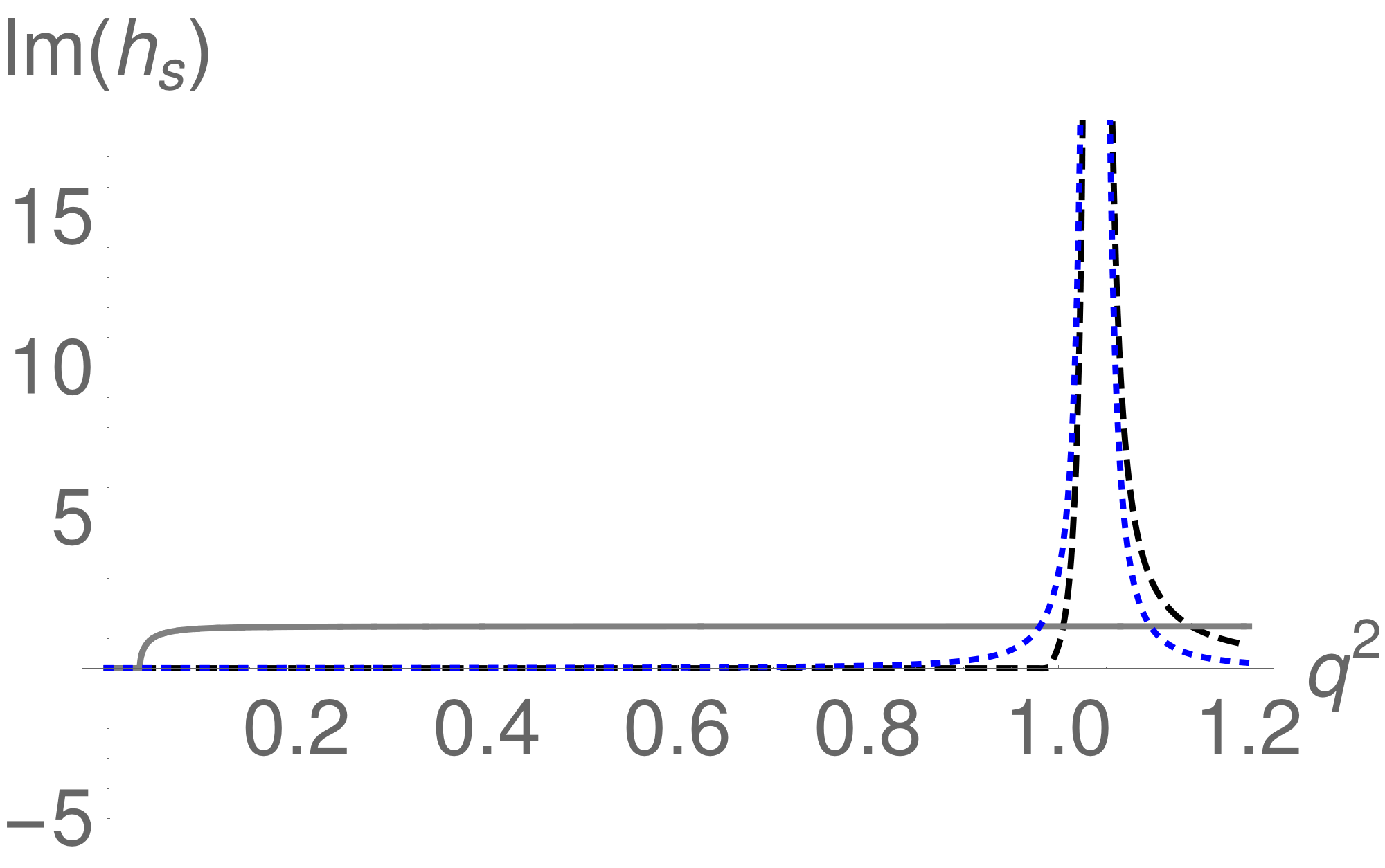}} 
\end{center}
\caption{Comparison of the perturbative result for the real and imaginary part of 
the function $h(q^2,m_s)$ 
(gray solid line) and the model (\ref{eq:hmod}) using (\ref{jud})
for its hadronic modification  (black dashed line) as a function of $q^2$ (in units of GeV$^2$).
On the right-hand side, we also compare to the simple 
Breit-Wigner ansatz (blue dotted line). The parameter values are chosen as follows:
$\sigma^2=2~\text{GeV}^2$, $a_s=1.4$, $b_s=0.1$, 
$\mu^2=1.5$~GeV$^2$. The parameter $n_\phi=2.37$ is tuned to reproduce the perturbative
result in the limit $q^2 \to -\infty$.
\label{fig:hs}}
\end{figure}

As explained above, the leading effect of hadronic operators with closed
quark loops can be described in terms of the loop functions 
\begin{align}
h(q^2,m_q) &= - \frac49 \left( \ln \frac{m_q^2}{\mu^2} - \frac23 - \zeta \right) 
- \frac49 \left( 2+ \zeta \right) \sqrt{1-\zeta} 
\, \ln \left[\frac{1+\sqrt{1-\zeta}}{\sqrt {-\zeta}}\right]  \,, 
\label{eq:hfunc}
\end{align}
appearing in (\ref{Yb},\ref{Yd}), with $\zeta=\frac{4 m_q^2}{q^2+i\epsilon}$. At asymptotic values in the deep Euclidean,
$q^2 \to -\infty$, these functions behave as 
\begin{align}
  h(q^2,m) & \stackrel{q^2 \to -\infty}{\longrightarrow} - \frac49  \ln\left( -\frac{q^2}{\mu^2} \right) 
  + \mbox{\rm \footnotesize finite terms}
\end{align}
We thus replace the perturbative function through a once-subtracted dispersion 
relation that involves the above model for the hadronic spectral function,
\begin{align*}
h(q^2,m_q) & \to h(-\sigma^2,m_q)\Big|_{\rm OPE} 
  + \frac49 \, \int_0^\infty ds \, \frac{\sigma^2+q^2}{\sigma^2+s} 
  \, \frac{j_q(s)}{s-q^2-i\epsilon} \,,
\qquad (m=0,m_s) \,,
\end{align*}
where the hadronic parameters in $j_q(s)$ depend on the light quark flavour,
as indicated. 
The default renormalization scale in the functions $h(s,m_q)$ 
is chosen as $\mu=1.5$~GeV. 
For the ansatz (\ref{jud}) and (\ref{js}), the parameters 
$n_V$ and $n_\phi$ are tuned by demanding that the OPE result for 
$h(q^2 \to -\infty, m_q)$ is reproduced, which implies 
\begin{align}
  \int_0^\infty ds \, \frac{j_q(s)-j_{\rm OPE}(s)}{s+\sigma^2} & \stackrel{!}{=} 0 \,. 
\end{align}
This yields $n_V\simeq1.94$ and $n_\phi\simeq 2.37$, 
{\cblue if the spectral parameters are fixed as explained in 
the previous subsection. We stress that at this point our goal is \emph{not}
to get a precise phenomenological determination of $n_V$ and $n_\phi$ itself,
which would better be achieved by standard sum-rule techniques applied 
to 2-point correlators (see also the 
discussion below). Rather, the so-obtained numerical values allow for a 
consistent comparison with the partonic prediction in the context of QCDF at LO.}
The numerical comparison between the original (partonic) functions and the result of
the simple hadronic model is illustrated in Figs.~\ref{fig:hud},\ref{fig:hs}.
Here we plot the real and 
imaginary part of the functions $h(q^2,m_q)$ and the above hadronic 
modification at negative and positive values of $q^2$. Zooming into the
region of small positive values of $q^2$ -- which is relevant for $D \to \rho$
decays -- we also show the contribution of a simple Breit-Wigner model 
for the low-lying resonances for comparison. We observe that
\begin{itemize}
  \item For negative values of $q^2$ and for $q^2 \gtrsim 5$~GeV$^2$, the perturbative 
    result approximates the hadronic model very well.
  \item For small positive values of $q^2$ the hadronic model exhibits the expected 
   oscillations around the perturbative estimate.
  \item In case of the real part (and also for the absolute value)
  of $h(q^2,m_{u,d})$ our model leads to somewhat larger
    values compared to a simple Breit-Wigner ansatz, which can be traced back to 
    the additional contributions from the higher resonances in the dispersion integral.
  \item By construction, the imaginary part of $h(q^2,m_q)$ in the vicinity of $q^2=M_V^2$
    looks similar in our model and for a Breit-Wigner resonance. However, for $q^2 \to 0$, 
    our model requires the imaginary part of $h(0^+,0)$ 
    to vanish, while the Breit-Wigner resonance formula yields a 
    finite result proportional to $\Gamma_\rho$.\footnote{In \cite{Fajfer:2015mia} the 
    rho-meson width is multiplied by hand with a factor $\sqrt{q^2}/m_\rho$ in the BW
    formula such that the imaginary part again vanishes for $q^2 \to 0$. However, we 
    remark that such an ansatz has wrong analytic properties for $q^2 < 0$.}
\end{itemize}

As discussed above the leading contribution from the quark-loop topologies 
enters through the function $Y^{(d)}(s)$ and involves the 
incomplete GIM cancellation between strange and up/down quarks in the difference 
$h(s,m_s)-h(s,m_d)$, for which we show the comparison between the perturbative
result and our hadronic model in Fig.~\ref{fig:plothsminushd} in the main body of
the manuscript.

As an aside, it is also instructive 
to perform a QCD-sum-rule analysis of our ansatz. 
In the spirit of the standard duality argument 
used in QCD sum rules, one
replaces the ``true'' hadronic spectrum (i.e.\ in our case the model for $j_{q}(s)$) 
by a single resonance plus continuum, leading to
\begin{align}
  \frac{n_V \, M_V^2}{M_V^2-q^2} & \simeq 
  \int_0^{s_0} ds \, \frac{\sigma^2+q^2}{\sigma^2+s} \, \frac{1}{s-q^2 - i\epsilon} \, \theta(s) \,.
  \label{sr1a}
\end{align}
Applying the standard Borel transformation, introducing the Borel mass parameter $M$, 
this yields 
\begin{align}
  e^{- M_V^2/M^2} \, n_V \, M_V^2 &= \int_0^{s_0} ds \, e^{-s/M^2} = M^2 \left( 1- e^{-s_0/M^2} \right)
  \cr 
  \leftrightarrow \quad 
  n_V \, M_V^2 &= M^2 \, e^{M_V^2/M^2} \left( 1- e^{-s_0/M^2} \right) \equiv 4 \pi^2 f_V^2 \,,
  \label{sr1b}
\end{align}
where we have identified the last term as the leading-order sum rule for the vector-meson 
decay constant, see e.g.\ \cite{Colangelo:2000dp}. Using $f_\rho =205$~MeV, this yields $n_V \simeq 2.75$
which is of the same order of magnitude as 
the values found for $n_V$ and $n_\phi$ 
by the procedure described in the previous paragraph. [The numerical difference
may be taken as a rough estimate for the intrinsic uncertainties
of our simplified model.]

\clearpage 

\clearpage

\subsection{Application to annihilation topology}

\label{app:model_anni}

\begin{figure}[t!pbh]
\begin{center}
\fbox{\includegraphics[width=0.46\textwidth]{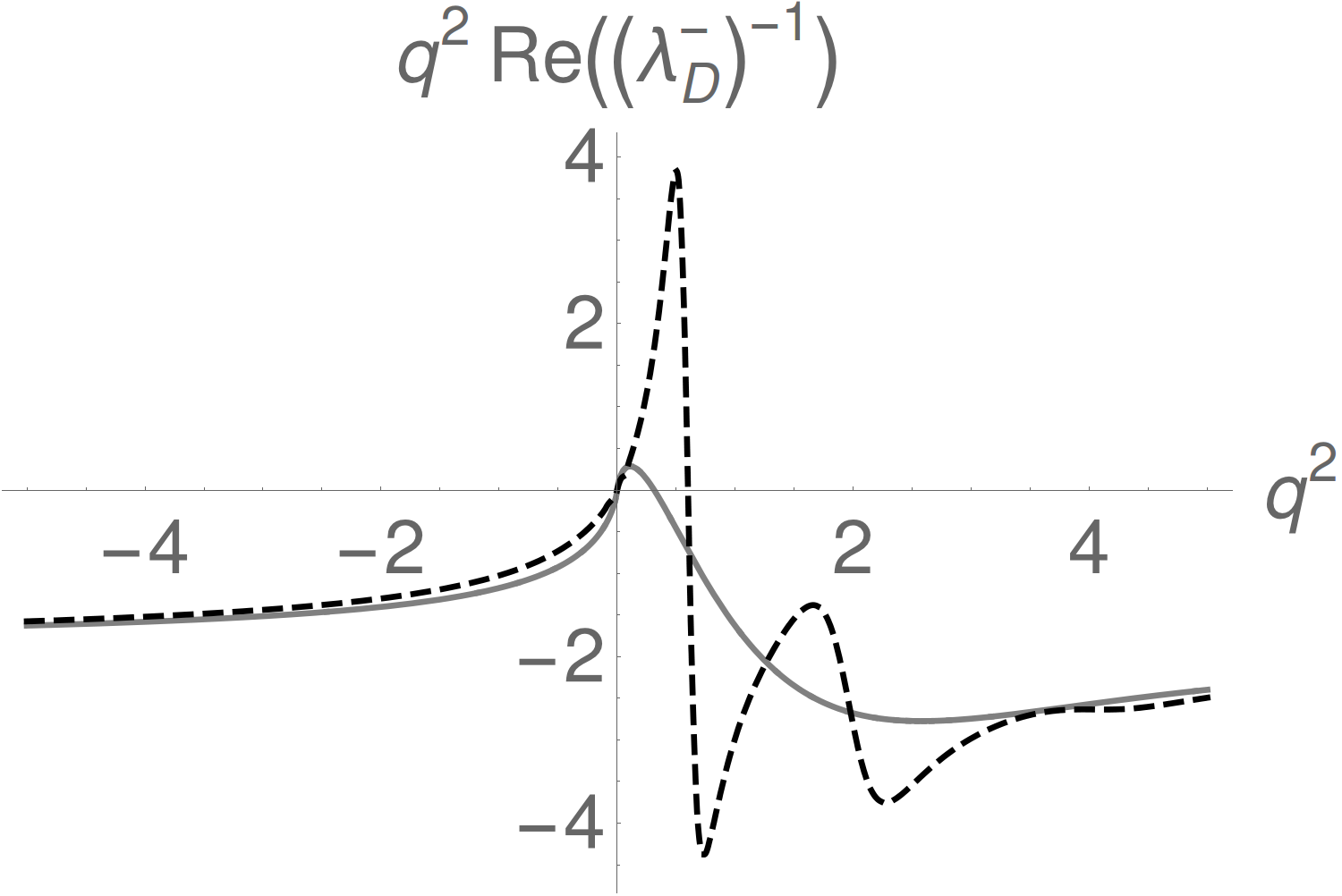}} \quad 
\fbox{\includegraphics[width=0.46\textwidth]{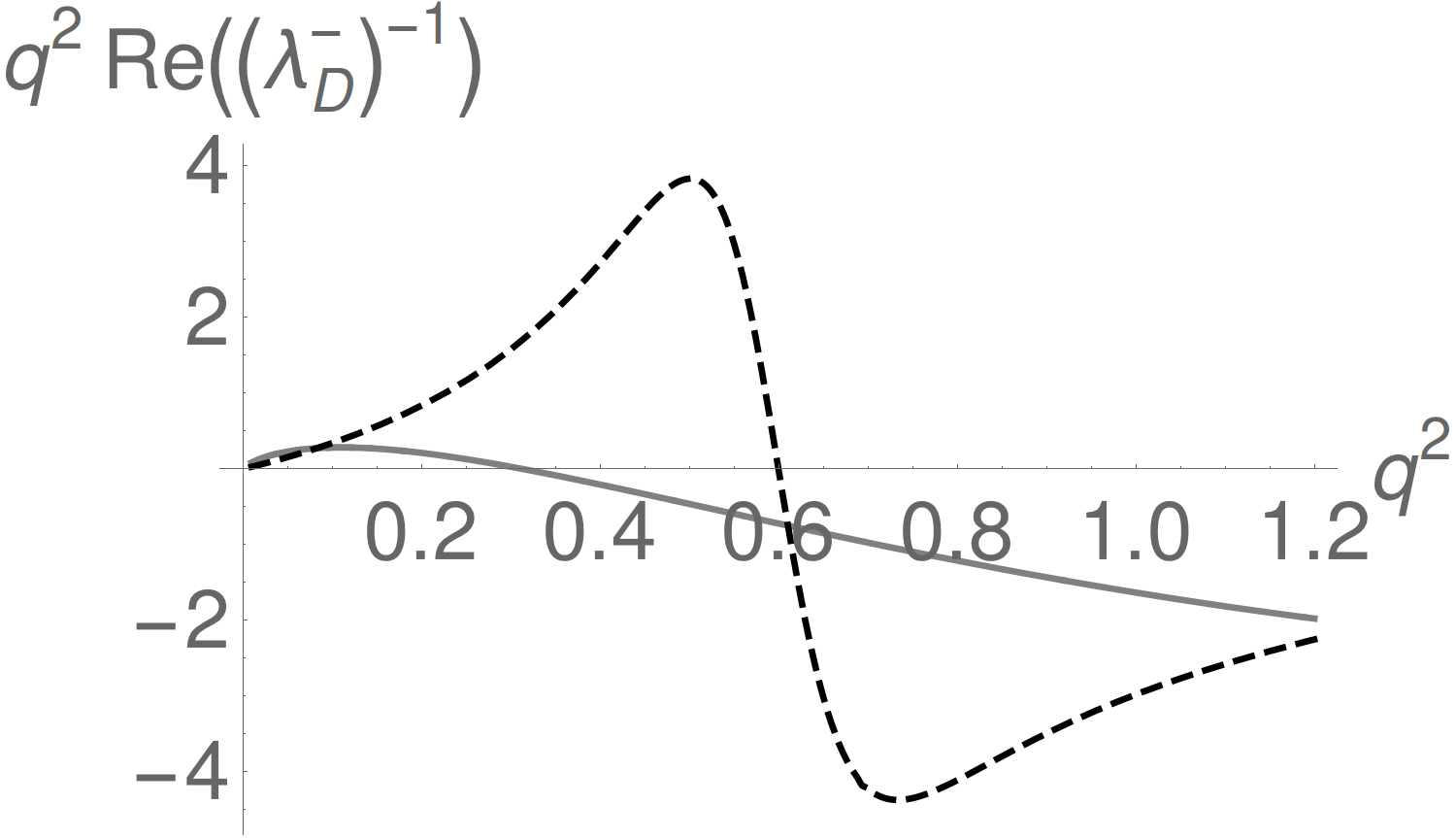}} \\[1em]
\fbox{\includegraphics[width=0.46\textwidth]{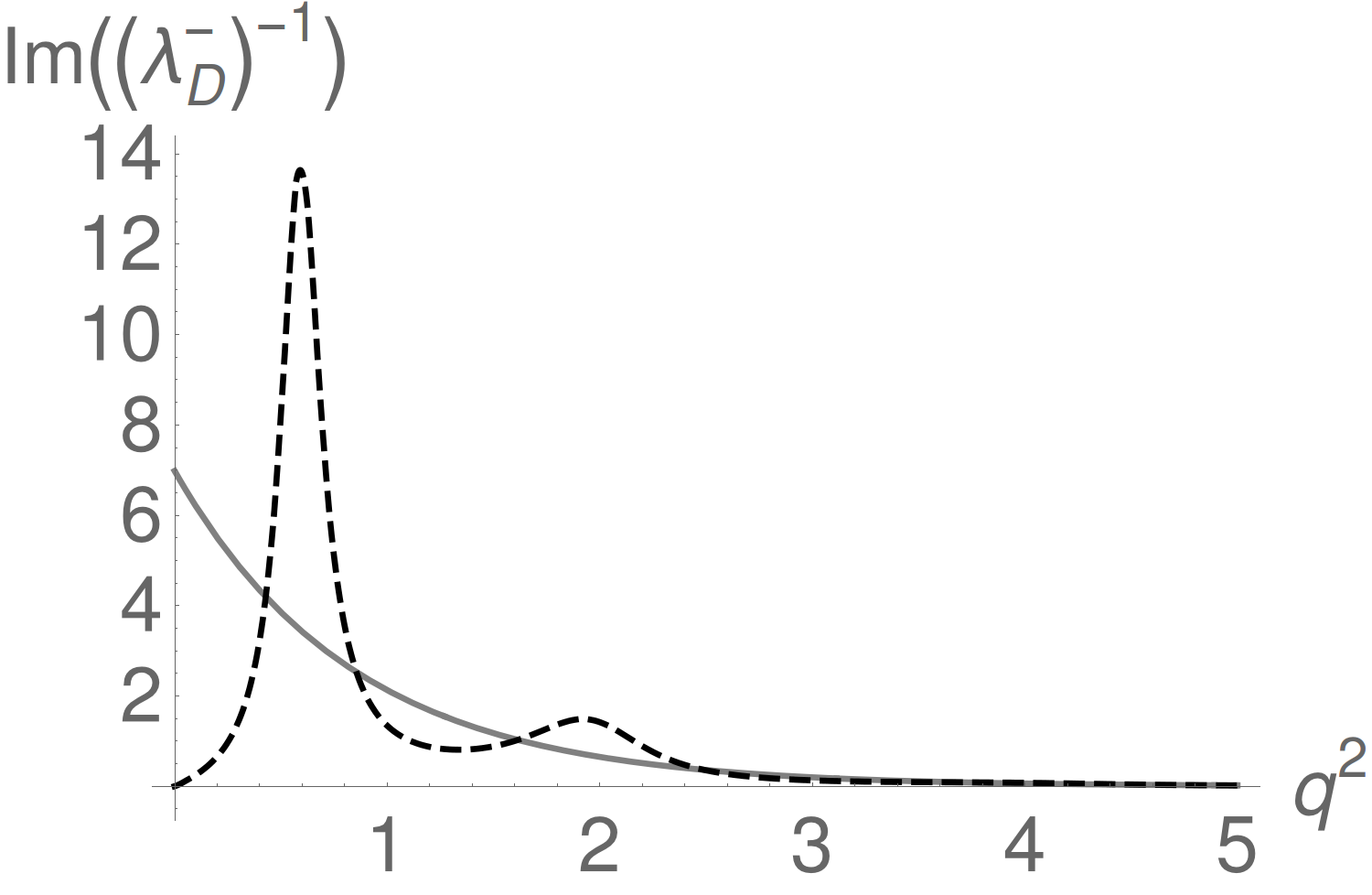}} \quad 
\fbox{\includegraphics[width=0.46\textwidth]{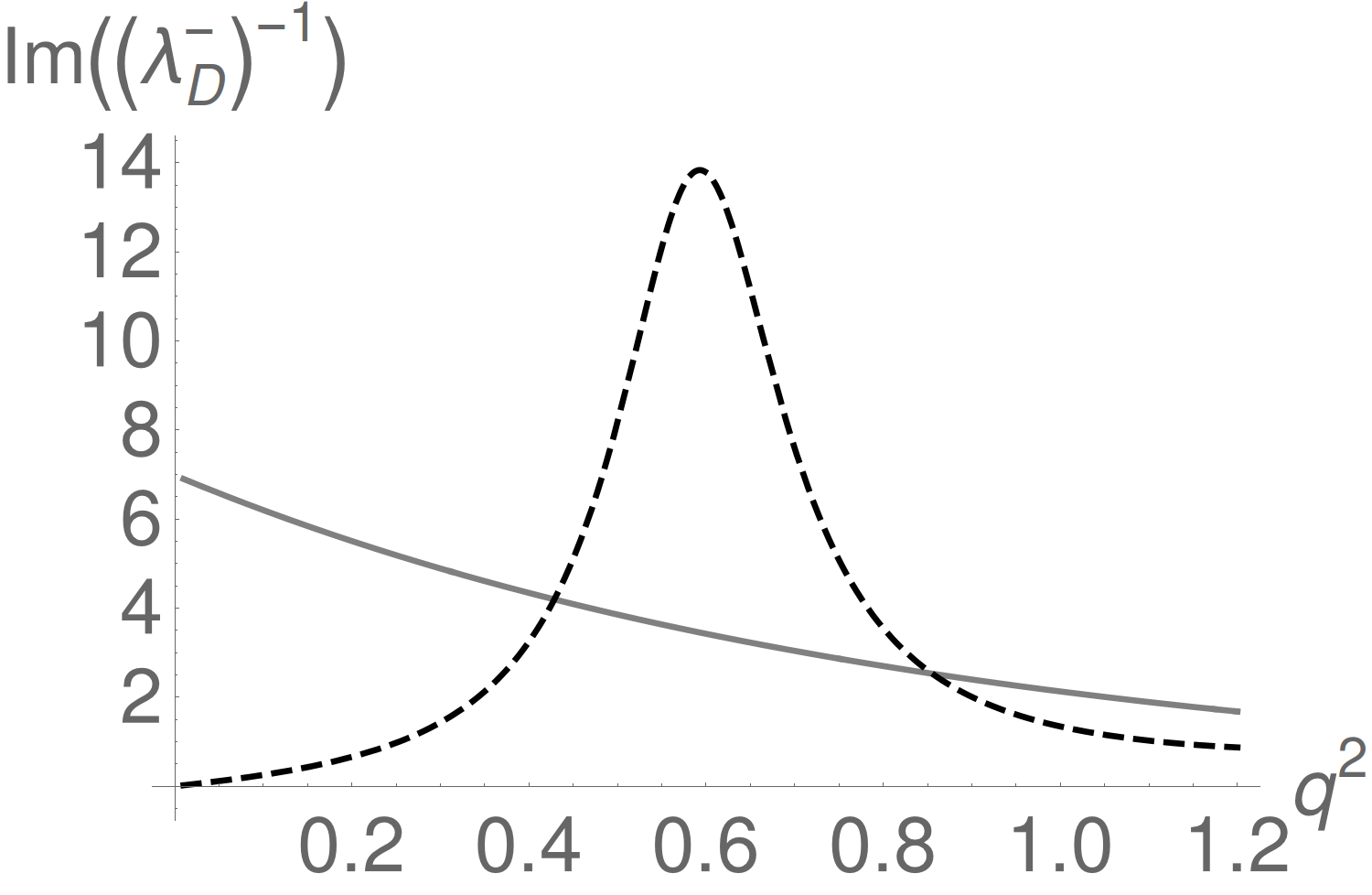}} 
\end{center}
\caption{Comparison of the perturbative result for the real and imaginary part of 
the function $\left(\lambda_D^-(q^2)\right)^{-1}$ 
(massless spectator, gray solid line) and the model (\ref{eq:lamminusmod}) for its hadronic modification  
(black dashed line) as a function of $q^2$ (in units of GeV$^2$).
Notice that the real part has been multiplied by $q^2$.
The parameter values are chosen as follows:
$\sigma^2=2~\text{GeV}^2$, $a=1$, $b=1/6$.
The $D$-meson LCDA is modelled by an exponential (\ref{eq:phim}) 
with $\omega_0=0.45$~GeV. 
The parameter $n_V=2.40$ is tuned to reproduce the result for deep Euclidean values of $q^2$.
\label{fig:plot9bis}}
\end{figure}

\begin{figure}[t!pbh]
\begin{center}
\fbox{\includegraphics[width=0.46\textwidth]{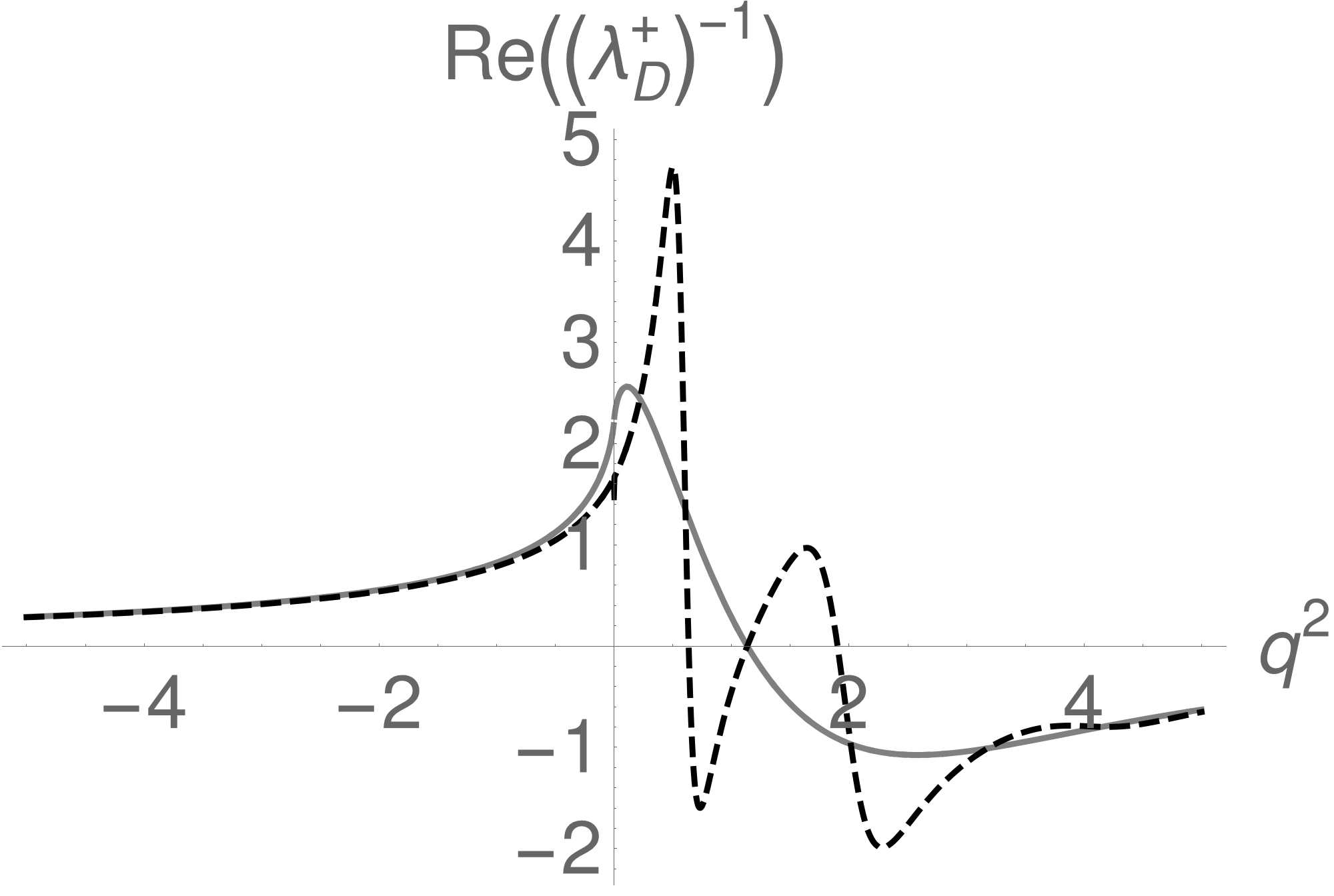}} \quad 
\fbox{\includegraphics[width=0.46\textwidth]{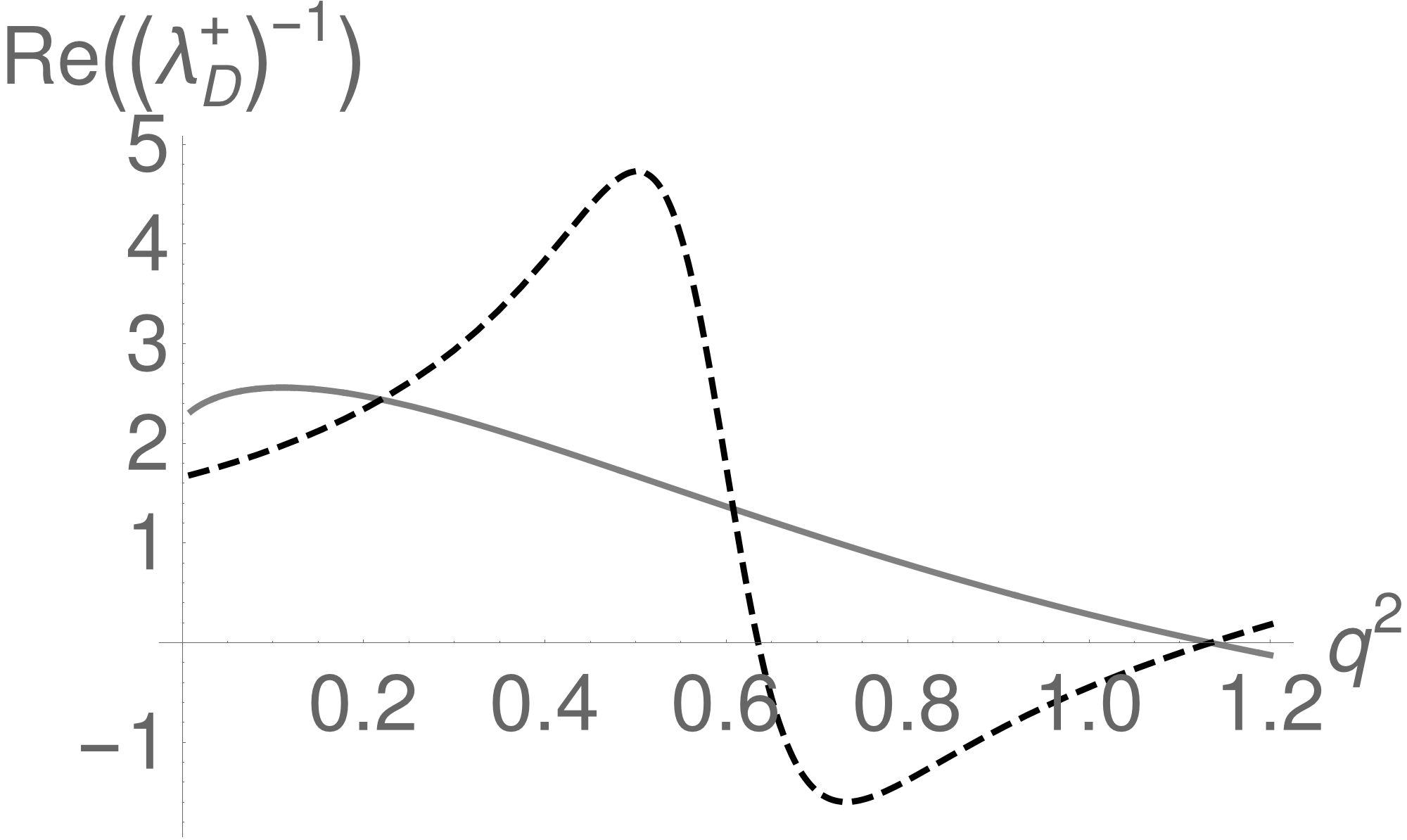}} \\[1em]
\cblue\fbox{\includegraphics[width=0.46\textwidth]{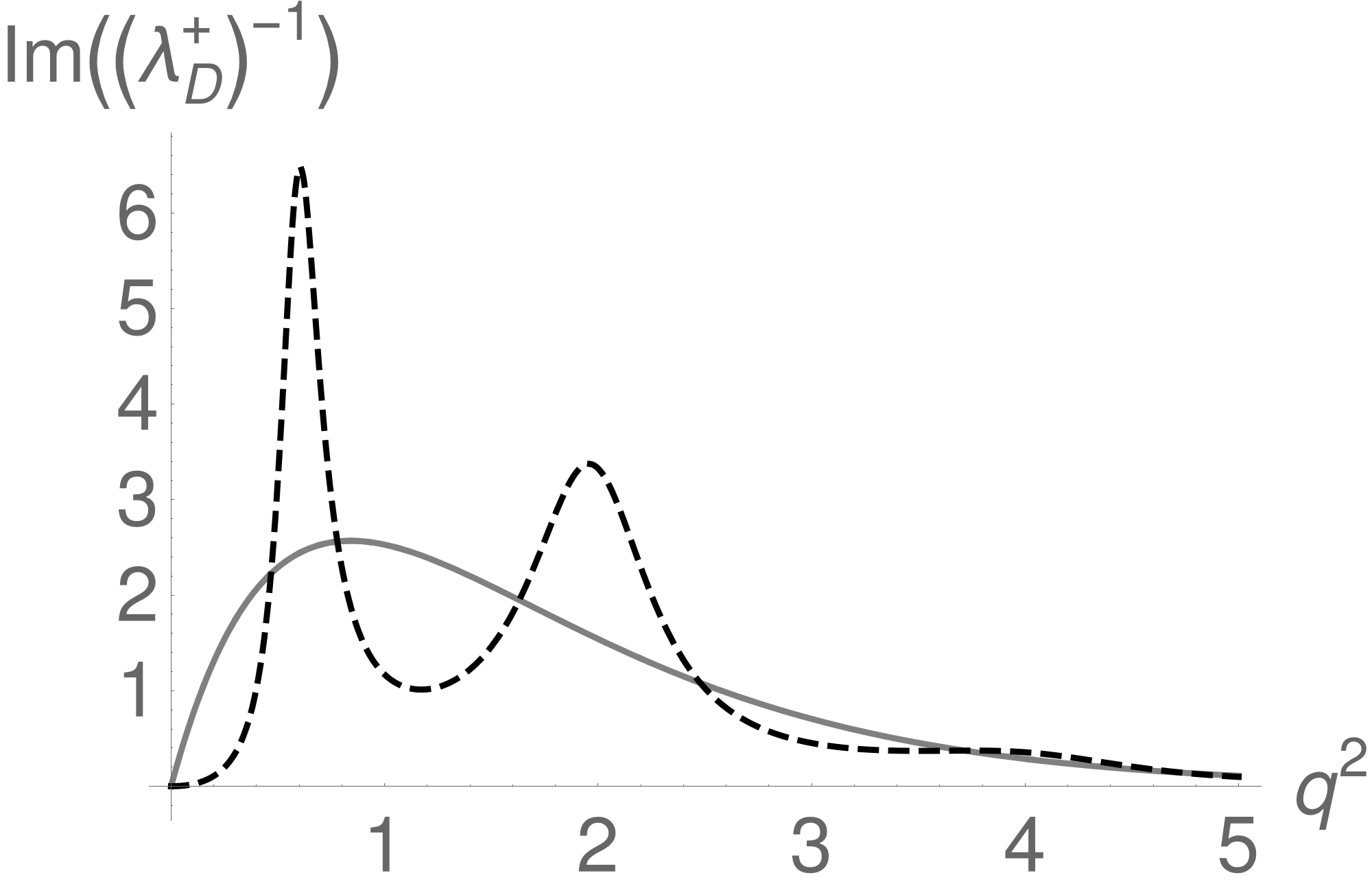}} \quad 
\fbox{\includegraphics[width=0.46\textwidth]{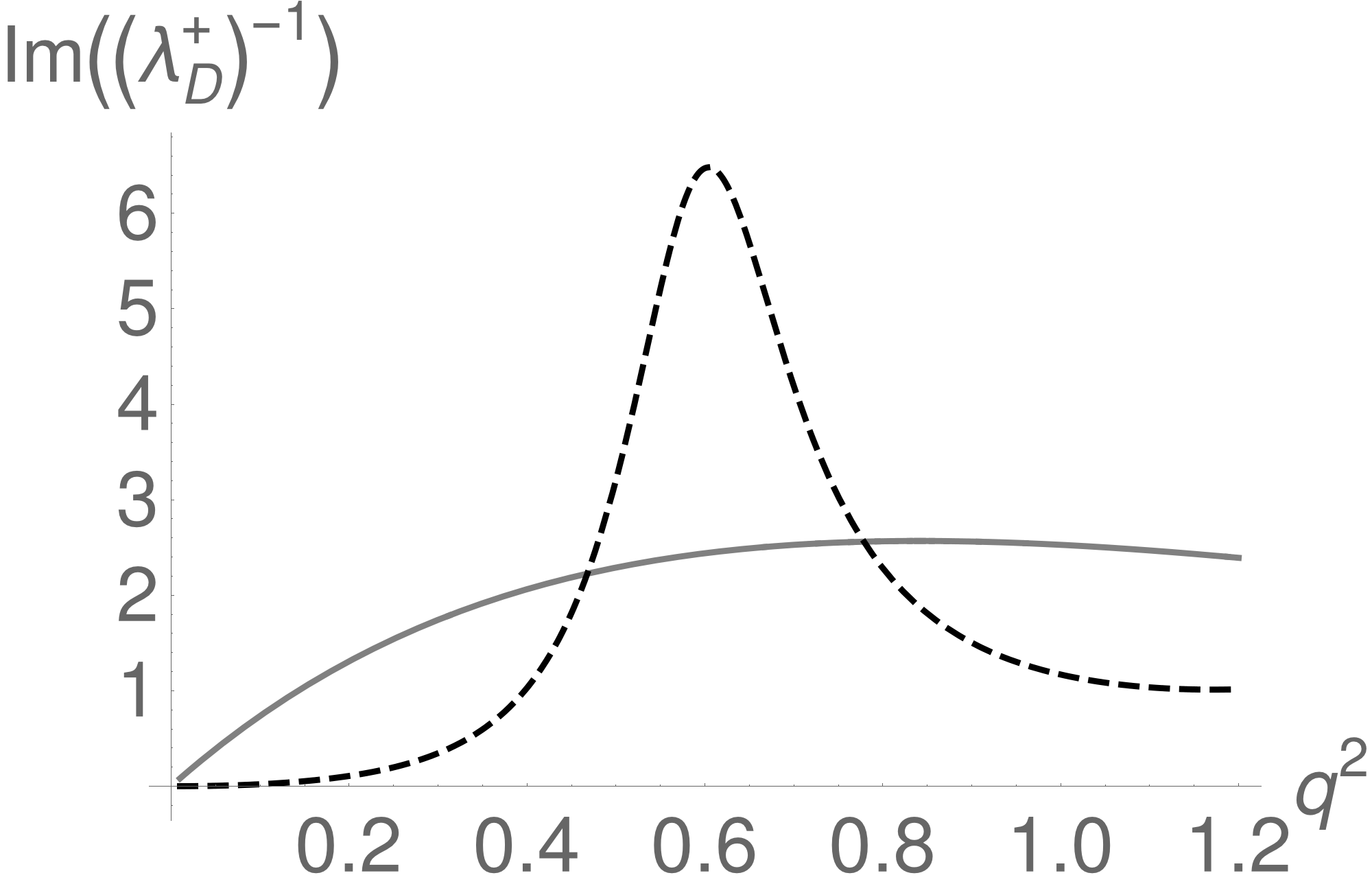}} 
\end{center}
\caption{Comparison of the perturbative result for the real and imaginary part of 
the function $\left(\lambda_D^+(q^2)\right)^{-1}$ 
(massless spectator, gray solid line) and the model (\ref{eq:lamplusmod}) for its hadronic modification  
(black dashed line) as a function of $q^2$ (in units of GeV$^2$).
The parameter values are chosen as follows:
$\sigma^2=2~\text{GeV}^2$, $a=1$, $b=1/6$.
The $D$-meson LCDA is modelled by an exponential with $\omega_0=0.45$~GeV. 
The parameter $n_V=1.75$ is tuned to reproduce the result for deep Euclidean values of $q^2$. 
\label{fig:plot5bis}}
\end{figure}

In the QCDF approach, the leading contributions to the annihilation topology are 
determined by the functions 
\begin{align}
  \left( \lambda_D^\pm(q^2) \right)^{-1} 
  &= \int_0^\infty \frac{d\omega}{\omega- q^2/m_D - i \epsilon} \, \phi_D^\pm(\omega)  \,.
\end{align} 
To the level of accuracy that we are working with, it is sufficient to neglect 3-particle
LCDAs in the $D$-meson. In that case the LCDAs $\phi_D^\pm(\omega)$ are not independent,
but fulfill a so-called Wandzura-Wilczek relation \cite{Beneke:2000wa},
\begin{align}
  \phi_D^+(\omega) & \simeq - \omega \, \frac{d\phi_D^-(\omega)}{d\omega} \,,
\end{align}
which for the above moment implies
\begin{align}
  - q^2 \, \frac{\partial}{\partial q^2} \left[ \lambda_D^-(q^2) \right]^{-1} 
  &= \left[ \lambda_D^+(q^2) \right]^{-1} \,. 
\label{lamrel}
\end{align}
Moreover, an alternative representation of the moments can be achieved in terms of 
the ``Wandzura-Wilczek'' wave function $\psi_D(x)$ as defined in \cite{Bell:2013tfa},
where 
\begin{align}
  \phi_D^+(\omega) &= \omega \, \int_\omega^\infty dx \, \psi_D(x) \,, 
  \qquad 
  \phi_D^-(\omega) = \int_\omega^\infty dx \, (x-\omega) \, \psi_D(x) \,,
\end{align}
and $x$ is twice the energy of the spectator quark in the $D$-meson.
With this, one has 
\begin{align}
  \left( \lambda_D^+(q^2) \right)^{-1} &= 
 \int_0^\infty dx \left[x+  \frac{q^2}{m_D} \, \ln\left( 1 - \frac{m_D \, x}{q^2+i\epsilon} \right) \right] 
  \psi_D(x)  \,,
  \cr 
   \left( \lambda_D^-(q^2) \right)^{-1} &= 
 \int_0^\infty dx \, x \,\ln\left( 1 - \frac{m_D \, x}{q^2+i\epsilon} \right) 
  \psi_D(x)  - \left( \lambda_D^+(q^2) \right)^{-1} \,.
\end{align}
Looking at the analytic properties of these expressions, one easily verifies that
\begin{align}
\frac{1}{\pi} {\rm Im} 
\left( \lambda_D^+(q^2) \right)^{-1} &= 
 \int_0^\infty dx \, \theta(q^2) \, \theta(m_D x - q^2) \, \frac{q^2}{m_D} \, \psi_D(x) = 
 \theta(q^2) \, \phi_D^+(q^2/m_D)  \,,
  \cr 
\frac{1}{\pi} {\rm Im}    \left( \lambda_D^-(q^2) \right)^{-1} &= 
 \int_0^\infty dx \, \theta(q^2) \, \theta(m_D x - q^2) \left(x-\frac{q^2}{m_D} \right) \psi_D(x) = 
 \theta(q^2) \, \phi_D^-(q^2/m_D) \,, \cr & 
\end{align}
and also (\ref{lamrel}) holds by simple differentiation.
As a naive ansatz we may simply replace $\theta(q^2)$ on the right-hand side of 
the above equations by the same function $j_{q}(q^2)$ that has been used to model
the hadronic spectrum for the function $h(q^2,m_q)$ in the previous subsection.
{\cblue This corresponds to the naive factorization into
the subprocesses $D \to \rho q\bar q$ (described in QCDF in terms of $\phi_D^\pm$) 
followed by $q\bar q \to \gamma^*$ (described by the resonance model).}\footnote{%
Obviously, this simple recipe would not work at higher orders in the strong coupling.
For a related discussion of the analytic properties of the spectator 
contributions from the operator ${\cal O}_8^g$ in the context of light-cone sum rules
can be found in \cite{Dimou:2012un}.}
With this, we take
\begin{align*}  
  \left( \lambda_D^\pm(q^2) \right)^{-1}  &  \to  
  \int_0^\infty  \frac{ds}{s-q^2-i\epsilon} \, 
  \phi_D^\pm(s/m_D) \,  j_{u/d}(s) 
  \,.
\end{align*}
As before, we tune the parameter $n_V$ (independently for both cases)
to reproduce the asymptotic limit $q^2 \ll - m_D \Lambda$, i.e.\ we
require 
\begin{align}
  \int_0^\infty  ds \, 
  \phi_D^\pm(s/m_D) \big( j_{u/d}(s) - 1 \big) &\stackrel{!}{=} 0 \,. 
\end{align}
Using an exponential model (\ref{eq:phim}) for the $D$-meson LCDAs
with $\omega_0=0.45$~GeV,
we obtain $n_V=2.40$ for $\lambda_D^-(q^2)$ and $n_V=1.75$ for $\lambda_D^+(q^2)$,
respectively.
The comparison between the perturbative result and the hadronic model
is shown in Figs.~\ref{fig:plot9bis},\ref{fig:plot5bis}, 
see also the discussion around  Figs.~\ref{fig:lammin},\ref{fig:lamplus} in the main text.

We may again analyze our model for $\lambda_D^\pm(q^2)$ 
in the framework of QCD sum rules.
With the analogous steps as in (\ref{sr1a}) this leads to 
\begin{align}
  \frac{n_V \, M_V^2}{M_V^2-q^2} \, \phi_D^\pm(M_V^2/m_D) & \simeq 
  \int_0^{s_0} ds \, \frac{1}{s-q^2 - i\epsilon} \, \phi_D^\pm(s/m_D) \,\theta(s) \,,
  \label{sr2a}
\end{align}
and after Borel transformation one has
\begin{align}
  e^{- M_V^2/M^2} \, n_V \, M_V^2\, \phi_D^\pm(M_V^2/m_D)
  &\simeq \int_0^{s_0} ds \, e^{-s/M^2} \, \phi_D^\pm(s/m_D) \,.  
  \label{sr2b}
\end{align}
In the heavy-mass limit, $m_D \cdot \langle \omega \rangle \gg s_0 \sim M^2$, one 
can expand the LCDAs,
$$\phi_D^-(\omega) \simeq \phi_D^-(0) \,, \qquad 
\phi_D^+(\omega) \simeq \omega \, \phi_D^+{}'(0) \,, $$
and (\ref{sr2b}) reduces to
\begin{align}
  M_V^2 \, n_V & =M^2 \, e^{M_V^2/M^2} \left( 1- e^{-s_0/M^2} \right) 
  \cr &= 4\pi^2 \, f_V^2 \qquad \left[\mbox{for $\lambda_D^-(q^2)$}\right],
\label{nVlammminus}
\end{align}
respectively,
\begin{align}
  M_V^2 \, n_V &= \frac{M^2}{M_V^2} \, e^{M_V^2/M^2} 
  \left( M^2 - e^{-s_0/M^2} \, (s_0 + M^2) \right)
  \cr & = 4\pi^2 f_V^2 \, \frac{M^2}{M_V^2}
  \frac{ 1 - e^{-s_0/M^2} \, (1+s_0/M^2)}{1- e^{-s_0/M^2}} 
  \qquad \left[\mbox{for $\lambda_D^+(q^2)$}\right].
\end{align}
Again, within the intrinsic uncertainties, the numerical values for $n_V$
fitted to the asymptotic behaviour of $\lambda_D^\pm(q^2)$ are consistent
with the above findings (without going into details of the sum-rule
analysis). In particular, {\cblue by comparing Eqs.~(\ref{nVlammminus}) 
and (\ref{sr1b})}, 
the value of $n_V$ -- to first approximation --
is expected to be universal\footnote{A similar argument has been used in 
\cite{DeFazio:2005dx,DeFazio:2007hw} to show the equivalence of 
the QCD factorization approach and the light-cone sum rule approach 
for the description of radiative QCD corrections to heavy-to-light 
form-factor ratios.} for the modelling 
of $h(s,m_q)$ and $\lambda_D^-(q^2)$.

\clearpage 

\clearpage 

\clearpage 

\section{More Input Parameters}

\label{app:input}

In Table~\ref{tab:rest} we summarize a number of input parameters
that have been used in the numerical analyses.
The parameters describing the $q^2$-dependence of the 
the $D \to \rho$ form factors are quoted after Eq.~(\ref{eq:ffmodel})
in the main body of the text.

\begin{table}[h!]
\centering
\renewcommand{\arraystretch}{1.0}
\begin{tabular}{|c|c|}
 \hline
 $\alpha_{\rm{em}}$ & $1 / 137$ \\ 
 $\bar{m}_c(\bar{m}_c)$ & $1.275 $ GeV \\ 
 $\bar{m}_s(2 \,\rm{GeV})$ & $93.5$ MeV \\ 
 \hline
 $M_{D^\pm}$ & $1.870$ GeV\\ 
 $M_{D^0}$ & $1.865$ GeV \\ 
 $\tau_{D^\pm}$ & $1.04$ ps\\ 
 $\tau_{D^0}$ & $0.41$ ps\\ 
 \hline
 $M_{D^{*\pm}}$ & $2.010$ GeV \\ 
 $M_{D^{*0}}$ & $2.007$ GeV\\ 
 \hline
\end{tabular}\qquad 
\begin{tabular}{|c|c|}
\hline
 $m_\rho$ & $0.775$ GeV \\ 
 $\Gamma_\rho$ & $0.149$ GeV \\
 \hline
 $m_\phi$ & $1.019$ GeV \\ 
 $\Gamma_\phi$ & $4.26$ MeV \\
 $m_K$ & $0.497$ GeV \\
 \hline
 $\lambda$ & $0.22506$\\ 
 $A$ & $0.811$ \\ 
 $\bar{\rho}$ & $0.124$ \\ 
 $\bar{\eta}$ & $0.356$ \\ 
 \hline
\end{tabular}
\caption{\label{tab:rest} Summary of input parameters not quoted in Table~\ref{tab:hadronic}.
All values taken from \cite{PDG}. (Here $\lambda$, $A$, $\bar\rho$, $\bar\eta$ are
the Wolfenstein parameters for the CKM matrix.)}
\end{table}

\bibliographystyle{JHEP-2}
\bibliography{bib}

\providecommand{\href}[2]{#2}\begingroup\raggedright\begin{thebibliography}{10}

\bibitem{Antonelli:2009ws}
M.~Antonelli {\em et.~al.}, {\it {Flavor Physics in the Quark Sector}},  {\em
  Phys. Rept.} {\bf 494} (2010) 197--414
  [\href{http://arXiv.org/abs/0907.5386}{{\tt 0907.5386}}].

\bibitem{Bediaga:2012py}
{\bf LHCb} Collaboration, R.~Aaij {\em et.~al.}, {\it {Implications of LHCb
  measurements and future prospects}},  {\em Eur. Phys. J.} {\bf C73} (2013),
  no.~4 2373 [\href{http://arXiv.org/abs/1208.3355}{{\tt 1208.3355}}].

\bibitem{Buchalla:2008jp}
M.~Artuso {\em et.~al.}, {\it {$B$, $D$ and $K$ decays}},  {\em Eur. Phys. J.}
  {\bf C57} (2008) 309--492 [\href{http://arXiv.org/abs/0801.1833}{{\tt
  0801.1833}}].

\bibitem{Fajfer:2001sa}
S.~Fajfer, S.~Prelovsek and P.~Singer, {\it {Rare charm meson decays $D \to P
  \ell^+ \ell^-$ and $c \to u \ell^+ \ell^-$ in SM and MSSM}},  {\em Phys.
  Rev.} {\bf D64} (2001) 114009
  [\href{http://arXiv.org/abs/hep-ph/0106333}{{\tt hep-ph/0106333}}].

\bibitem{Burdman:2001tf}
G.~Burdman, E.~Golowich, J.~L. Hewett and S.~Pakvasa, {\it {Rare charm decays
  in the standard model and beyond}},  {\em Phys. Rev.} {\bf D66} (2002) 014009
  [\href{http://arXiv.org/abs/hep-ph/0112235}{{\tt hep-ph/0112235}}].

\bibitem{Fajfer:2005ke}
S.~Fajfer and S.~Prelovsek, {\it {Effects of littlest Higgs model in rare D
  meson decays}},  {\em Phys. Rev.} {\bf D73} (2006) 054026
  [\href{http://arXiv.org/abs/hep-ph/0511048}{{\tt hep-ph/0511048}}].

\bibitem{Fajfer:2007dy}
S.~Fajfer, N.~Kosnik and S.~Prelovsek, {\it {Updated constraints on new physics
  in rare charm decays}},  {\em Phys. Rev.} {\bf D76} (2007) 074010
  [\href{http://arXiv.org/abs/0706.1133}{{\tt 0706.1133}}].

\bibitem{Paul:2011ar}
A.~Paul, I.~I. Bigi and S.~Recksiegel, {\it {On $D\to X_u \ell^+ \ell^-$ within
  the Standard Model and Frameworks like the Littlest Higgs Model with T
  Parity}},  {\em Phys. Rev.} {\bf D83} (2011) 114006
  [\href{http://arXiv.org/abs/1101.6053}{{\tt 1101.6053}}].

\bibitem{Cappiello:2012vg}
L.~Cappiello, O.~Cata and G.~D'Ambrosio, {\it {Standard Model prediction and
  new physics tests for $D^0 \to h^+ h^- \ell^+ \ell^- (h=\pi,K:
  \ell=e,\mu)$}},  {\em JHEP} {\bf 04} (2013) 135
  [\href{http://arXiv.org/abs/1209.4235}{{\tt 1209.4235}}].

\bibitem{Isidori:2012yx}
G.~Isidori and J.~F. Kamenik, {\it {Shedding light on CP violation in the charm
  system via $D \to V \gamma$ decays}},  {\em Phys. Rev. Lett.} {\bf 109}
  (2012) 171801 [\href{http://arXiv.org/abs/1205.3164}{{\tt 1205.3164}}].

\bibitem{Lyon:2012fk}
J.~Lyon and R.~Zwicky, {\it {Anomalously large ${\cal O}_8$ and long-distance
  chirality from $A_{\rm CP}[D^0 \to (\rho^0,\omega) \gamma](t)$}},
  \href{http://arXiv.org/abs/1210.6546}{{\tt 1210.6546}}.

\bibitem{Fajfer:2012nr}
S.~Fajfer and N.~Kosnik, {\it {Resonance catalyzed CP asymmetries in $D\to P
  \ell^+\ell^-$}},  {\em Phys. Rev.} {\bf D87} (2013), no.~5 054026
  [\href{http://arXiv.org/abs/1208.0759}{{\tt 1208.0759}}].

\bibitem{Fajfer:2015mia}
S.~Fajfer and N.~Kosnik, {\it {Prospects of discovering new physics in rare
  charm decays}},  {\em Eur. Phys. J.} {\bf C75} (2015), no.~12 567
  [\href{http://arXiv.org/abs/1510.00965}{{\tt 1510.00965}}].

\bibitem{deBoer:2015boa}
S.~de~Boer and G.~Hiller, {\it {Flavor and new physics opportunities with rare
  charm decays into leptons}},  {\em Phys. Rev.} {\bf D93} (2016), no.~7 074001
  [\href{http://arXiv.org/abs/1510.00311}{{\tt 1510.00311}}].

\bibitem{deBoer:2017que}
S.~de~Boer and G.~Hiller, {\it {Rare radiative charm decays within the standard
  model and beyond}},  \href{http://arXiv.org/abs/1701.06392}{{\tt
  1701.06392}}.

\bibitem{Biswas:2017eyn}
A.~Biswas, S.~Mandal and N.~Sinha, {\it {Searching for New physics in Charm
  Radiative decays}},  \href{http://arXiv.org/abs/1702.05059}{{\tt
  1702.05059}}.

\bibitem{Greub:1996wn}
C.~Greub, T.~Hurth, M.~Misiak and D.~Wyler, {\it {The $c \to u \gamma$
  contribution to weak radiative charm decay}},  {\em Phys. Lett.} {\bf B382}
  (1996) 415--420 [\href{http://arXiv.org/abs/hep-ph/9603417}{{\tt
  hep-ph/9603417}}].

\bibitem{deBoer:2016dcg}
S.~de~Boer, B.~Müller and D.~Seidel, {\it {Higher-order Wilson coefficients
  for $c \to u$ transitions in the Standard Model}},
  \href{http://arXiv.org/abs/1606.05521}{{\tt 1606.05521}}.

\bibitem{Beneke:2001at}
M.~Beneke, T.~Feldmann and D.~Seidel, {\it {Systematic approach to exclusive $B
  \to V \ell^+ \ell^-$, $V \gamma$ decays}},  {\em Nucl. Phys.} {\bf B612}
  (2001) 25--58 [\href{http://arXiv.org/abs/hep-ph/0106067}{{\tt
  hep-ph/0106067}}].

\bibitem{Bosch:2001gv}
S.~W. Bosch and G.~Buchalla, {\it {The Radiative decays $B \to V \gamma$ at
  next-to-leading order in QCD}},  {\em Nucl. Phys.} {\bf B621} (2002) 459--478
  [\href{http://arXiv.org/abs/hep-ph/0106081}{{\tt hep-ph/0106081}}].

\bibitem{Ali:2001ez}
A.~Ali and A.~Y. Parkhomenko, {\it {Branching ratios for $B \to K^* \gamma$ and
  $B \to \rho \gamma$ decays in next-to-leading order in the large energy
  effective theory}},  {\em Eur. Phys. J.} {\bf C23} (2002) 89--112
  [\href{http://arXiv.org/abs/hep-ph/0105302}{{\tt hep-ph/0105302}}].

\bibitem{Kagan:2001zk}
A.~L. Kagan and M.~Neubert, {\it {Isospin breaking in $B \to K^* \gamma$
  decays}},  {\em Phys. Lett.} {\bf B539} (2002) 227--234
  [\href{http://arXiv.org/abs/hep-ph/0110078}{{\tt hep-ph/0110078}}].

\bibitem{Feldmann:2002iw}
T.~Feldmann and J.~Matias, {\it {Forward backward and isospin asymmetry for $B
  \to K^* l^+ l^-$ decay in the standard model and in supersymmetry}},  {\em
  JHEP} {\bf 01} (2003) 074 [\href{http://arXiv.org/abs/hep-ph/0212158}{{\tt
  hep-ph/0212158}}].

\bibitem{Beneke:2004dp}
M.~Beneke, T.~Feldmann and D.~Seidel, {\it {Exclusive radiative and electroweak
  $b \to d$ and $b \to s$ penguin decays at NLO}},  {\em Eur. Phys. J.} {\bf
  C41} (2005) 173--188 [\href{http://arXiv.org/abs/hep-ph/0412400}{{\tt
  hep-ph/0412400}}].

\bibitem{Ali:2007sj}
A.~Ali, B.~D. Pecjak and C.~Greub, {\it {$B \to$ V $\gamma$ Decays at NNLO in
  SCET}},  {\em Eur. Phys. J.} {\bf C55} (2008) 577--595
  [\href{http://arXiv.org/abs/0709.4422}{{\tt 0709.4422}}].

\bibitem{Bartsch:2009qp}
M.~Bartsch, M.~Beylich, G.~Buchalla and D.~N. Gao, {\it {Precision Flavour
  Physics with $B \to K \nu \bar\nu$ and $B \to K l^+ l^-$}},  {\em JHEP} {\bf
  11} (2009) 011 [\href{http://arXiv.org/abs/0909.1512}{{\tt 0909.1512}}].

\bibitem{Jager:2012uw}
S.~J{\"a}ger and J.~Martin~Camalich, {\it {On $B \to V \ell \ell$ at small
  dilepton invariant mass, power corrections, and new physics}},  {\em JHEP}
  {\bf 05} (2013) 043 [\href{http://arXiv.org/abs/1212.2263}{{\tt 1212.2263}}].

\bibitem{Buchalla:1998mt}
G.~Buchalla and G.~Isidori, {\it {Nonperturbative effects in $\bar B \to X_s
  l^+ l^-$ for large dilepton invariant mass}},  {\em Nucl. Phys.} {\bf B525}
  (1998) 333--349 [\href{http://arXiv.org/abs/hep-ph/9801456}{{\tt
  hep-ph/9801456}}].

\bibitem{Grinstein:2004vb}
B.~Grinstein and D.~Pirjol, {\it {Exclusive rare $B \to K^*\ell^+\ell^-$ decays
  at low recoil: Controlling the long-distance effects}},  {\em Phys. Rev.}
  {\bf D70} (2004) 114005 [\href{http://arXiv.org/abs/hep-ph/0404250}{{\tt
  hep-ph/0404250}}].

\bibitem{Beylich:2011aq}
M.~Beylich, G.~Buchalla and T.~Feldmann, {\it {Theory of $B \to K^{(*)}\ell^+
  \ell^-$ decays at high $q^2$: OPE and quark-hadron duality}},  {\em Eur.
  Phys. J.} {\bf C71} (2011) 1635 [\href{http://arXiv.org/abs/1101.5118}{{\tt
  1101.5118}}].

\bibitem{Khodjamirian:2010vf}
A.~Khodjamirian, T.~Mannel, A.~A. Pivovarov and Y.~M. Wang, {\it {Charm-loop
  effect in $B \to K^{(*)} \ell^{+} \ell^{-}$ and $B\to K^*\gamma$}},  {\em
  JHEP} {\bf 09} (2010) 089 [\href{http://arXiv.org/abs/1006.4945}{{\tt
  1006.4945}}].

\bibitem{Lyon:2014hpa}
J.~Lyon and R.~Zwicky, {\it {Resonances gone topsy turvy - the charm of QCD or
  new physics in $b \to s \ell^+ \ell^-$?}},
  \href{http://arXiv.org/abs/1406.0566}{{\tt 1406.0566}}.

\bibitem{Aaij:2013pta}
{\bf LHCb} Collaboration, R.~Aaij {\em et.~al.}, {\it {Observation of a
  resonance in $B^+ \to K^+ \mu^+\mu^-$ decays at low recoil}},  {\em Phys.
  Rev. Lett.} {\bf 111} (2013), no.~11 112003
  [\href{http://arXiv.org/abs/1307.7595}{{\tt 1307.7595}}].

\bibitem{Beneke:1999br}
M.~Beneke, G.~Buchalla, M.~Neubert and C.~T. Sachrajda, {\it {QCD factorization
  for $B \to \pi \pi$ decays: Strong phases and CP violation in the heavy quark
  limit}},  {\em Phys. Rev. Lett.} {\bf 83} (1999) 1914--1917
  [\href{http://arXiv.org/abs/hep-ph/9905312}{{\tt hep-ph/9905312}}].

\bibitem{Beneke:2001ev}
M.~Beneke, G.~Buchalla, M.~Neubert and C.~T. Sachrajda, {\it {QCD factorization
  in $B \to \pi K, \pi \pi$ decays and extraction of Wolfenstein parameters}},
  {\em Nucl. Phys.} {\bf B606} (2001) 245--321
  [\href{http://arXiv.org/abs/hep-ph/0104110}{{\tt hep-ph/0104110}}].

\bibitem{Blok:1997hs}
B.~Blok, M.~A. Shifman and D.-X. Zhang, {\it {An Illustrative example of how
  quark hadron duality might work}},  {\em Phys. Rev.} {\bf D57} (1998)
  2691--2700 [\href{http://arXiv.org/abs/hep-ph/9709333}{{\tt
  hep-ph/9709333}}]. [Erratum: Phys. Rev.D59,019901(1999)].

\bibitem{Shifman:2000jv}
M.~A. Shifman, {\it {Quark hadron duality}},  in {\em {Proceedings, 8th
  International Symposium on Heavy Flavor Physics (Heavy Flavors 8)}},
  p.~hf8/013, 2000.
\newblock \href{http://arXiv.org/abs/hep-ph/0009131}{{\tt hep-ph/0009131}}.

\bibitem{Shifman:2003de}
M.~Shifman, {\it {The quark hadron duality}},  {\em eConf} {\bf C030614} (2003)
  001.

\bibitem{Chetyrkin:1997gb}
K.~G. Chetyrkin, M.~Misiak and M.~M{\"u}nz, {\it {$|\Delta F| = 1$ nonleptonic
  effective Hamiltonian in a simpler scheme}},  {\em Nucl. Phys.} {\bf B520}
  (1998) 279--297 [\href{http://arXiv.org/abs/hep-ph/9711280}{{\tt
  hep-ph/9711280}}].

\bibitem{Seidel:2004jh}
D.~Seidel, {\it {Analytic two loop virtual corrections to $b \to d \ell^+
  \ell^-$}},  {\em Phys. Rev.} {\bf D70} (2004) 094038
  [\href{http://arXiv.org/abs/hep-ph/0403185}{{\tt hep-ph/0403185}}].

\bibitem{Beneke:2000wa}
M.~Beneke and T.~Feldmann, {\it {Symmetry breaking corrections to heavy to
  light B meson form-factors at large recoil}},  {\em Nucl. Phys.} {\bf B592}
  (2001) 3--34 [\href{http://arXiv.org/abs/hep-ph/0008255}{{\tt
  hep-ph/0008255}}].

\bibitem{Buchalla:2010jv}
G.~Buchalla, {\it {Precision flavour physics with $B\to K\nu\bar\nu$ and $B\to
  Kl^+l^-$}},  {\em Nucl. Phys. Proc. Suppl.} {\bf 209} (2010) 137--142
  [\href{http://arXiv.org/abs/1010.2674}{{\tt 1010.2674}}].

\bibitem{CLEO:2011ab}
{\bf CLEO} Collaboration, S.~Dobbs {\em et.~al.}, {\it {First Measurement of
  the Form Factors in the Decays $D^0 \to \rho^- e^+ \nu_e$ and $D^+ \to \rho^0
  e^+ \nu_e$}},  {\em Phys. Rev. Lett.} {\bf 110} (2013), no.~13 131802
  [\href{http://arXiv.org/abs/1112.2884}{{\tt 1112.2884}}].

\bibitem{Fajfer:2005ug}
S.~Fajfer and J.~F. Kamenik, {\it {Charm meson resonances and $D \to V$
  semileptonic form-factors}},  {\em Phys. Rev.} {\bf D72} (2005) 034029
  [\href{http://arXiv.org/abs/hep-ph/0506051}{{\tt hep-ph/0506051}}].

\bibitem{Buchalla:1995vs}
G.~Buchalla, A.~J. Buras and M.~E. Lautenbacher, {\it {Weak decays beyond
  leading logarithms}},  {\em Rev. Mod. Phys.} {\bf 68} (1996) 1125--1144
  [\href{http://arXiv.org/abs/hep-ph/9512380}{{\tt hep-ph/9512380}}].

\bibitem{Khodjamirian:1995uc}
A.~Khodjamirian, G.~Stoll and D.~Wyler, {\it {Calculation of long distance
  effects in exclusive weak radiative decays of B meson}},  {\em Phys. Lett.}
  {\bf B358} (1995) 129--138 [\href{http://arXiv.org/abs/hep-ph/9506242}{{\tt
  hep-ph/9506242}}].

\bibitem{Gambino:2003zm}
P.~Gambino, M.~Gorbahn and U.~Haisch, {\it {Anomalous dimension matrix for
  radiative and rare semileptonic B decays up to three loops}},  {\em Nucl.
  Phys.} {\bf B673} (2003) 238--262
  [\href{http://arXiv.org/abs/hep-ph/0306079}{{\tt hep-ph/0306079}}].

\bibitem{Greub:2008cy}
C.~Greub, V.~Pilipp and C.~Schupbach, {\it {Analytic calculation of two-loop
  QCD corrections to $b \to s \ell^+ \ell^-$ in the high $q^2$ region}},  {\em
  JHEP} {\bf 12} (2008) 040 [\href{http://arXiv.org/abs/0810.4077}{{\tt
  0810.4077}}].

\bibitem{Greub:1996tg}
C.~Greub, T.~Hurth and D.~Wyler, {\it {Virtual ${\cal O}(\alpha_s)$ corrections
  to the inclusive decay $b \to s \gamma$}},  {\em Phys. Rev.} {\bf D54} (1996)
  3350--3364 [\href{http://arXiv.org/abs/hep-ph/9603404}{{\tt
  hep-ph/9603404}}].

\bibitem{Grozin:1996pq}
A.~G. Grozin and M.~Neubert, {\it {Asymptotics of heavy meson form-factors}},
  {\em Phys. Rev.} {\bf D55} (1997) 272--290
  [\href{http://arXiv.org/abs/hep-ph/9607366}{{\tt hep-ph/9607366}}].

\bibitem{Braun:2012kp}
V.~M. Braun and A.~Khodjamirian, {\it {Soft contribution to $B\to \gamma \ell
  \nu_\ell$ and the $B$-meson distribution amplitude}},  {\em Phys. Lett.} {\bf
  B718} (2013) 1014--1019 [\href{http://arXiv.org/abs/1210.4453}{{\tt
  1210.4453}}].

\bibitem{Beneke:2011nf}
M.~Beneke and J.~Rohrwild, {\it {$B$-meson distribution amplitude from $B \to
  \gamma \ell \nu$}},  {\em Eur. Phys. J.} {\bf C71} (2011) 1818
  [\href{http://arXiv.org/abs/1110.3228}{{\tt 1110.3228}}].

\bibitem{PDG}
{\bf Particle Data Group} Collaboration, C.~Patrignani {\em et.~al.}, {\it
  {Review of Particle Physics}},  {\em Chin. Phys.} {\bf C40} (2016), no.~10
  100001.

\bibitem{Ball:2006nr}
P.~Ball and R.~Zwicky, {\it {$|V_{td} / V_{ts}|$ from $B \to V \gamma$}},  {\em
  JHEP} {\bf 04} (2006) 046 [\href{http://arXiv.org/abs/hep-ph/0603232}{{\tt
  hep-ph/0603232}}].

\bibitem{Vladikas:2015bra}
A.~Vladikas, {\it {FLAG: Lattice QCD Tests of the Standard Model and Foretaste
  for Beyond}},  {\em PoS} {\bf FPCP2015} (2015) 016
  [\href{http://arXiv.org/abs/1509.01155}{{\tt 1509.01155}}].

\bibitem{Kruger:2005ep}
F.~Kr{\"u}ger and J.~Matias, {\it {Probing new physics via the transverse
  amplitudes of $B^0\to K^{*0} (\to K^- \pi^+) l^+l^-$ at large recoil}},  {\em
  Phys. Rev.} {\bf D71} (2005) 094009
  [\href{http://arXiv.org/abs/hep-ph/0502060}{{\tt hep-ph/0502060}}].

\bibitem{Abdesselam:2016yvr}
{\bf Belle} Collaboration, A.~Abdesselam {\em et.~al.}, {\it {Observation of
  $D^0\to \rho^0\gamma$ and search for $CP$ violation in radiative charm
  decays}},  {\em Phys. Rev. Lett.} {\bf 118} (2017), no.~5 051801
  [\href{http://arXiv.org/abs/1603.03257}{{\tt 1603.03257}}].

\bibitem{Cata:2005zj}
O.~Cata, M.~Golterman and S.~Peris, {\it {Duality violations and spectral sum
  rules}},  {\em JHEP} {\bf 08} (2005) 076
  [\href{http://arXiv.org/abs/hep-ph/0506004}{{\tt hep-ph/0506004}}].

\bibitem{Cata:2008ye}
O.~Cata, M.~Golterman and S.~Peris, {\it {Unraveling duality violations in
  hadronic $\tau$ decays}},  {\em Phys. Rev.} {\bf D77} (2008) 093006
  [\href{http://arXiv.org/abs/0803.0246}{{\tt 0803.0246}}].

\bibitem{Cata:2008ru}
O.~Cata, M.~Golterman and S.~Peris, {\it {Possible duality violations in $\tau$
  decay and their impact on the determination of $\alpha_s$}},  {\em Phys.
  Rev.} {\bf D79} (2009) 053002 [\href{http://arXiv.org/abs/0812.2285}{{\tt
  0812.2285}}].

\bibitem{Boito:2011qt}
D.~Boito, O.~Cata, M.~Golterman, M.~Jamin, K.~Maltman, J.~Osborne and S.~Peris,
  {\it {A new determination of $\alpha_s$ from hadronic $\tau$ decays}},  {\em
  Phys. Rev.} {\bf D84} (2011) 113006
  [\href{http://arXiv.org/abs/1110.1127}{{\tt 1110.1127}}].

\bibitem{Colangelo:2000dp}
P.~Colangelo and A.~Khodjamirian, {\it {QCD sum rules, a modern perspective}},
  \href{http://arXiv.org/abs/hep-ph/0010175}{{\tt hep-ph/0010175}}.

\bibitem{Bell:2013tfa}
G.~Bell, T.~Feldmann, Y.-M. Wang and M.~W.~Y. Yip, {\it {Light-Cone
  Distribution Amplitudes for Heavy-Quark Hadrons}},  {\em JHEP} {\bf 11}
  (2013) 191 [\href{http://arXiv.org/abs/1308.6114}{{\tt 1308.6114}}].

\bibitem{Dimou:2012un}
M.~Dimou, J.~Lyon and R.~Zwicky, {\it {Exclusive Chromomagnetism in
  heavy-to-light FCNCs}},  {\em Phys. Rev.} {\bf D87} (2013), no.~7 074008
  [\href{http://arXiv.org/abs/1212.2242}{{\tt 1212.2242}}].

\bibitem{DeFazio:2005dx}
F.~De~Fazio, T.~Feldmann and T.~Hurth, {\it {Light-cone sum rules in
  soft-collinear effective theory}},  {\em Nucl. Phys.} {\bf B733} (2006) 1--30
  [\href{http://arXiv.org/abs/hep-ph/0504088}{{\tt hep-ph/0504088}}]. [Erratum:
  Nucl. Phys.B800,405(2008)].

\bibitem{DeFazio:2007hw}
F.~De~Fazio, T.~Feldmann and T.~Hurth, {\it {SCET sum rules for $B\to P$ and $B
  \to V$ transition form factors}},  {\em JHEP} {\bf 02} (2008) 031
  [\href{http://arXiv.org/abs/0711.3999}{{\tt 0711.3999}}].

\end{thebibliography}\endgroup
%
%

\end{document}